\documentclass[11pt,prd,aps,amsfonts,eqsecnum,superscriptaddress,nofootinbib,longbibliography,notitlepage]{revtex4-1}

\usepackage{lmodern}    
\usepackage{microtype}  
\usepackage[english]{babel}
\usepackage{setspace}
\usepackage{enumitem}
\usepackage{fancyhdr}
\usepackage{dsfont}

\usepackage{caption}
\usepackage{graphicx}
\usepackage{xcolor}
\usepackage{subcaption}
\usepackage{rotating}
\usepackage{amsmath, amssymb, graphics, amsthm, isomath}
\usepackage{physics}
\usepackage{verbatim}
\usepackage{tikz}
\usepackage{tikz-cd}
\usepackage{circuitikz}
\usetikzlibrary{patterns}
\usetikzlibrary{through,hobby}
\usetikzlibrary{decorations.markings}
\usetikzlibrary{fadings,shadings}
\usetikzlibrary{plotmarks}
\usetikzlibrary{positioning}
\usetikzlibrary{decorations,arrows}
\usetikzlibrary{decorations.pathreplacing}
\usetikzlibrary{chains, scopes, positioning, backgrounds, shapes, fit, shadows, calc, arrows.meta, decorations.pathreplacing}
\usepackage{blkarray}
\usepackage{svg}
\usepackage{mathrsfs}
\usepackage{algorithm}
\usepackage{algpseudocode}

\usepackage{xcolor}
\definecolor{mycitecolor}{rgb}{0.0, 0.45, 0.85}   

\DeclareMathOperator{\dom}{dom}

\usepackage[
  colorlinks=true,
  linkcolor=mycitecolor,
  citecolor=mycitecolor,
  urlcolor=mycitecolor,
  hyperindex=true,
  linktocpage=true
]{hyperref}

\usepackage[capitalise,compress]{cleveref}
\usepackage{tcolorbox}

\newtcolorbox{shadedtheorem}{
  colback=gray!15,    
  colframe=white,     
  boxrule=0pt,        
  arc=0pt,            
  left=5pt,           
  right=5pt,
  top=5pt,
  bottom=5pt
}

\newtheoremstyle{upright-hang}
  {6pt}   
  {6pt}   
  {\normalfont\hangindent=1.5em\hangafter=1\relax} 
  {0pt}   
  {\bfseries} 
  {.}     
  {0.5em} 
  {}      
\usepackage{thm-restate}
\theoremstyle{upright-hang}
\newtheorem{thm}{Theorem}
\numberwithin{thm}{section}

\renewcommand{\thesection}{\arabic{section}}
\renewcommand{\thesubsection}{\thesection.\arabic{subsection}}
\renewcommand{\thesubsubsection}{\thesubsection.\arabic{subsubsection}}

\makeatletter
\renewcommand{\p@subsection}{}
\renewcommand{\p@subsubsection}{}
\makeatother

\makeatletter
\fancypagestyle{plain}{%
  \fancyhf{}%
  \fancyfoot[C]{\thepage}%
}
\fancypagestyle{revtexfooter}{%
  \fancyhf{}%
  \fancyfoot[C]{\thepage}%
}
\makeatother
\AtBeginDocument{\setstretch{1.25}\pagestyle{revtexfooter}}
\makeatletter
\renewcommand\frontmatter@title@format{\LARGE\bfseries\centering\parskip\z@skip}

\def\frontmatter@affiliationfont{\normalfont\selectfont}
\newcommand{\AndySectionStar}[1]{%
  \par\addpenalty\@secpenalty
  \addvspace{3.5ex \@plus 1ex \@minus .2ex}%
  \noindent{\normalfont\Large\bfseries #1\par}%
  \nobreak\vspace{2.3ex \@plus .2ex}%
  \@afterheading}
\newcommand{\AndySubsectionStar}[1]{%
  \par\addpenalty\@secpenalty
  \addvspace{3.25ex \@plus 1ex \@minus .2ex}%
  \noindent{\normalfont\large\bfseries #1\par}%
  \nobreak\vspace{1.5ex \@plus .2ex}%
  \@afterheading}
\newcommand{\AndySubsubsectionStar}[1]{%
  \par\addpenalty\@secpenalty
  \addvspace{3.25ex \@plus 1ex \@minus .2ex}%
  \noindent{\normalfont\normalsize\bfseries #1\par}%
  \nobreak\vspace{1.5ex \@plus .2ex}%
  \@afterheading}
\renewcommand\@seccntformat[1]{\csname the#1\endcsname\quad}
\def\@hangfrom@section#1#2#3{\@hangfrom{#1#2}#3}
\def\@hangfroms@section#1#2{#1#2}
\renewcommand\section{\@ifstar{\AndySectionStar}{\@startsection {section}{1}{\z@}%
  {-3.5ex \@plus -1ex \@minus -.2ex}%
  {2.3ex \@plus .2ex}%
  {\normalfont\Large\bfseries}}}
\renewcommand\subsection{\@ifstar{\AndySubsectionStar}{\@startsection{subsection}{2}{-\parindent}%
  {-3.25ex\@plus -1ex \@minus -.2ex}%
  {1.5ex \@plus .2ex}%
  {\normalfont\large\bfseries}}}
\renewcommand\subsubsection{\@ifstar{\AndySubsubsectionStar}{\@startsection{subsubsection}{3}{-\parindent}%
  {-3.25ex\@plus -1ex \@minus -.2ex}%
  {1.5ex \@plus .2ex}%
  {\normalfont\normalsize\bfseries}}}
\renewcommand*\l@section[2]{%
  \ifnum \c@tocdepth >\z@
    \addpenalty\@secpenalty
    \addvspace{0.75em \@plus\p@}%
    \begingroup
      \parindent\z@
      \rightskip\@pnumwidth
      \parfillskip-\@pnumwidth
      \leavevmode\bfseries
      \setlength\@tempdima{2.3em}%
      \advance\leftskip\@tempdima
      \hskip-\leftskip
      #1\nobreak\hfil\nobreak\hb@xt@\@pnumwidth{\hss #2}\par
    \endgroup
  \fi}
\renewcommand*\l@subsection{\@dottedtocline{2}{1.5em}{3.2em}}
\renewcommand*\l@subsubsection{\@dottedtocline{3}{3.8em}{4.1em}}
\makeatother

\usepackage{mathtools}
\tikzstyle{densely dashed}= [dash pattern=on 4pt off 3pt]

\usepackage{dsfont}

\allowdisplaybreaks

\begin{document}

\title{Entropic-ordered superfluid}

\author{Isabella Zane}
\email{isabella.zane@colorado.edu}
\affiliation{Department of Physics and Center for Theory of Quantum Matter, University of Colorado, Boulder, CO 80309, USA}

\author{Andrew Lucas}
\email{andrew.j.lucas@colorado.edu}
\affiliation{Department of Physics and Center for Theory of Quantum Matter, University of Colorado, Boulder, CO 80309, USA}

\begin{abstract}
    We present a minimal classical lattice model of superfluidity at arbitrarily high temperatures, using a modified three-dimensional XY model with ``soft" rotors.    The global U(1) symmetry is spontaneously broken above a positive critical temperature due to entropic order.  We study hydrodynamics in this high-temperature regime and verify consistency with the fluctuation-dissipation theorem, explaining how the large noise variance at very high temperatures is not sufficient to destroy superfluidity or the long lifetime of Goldstone modes.
\end{abstract}

\maketitle

\tableofcontents

\section{Introduction}
\label{sec:introduction}
At high temperature, entropy dominates the minimization of the free energy, and conventional statistical mechanics associates this regime with disordered phases. With that said, it is well-known that sometimes one can heat from a disordered phase into an ordered phase at higher temperatures \cite{Plazanet2004,Radzihovsky2001}, as in the Pomeranchuk effect in liquid helium \cite{Pomeranchuk1950}, or in Rochelle salts \cite{Kao2004}, or in the theory of ``order by disorder" in magnets \cite{Villain1980}. However, in each of these cases, once the system is sufficiently heated, the system again disorders.  In fact, classical ``no-go" theorems for interacting lattice models of spins state that for many-body systems which have bounded, local interactions, there is a critical temperature such that above it, no long-range order can persist \cite{Dobrushin1970,Simon1993}. So for conventional systems in statistical mechanics, we do not expect to have long-range order at arbitrarily high temperatures.

Recent work \cite{Han2026} showed how to avoid this no-go theorem in classical lattice models with interacting ``bosons" (i.e. integer-valued degrees of freedom).   In these models, typical high-energy states can be ordered, allowing an ordered phase to persist to arbitrarily high temperatures.  The explicit mechanism, which was called ``entropic order", is analogous to the Pomeranchuk effect:  by ordering one degree of freedom, the energy cost of bosonic fluctuations is reduced, meaning that the entropy of the ordered phase is larger.   Since bosons can have unbounded fluctuations, this effect never disappears at high temperature.  This work provided a unified principle explaining high-temperature order that has been observed in quantum field theory \cite{Weinberg1974,Chai2020PRD,Chai2020PRL,KomargodskiPopov2025,Hawashin2025,Buchel2021a,Buchel2021b,Agrawal2021,Chaudhuri2021,Nakayama2021,Bajc2021,ChaudhuriRabinovici2021,Chai2021biconical,Liendo2023,Hawashin:2025ikp,Bajc:2026ppk,Smolkin:2026wij,Smolkin:2026bxa}.  

It was already shown in \cite{Han2026} that essentially any phase of matter can be stabilized at high-enough temperature, given a creative enough lattice model. However, so far, the focus has been on static (equilibrium) properties of theories with entropic order. The purpose of this paper is to understand the implications of entropic order on dynamics. We will explore a minimalist example, corresponding to  the hydrodynamics of an entropic-ordered superfluid that persists to arbitrarily high temperature. The reason this question is intriguing is that at finite temperature, hydrodynamics is subject to a fluctuation-dissipation theorem \cite{Kubo1966}, which (naively) states that the noise strength is increasingly strong  compared to dissipation at increasing temperatures.  When the entropic order persists to arbitrarily high temperature, it is important to understand whether there is a pathological nature to the dynamics, or whether hydrodynamic modes such as Goldstone bosons \cite{Goldstone1961, GoldstoneSalamWeinberg1962, Nambu1960} continue to be well-defined and long-lived. Using the effective field theory of hydrodynamics \cite{Crossley2017,Glorioso2018,Dubovsky2012,Glorioso2017}, we expect that these Goldstones have dispersion relations of the form $\omega = \pm vk - \mathrm{i}\Gamma k^2 + \cdots $; does the fluctuation-dissipation theorem imply that $\Gamma$ diverges as $T\rightarrow \infty$?  This could mean that from a dynamical perspective, the high-temperature superfluid is pathological, even if from a static perspective it exists!


In order to answer this question, we will introduce a new and simpler lattice model for entropic-ordered superfluidity, relative to the earlier work \cite{Han2026}. In particular, in this previous work on classical entropic order, the entropic order is always driven by coupling to integer-valued degrees of freedom.  This complicates the analysis of hydrodynamics classically, however, since the microscopic dynamics in continuous time would require a Markov process governing the discrete jumps in the integer-valued field.  We will find it much more natural to instead build a model of entropic-ordered superfluidity made entirely out of continuous variables, which can arise as a symplectic (Hamiltonian) dynamical system on phase space.  Clearly, in this setting, it will be much simpler to directly analyze the classical hydrodynamics in the superfluid phase.  To achieve such a construction, we will build an XY-model \cite{Stanley1968,Campostrini2001} on a three-dimensional cubic lattice (just for simplicity) made out of ``soft" rotors: rather than simply adding a single angular degree of freedom $\theta_v$ at each lattice site, we will also add a ``length" $r_v$ to the rotor.  The fluctuations in $r_v$ can drive entropic-ordered superfluidity, where the global angular shift symmetry $\theta_v \rightarrow \theta_v + c$ is spontaneously broken. In this high-temperature superfluid phase, we will find that long-range order in $\theta_v$ will enhance fluctuations in $r_v$, meaning that as in \cite{Han2026}, the ordered phase will in fact have the highest entropy, and therefore exist at high enough temperature.

The remainder of the paper is organized as follows.  In Section \ref{sec:soft_rotor_model_and_MF}, we introduce a minimal model of an entropic-ordered superfluid, consisting of a large $N$ number of coupled soft rotors, and study the phase diagram as well as dynamics in a mean-field limit of the model.  Section \ref{sec:Cluster_lattice} then describes the overdamped relaxational dynamics of a $d$-dimensional lattice of these large-$N$ ``clusters"; superfluidity is a stable phase in $d>2$, as per the Mermin-Wagner Theorem \cite{MerminWagner1966,Hohenberg1967}. In Section \ref{sec:Superfluid_model}, we generalize our model of superfluidity to include both a conserved U(1) charge along with a dynamical Goldstone boson, and discuss the peculiar thermodynamics of the resulting models; Section \ref{sec:dynamics_of_propagating_SF_model} returns to the superfluid hydrodynamics of this model and demonstrates the existence of a superfluid sound mode.

\section{The soft rotor model and a mean-field analysis}
\label{sec:soft_rotor_model_and_MF}
In this section, we will introduce the simplest possible version of our entropic-ordered superfluid by considering a model of all-to-all interacting soft rotors.  
\subsection{Review of the XY model}
\label{subsec:XY_model_review}

First, let us review the mean-field (all-to-all connected) XY-model \cite{Dauxois2002HMF,Antoni1995}. The Hamiltonian is given by 
\begin{align}
\label{eq:XY_model_Hamiltonian}
    H = -\frac{J}{2N}\sum_{i \neq j}\vec{s}_i\cdot \vec{s}_j = -\frac{J}{2N}\sum_{i\neq j}\cos(\theta_i-\theta_j)
\end{align}
where $|\vec{s}_i| =1$ and $\vec{s}_i = (\cos\theta_i,\sin\theta_i)$. The order parameter is given by the  magnetization $m = |\bar{s}|$ where $\vec{\bar{s}} = \frac{1}{N}\sum_i \vec{s_i}$. In this model, the thermal fluctuations compete against the long-range ferromagnetic interactions (strength $J$). The free energy density, defined here as the free energy per rotor, can be written in the form
\begin{align}
    f(m)=u(m)-\frac{1}{\beta}s(m), \qquad u(m)=-\frac{J}{2}m^2,\qquad s(m)=\min_h\left[-hm+\log\left(2\pi I_0(h)\right)\right],
\end{align}
where the auxiliary field is fixed by the saddle point condition $m=I_1(h)/I_0(h)$, which can be solved to determine $h(m)$. This form makes the competition between energy and entropy explicit. The energy density is minimized at $m=1$, favoring an ordered state. By contrast, the entropy is largest at $m=0$. Next, $\mathrm{d}s/\mathrm{d}m=-h(m)$ and, since the saddle-point condition implies $h>0$ for $m>0$, we have that $s(m)$ decreases monotonically with $m$. Thus entropy favors the disordered state, while the interaction energy favors alignment. To see explicitly how this energy-entropy competition determines the equilibrium state, we use that in equilibrium \begin{equation}
    0 = \frac{\mathrm{d}f}{\mathrm{d}m} = -Jm + \frac{h(m)}{\beta}.
\end{equation}
Plugging back into $f(m)$, we find
\begin{align}
    f(m) = \frac{J}{2}m^2 - \frac{1}{\beta}\log(2\pi I_0(\beta J m))
\end{align}
The equilibrium magnetization is then obtained by minimizing this expression with respect to $m$. At high temperatures, $\beta \to 0$, and the small argument approximation of the modified Bessel function of the first kind is,
\begin{align}
    I_0(x) = 1 + \frac{x^2}{4} + \mathcal{O}(x^4)
\end{align}
Using also $\log(1+x) \approx x$ for small $x$, at high temperatures, the free energy becomes 
\begin{align}
    f(m) \approx \frac{Jm^2}{2} \left(1 - \frac{\beta J}{2}\right) - \frac{\log 2\pi}{\beta} 
\end{align}
Notice that the quadratic coefficient changes sign at $\beta_c = 2/J$. So for $\beta < \beta_c$, the disordered state $m=0$ is the stable free energy minimum. Above $\beta_c$, $m=0$ becomes unstable and the minimum shifts to $m>0$. Thus, the critical temperature for this phase transition is given by $T_c = J/2$. Now let us consider the low temperature $(\beta \to \infty)$ limit. For large $x$, 
\begin{align}
    I_0(x) \sim \frac{e^x}{\sqrt{2\pi x}}
\end{align}
The free energy density at low temperature is 
\begin{align}
    f(m) = \frac{J}{2}m^2 -Jm + \mathcal{O}(\beta^{-1})
\end{align}
This is minimized at $m=1$. So for low temperatures, the interaction strength wins and, despite the entropic penalty, free energy is minimized by alignment.

\subsection{Our model}
\label{subsec:Introducing_soft_rotor_model}
We now turn to a soft rotor modification of the XY model that could realize entropic order. By ``soft rotor" we mean that the length of each rotor is no longer fixed. We can describe this system by the following Hamiltonian where each rotor is described by polar coordinates $(r_i,\theta_i)$ where $r_i\geq0, \;\theta_i \in [0,2\pi)$. 
\begin{align}
\label{eq:microscopic_overdamped_Hamiltonian}
    H = \frac{1}{2N}\sum_{i \neq j}r_ir_j(a-\cos(\theta_i-\theta_j)) + b\sum_i r_i
\end{align}
In order to have a well-defined theory, the partition function \begin{equation}
    Z(\beta) = \int \prod_i r_i \mathrm{d}r_i \mathrm{d}\theta_i \; \mathrm{e}^{-\beta H}
\end{equation} must be well-defined and finite, and we take \begin{equation}
    a\geq 1, \;\;\; b>0.  \label{eq:ab_constraints}
\end{equation}
If $b=0$, then this system would have an infinitely-degenerate ground state, since a single rotor can be moved away from the origin at no cost, and the critical temperature found below would vanish; if $a<1$ then there is no energy penalty for unbounded fluctuations $r\rightarrow \infty$, causing the partition function to diverge.

Assuming \eqref{eq:ab_constraints}, we can easily see that the ground state of $H$ is unique and corresponds to all $r_i=0$.\footnote{We remind the reader that the coordinate degeneracy in polar coordinates means that this point is indeed a unique point in the classical configuration space.}  Therefore, at $T=0$, the system is in a disordered phase -- the U(1) symmetry is not broken.

When $T>0$, thermal fluctuations introduce noise into the system which causes the length of the rotors to be kicked around. Like in the Curie-Weiss XY-model, the thermal fluctuations compete against the interaction term which favors alignment. However, since the magnitude of each rotor is allowed to fluctuate, the effective interaction strength is no longer fixed. This changes the straightforward logic that thermal fluctuations inevitably lead to  a disordered phase; we will in fact see that these fluctuations lead to the entropic-ordered superfluid. Indeed, when the angles align [$\cos(\theta_i-\theta_j)\approx 1$], the radial interaction term is minimized which decreases the energetic penalty for large radial fluctuations in $r_i$. When $r_i$ fluctuate more strongly, a larger volume of phase space is explored.

\subsection{Phase diagram of the model}
\label{subsec:phase_diagram_of_model}
To quantitatively determine when the enhanced phase space explored by $r_i$ is enough to lead to entropic order, we first determine the stationary mean-field branches exactly in the $N\rightarrow\infty$ limit. Their thermodynamic stability will be established from the free energy at the end of this subsection. Each rotor effectively sees only the average (mean-field) configuration of all other rotors. It is useful to associate to each rotor a two-component vector
\begin{equation}
    \vec{s}_i = r_i(\cos\theta_i,\sin\theta_i).
\end{equation}
Similarly, we define the mean-field
\begin{align}
    \bar{r} = \frac{1}{N}\sum_i r_i
    \quad \text{and}\quad
    \vec{\bar{s}} = \frac{1}{N}\sum_i \vec{s}_i
\end{align}
Here, $\bar{r}$ is the average rotor length, while $\vec{\bar{s}}$ is the vector order parameter. Its magnitude $\bar{s}=|\vec{\bar{s}}|$ measures the degree of angular ordering, and its phase $\bar{\theta}=\arg(\vec{\bar{s}})$ gives the average direction of the rotors. In the disordered phase, the angular average vanishes, so $\vec{\bar{s}}=0$.  In the ordered phase, we have $\vec{\bar{s}}\ne \vec 0$, and without loss of generality we assume that $\vec{\bar{s}} = (\bar s, 0)$.  Then as $N\rightarrow \infty$, each rotor $i$ sees an effective single-particle Hamiltonian
\begin{align}
    H_i = (a\bar{r}+b)r_i - \bar{s}r_i\cos\theta_i.
\end{align}

We now deduce the Gibbs state by self-consistently determining $\bar r$ and $\bar s$:\begin{subequations}
    \begin{align}
    \bar{r} &= \frac{1}{\bar Z}\int  r \mathrm{d}r \mathrm{d}\theta \; r \mathrm{e}^{-\beta [(a\bar r +b)r-\bar s r\cos\theta]},  \\  \bar{s} &= \frac{1}{\bar Z}\int  r \mathrm{d}r \mathrm{d}\theta \; r\cos\theta \mathrm{e}^{-\beta [(a\bar r +b)r-\bar s r\cos\theta]},
\end{align}
\end{subequations} 
where \begin{equation}
    \bar Z = \int r \mathrm{d}r \mathrm{d}\theta \; \mathrm{e}^{-\beta [(a\bar r +b)r-\bar s r\cos\theta]}.
\end{equation} 
Performing the integrations explicitly, the self-consistency conditions reduce to
\begin{subequations}
\label{eq:self_consistency_overdamped}
\begin{align}
    \bar{r} &= \frac{2(a\bar{r}+b)^2+\bar{s}^2}{\beta(a\bar{r}+b)\left((a\bar{r}+b)^2-\bar{s}^2\right)}, \\
    \bar{s} &= \frac{3\bar{s}}{\beta \left((a\bar{r}+b)^2-\bar{s}^2\right)}.
\end{align}\end{subequations}
The equation for $\bar{s}$ implies two possible stationary branches. Either $\bar{s}=0$, corresponding to the isotropic disordered phase, or
\begin{align} \label{eq:nicerelation}
    (a\bar{r}+b)^2-\bar{s}^2 = \frac{3}{\beta}
\end{align}
which permits solutions with $\bar{s}\neq 0$. In the latter case, the angular order parameter acquires a nonzero expectation value, and the $\mathrm{U}(1)$ symmetry is spontaneously broken. For the isotropic branch, the self-consistency equations give
\begin{align}
\label{eq:isotropic_solution_overdamped}
    \bar{r}_{\mathrm{eq}} =\frac{-b+ \sqrt{8aT+b^2}}{2a}, \quad \bar{s}_{\mathrm{eq}} =0
\end{align}
For the symmetry-broken stationary branch, when $a>1$, one obtains
\begin{subequations}
\begin{align}
    \bar{r}_{\mathrm{eq}}  &= \frac{b-2ab + \sqrt{b^2+4aT(a-1)}}{2a(a-1)}\\
    \bar{s}_{\mathrm{eq}} &= \sqrt{\frac{b^2-2T(2a^2-5a+3)-b\sqrt{b^2+4aT(a-1)}}{2(a-1)^2}}
         \end{align}
\end{subequations}
We will comment on the $a=1$ case at the end of this section. The ordered stationary solution above exists when $1<a<3/2$ and 
\begin{align}
\label{eq:critical_temp_overdamped}
    T > T_{\mathrm{c}} = \frac{3b^2}{(3-2a)^2}.
\end{align}
Solving the self-consistency equations determines the stationary mean-field branches, but does not by itself establish which branch globally minimizes the free energy. In Appendix \ref{sec:Appendix_free_energy_overdamped}, we calculate the free energy density as a function of the order parameter $\bar{s}$ and show that the transition at $T=T_c$ is continuous. The disordered solution $\bar{s}=0$ is stable for $T<T_c$, becomes unstable at $T>T_c$, and is replaced continuously by the stable symmetry-broken branch with $\bar{s}\neq 0$.

Having established the thermodynamic stability of these branches, we can now interpret the dependence of the phase boundary on the parameters $a$ and $b$. The parameter $b$ must be sufficiently small relative to $T$ to keep the energy cost of large radius favorable. Since the minimal energetic penalty for large radii in the interaction term is given by the factor $(a-1)$, the parameter $a$ must be small enough to keep the minimal energy cost from suppressing $r_i$ too much. So if we are above the critical value $a_{\mathrm{c}} = 3/2$, then the energetic penalty is too high and the state that has the most available phase space is the pure disordered state.

Finally, the temperature dependence of the self-consistent solution makes the entropic-ordering mechanism explicit. At large $T$, the isotropic branch scales as
\begin{align}
    \bar{r}_{\mathrm{eq}} \sim \sqrt{\frac{2T}{a}}, \quad \bar{s}_{\mathrm{eq}}=0,
\end{align}
whereas in the symmetry-broken branch,
\begin{align}
    \bar{r}_{\mathrm{eq}} \sim \sqrt{\frac{T}{a(a-1)}}, \quad 
    \bar{s}_{\mathrm{eq}} \sim \sqrt{\frac{(3-2a)T}{a-1}} .
\end{align}
Thus, whenever the symmetry-broken solution exists, both the average rotor length and the angular order parameter grow as $\sqrt{T}$. This scaling shows that increasing temperature does not simply disorder the angles, but also increases the typical rotor length in a way that supports angular ordering.

Finally, we comment on the marginal case $a=1$. Taking the $a\to 1$ limit directly in \eqref{eq:self_consistency_overdamped}, and solving the self consistency equations yields the following anisotropic stationary solution
\begin{align}
    \bar{r} = \frac{T}{b}-b, \qquad \bar{s}_{\text{eq}}= \frac{\sqrt{T(T-3b^2)}}{b}
\end{align}
A symmetry broken stationary branch exists when 
\begin{align}
    b>0, \quad T>T_c = 3b^2
\end{align}
The isotropic solution is unchanged and is obtained by setting $a=1$ in \eqref{eq:isotropic_solution_overdamped}. Notice that the ordered branch exhibits different scaling in the $a=1$ case; $\bar{r},\bar{s}_{\text{eq}} \sim T$ compared to the $a>1$ scaling where $\bar{r},\bar{s}_{\text{eq}}\sim \sqrt{T}$. We therefore exclude the end point $a=1$ for simplicity and restrict to the $a>1$ regime in what follows.

\subsection{Dynamics}
\label{subsec:dynamics_of_soft_rotor_model}
Since we are ultimately interested in understanding the dynamics of entropic-ordered superfluids, we must now describe how to calculate the dynamics of this XY model out of equilibrium.  For the discussion, it is useful to work in rectangular, not polar, coordinates for each soft rotor.  We take the microscopic dynamics to be overdamped Langevin dynamics:
\begin{subequations}\label{eq:langevin}
\begin{align}
\label{eq:Langevin_eqns_overdamped}
    \mathrm{d}x_i &= -\gamma \frac{\partial H}{\partial x_i}\mathrm{d}t + \sqrt{2T\gamma}\mathrm{d}W_i^x \\
    \mathrm{d}y_i &= -\gamma \frac{\partial H}{\partial y_i}\mathrm{d}t + \sqrt{2T\gamma}\mathrm{d}W_i^y,
\end{align}
\end{subequations}
with independent Brownian motions $W_i^x$ and $W_i^y$.  A priori, in the large $N$ limit, we must solve a very high-dimensional Fokker-Planck equation for the relative probabilities of all of the rotors' configurations.  Thankfully, there is an extensive mathematical literature \cite{McKean1966,McKean1967PropagationChaos,Sznitman1991,JabinWang2018,WangZhaoZhu2024GaussianFluctuations} which shows that in the large $N$ limit of interest in this paper, this equation can instead be replaced with a nonlinear Fokker-Planck-like equation.

 To see how this works in the context of our model, denote \begin{equation}
    P_N(x,y,t) = \frac{1}{N}\sum_{i}\delta(x-x_i(t))\delta(y-y_i(t))
\end{equation} 
as the probability density of the $N$ rotors' configurations.  As $N\rightarrow \infty$ this will increasingly accurately approximate a smooth distribution function, i.e. $P_N\to P$. Moreover, we assert that $P(x,y,t)$ obeys a deterministic and closed nonlinear equation as $N\rightarrow\infty$: 
\begin{align}
\label{eq:Nonlin_PDE}
    \partial_t P = \gamma\partial_x( P(\partial_x \bar{H}) [P])+\gamma\partial_y( P (\partial_y \bar{H})[P] )+ \gamma T \partial_x^2P + \gamma T\partial_y^2P,
\end{align}
where, with $r=\sqrt{x^2+y^2}$,
\begin{align} \label{eq:barH}
    \bar{H}[x,y,P] := (a\bar{r}[P]+b)r -\bar{x}[P]x-\bar{y}[P]y
\end{align}
The moments are defined by
\begin{subequations}
    \begin{align}
    \bar{r}[P] &= \int \mathrm{d}^2x 
    \,P r, \\
    \bar{x}[P] &= \int \mathrm{d}^2x \,P x, \\
    \bar{y}[P] &=\int \mathrm{d}^2x\, P y.
\end{align}
\end{subequations}
In evaluating $\partial_x\bar H$ and $\partial_y\bar H$ in \eqref{eq:Nonlin_PDE}, notice that $\bar r[P]$, $\bar x[P]$ and $\bar y[P]$ do not have explicit $(x,y)$-coordinate dependence, therefore the derivative does not act on them: e.g. \begin{equation}
    \partial_x \bar H = (a\bar r+b)\frac{x}{\sqrt{x^2+y^2}} - \bar x.
\end{equation}
The intuition behind \eqref{eq:Nonlin_PDE} is as follows:  each individual rotor sees an effective mean-field Hamiltonian of the form \eqref{eq:barH} which is effectively a single-particle potential for the individual particle.  On the other hand, we must self-consistently determine $\bar H$ based on the position of every other particle, which is deduced in terms of $P$ itself, making the equation nonlinear in $P$.

We now solve for the stationary solutions of \eqref{eq:Nonlin_PDE}. For a fixed
mean-field Hamiltonian $\bar{H}$, the stationary solution has the Gibbs form. 
In the present problem, of course, $\bar{H}$ is not fixed externally but instead depends on the distribution $P$ through the collective moments. Therefore equilibrium
is determined by the nonlinear fixed-point equation
\begin{align}
    P_{\text{eq}}= \frac{\mathrm{e}^{-\beta \bar{H}[P_{\text{eq}}]}}{Z[P_{\text{eq}}]}.
\end{align}
This exactly matches the solution that we obtained in Section \ref{subsec:phase_diagram_of_model}.

\subsection{Linearized equations of motion}
\label{subsec:linear_MF_PDE}
We now linearize the deterministic mean-field PDE \eqref{eq:Nonlin_PDE} about the stationary solution $P_{\text{eq}}$. In what follows, we focus on the dynamics in the ordered phase where $\bar{s}_{\mathrm{eq}}\neq 0$. This is the regime corresponding to the superfluid phase; the disordered branch has no broken $\mathrm{U}(1)$ symmetry and there will be no slow degrees of freedom remaining.

It will be useful to write the linearized equation in weak form using the natural weighted $\mathrm{L}^2(P_{\text{eq}})$ inner product associated with the equilibrium measure: 
\begin{align} \label{eq:innerproduct}
    \langle f|g\rangle = \int \mathrm{d}^2 x\beta P_{\mathrm{eq}} \; fg
\end{align}
Now, suppose that the system is perturbed by a small $\phi$ from the stationary distribution. Up to normalization, the perturbed distribution is
\begin{align}
    P = \mathrm{e}^{-\beta (\bar{H}_{\mathrm{eq}}-\phi)} \approx \mathrm{e}^{-\beta \bar{H}_{\mathrm{eq}}} \left(1+\beta \phi\right)
\end{align}
The component of $\phi$ along the constant function changes only the total normalization of the distribution. Indeed, for the linearized ansatz,
\begin{align}
    \int \mathrm{d}^2x P \approx \int \mathrm{d}^2x e^{-\beta \bar{H}_{\mathrm{eq}}} + \beta \int \mathrm{d}^2x\, e^{-\beta \bar{H}_{\mathrm{eq}}}\phi  = Z_{\mathrm{eq}}(1+\langle \phi|1\rangle )
\end{align}
Therefore, $\langle \phi|1\rangle$ represents an unphysical mode rather than a genuine fluctuation of the state. We may therefore quotient out this direction, or equivalently impose the normalization condition
\begin{align}
\langle \phi|1\rangle =0
\end{align}
without loss of generality. With this convention, the perturbed distribution is $P\approx P_{\mathrm{eq}}(1+\beta \phi)$. We now look at how the collective variables shift under the perturbed measure. Notice that 
\begin{align}
   \beta  \bar{r}_{\mathrm{eq}} =  \langle r|1\rangle,
\end{align}
with analogous formulas $\beta \bar{x}_{\mathrm{eq}} = \langle x|1\rangle$ and $\beta \bar{y}_{\mathrm{eq}} = \langle y|1\rangle$. The perturbed values of the collective variables about the stationary distribution are then given by:
\begin{subequations}
\begin{align}
    \bar{r}[P] &= \frac{\int \mathrm{d}^2x P_{\mathrm{eq}}(1+\beta \phi)r}{\int \mathrm{d}^2xP_{\mathrm{eq}}(1+\beta \phi)} \approx \bar{r}_{\mathrm{eq}} + \langle \phi|r\rangle, \\
    \bar{x}[P]&= \bar{x}_{\mathrm{eq}} + \langle \phi|x\rangle, \\
    \bar{y}[P]&= \bar{y}_{\mathrm{eq}} + \langle \phi|y\rangle .
\end{align}
\end{subequations}
We are now able to compute $\bar{H}[P]$ by plugging in the values of the collective variables with respect to $P$. We find
\begin{align} \label{eq:barH_phi}
   \bar{H}[P]=\bar{H}_{\mathrm{eq}}+ ar \langle \phi|r\rangle  - x\langle \phi|x\rangle  - y\langle \phi|y\rangle =  \bar{H}_{\mathrm{eq}} + \delta \bar{H}(\phi).
\end{align}
For a perturbation with $\langle\phi|1\rangle\neq0$, the collective variables shift by the centered coordinates, $\tilde{r}:=r-\bar{r}_{\mathrm{eq}}$, $\tilde{x}:=x-\bar{x}_{\mathrm{eq}}$ and $\tilde{y}:=y - \bar{y}_{\text{eq}}$, so that  
\begin{align}
\label{eq:deltaH_centered}
    \delta\bar{H}(\phi) = ar\langle\phi|\tilde{r}\rangle - x\langle\phi|\tilde{x}\rangle - y\langle\phi|\tilde{y}\rangle .
\end{align}
This reduces to \eqref{eq:barH_phi} when $\langle\phi|1\rangle=0$, and it gives $\delta\bar{H}(1)=0$. We now linearize the following mean-field PDE about the equilibrium distribution,
\begin{align}
    \partial_t P=\gamma\nabla\cdot(P\nabla \bar H[P])+\gamma T\nabla^2P 
\end{align}
Substituting $P=P_{\rm eq}(1+\beta\phi)$ and $\bar H[P]=\bar H_{\rm eq}+\delta \bar H(\phi)$, 
and keeping only terms linear in $\phi$, we obtain 
\begin{align}
    \beta P_{\rm eq}\partial_t\phi=\gamma\nabla\cdot(P_{\rm eq}\nabla\phi)+\gamma\nabla\cdot(P_{\rm eq}\nabla\delta \bar H(\phi))
\end{align}
To express this equation in weak form, we multiply by a test function $\Psi$ and integrate by parts to get
\begin{align}
    \int d^2x \, \beta P_{\mathrm{eq}}\Psi\partial_t\phi 
    &= -\gamma \int d^2x P_{\mathrm{eq}} \nabla \Psi\cdot \nabla \phi -\gamma \int d^2x P_{\mathrm{eq}}\nabla \Psi \cdot \nabla \delta \bar{H}(\phi)
\end{align}
This becomes
\begin{align}
    \langle \Psi|\partial_t\phi\rangle=-\langle \Psi|W|\phi\rangle,
\end{align}
where the linearized operator $W$ is defined weakly by
\begin{align}
    \langle \Psi|W|\phi\rangle=\frac{\gamma}{\beta}\langle \nabla\Psi|\nabla\phi\rangle+\frac{\gamma}{\beta}\langle \nabla\Psi|\nabla\delta \bar H(\phi)\rangle 
\end{align}

We now identify the zero modes of the linearized mean-field operator $W$:
\begin{align}
    \langle \Psi |W|\phi\rangle=0
\end{align}
for all test functions $\Psi$.  A trivial solution is the unphysical mode $|\phi\rangle = |1\rangle$, as is easy to see by inspection. All remaining zero modes must satisfy
\begin{align}
    \phi + \delta \bar{H}(\phi) = \text{const.}
\end{align}
which implies that $\phi \in \text{span}\{1,r,x,y\}$. To see this, define 
$\mathcal{V}:= \text{span}\{1,y,r,x\}$ and decompose $\phi = \phi_{\parallel}+\phi_{\perp}$ such that $\phi_{\parallel} \in \mathcal{V}$ and $\phi_{\perp} \in \mathcal{V}^{\perp}$. Indeed, from \eqref{eq:barH_phi}, any orthogonal vector $\phi_\perp$ has $\delta \bar H(\phi_\perp) = 0$, and $\langle \phi_\perp|W|\phi_\perp \rangle  \propto \int \mathrm{d}^2x P_{\mathrm{eq}} |\nabla \phi_\perp|^2 >0$.

We now choose the ordered frame so that $\bar{y}_{\mathrm{eq}}=0$. Then the equilibrium distribution is symmetric under $y\to -y$, so
\begin{align}
    \langle r|y\rangle = \langle x|y\rangle = \langle 1|y\rangle = 0.
\end{align}
Consider $|\phi\rangle  = |y\rangle $ as a candidate zero mode:
\begin{align}
    \delta \bar{H}( |y\rangle ) = -y \langle y|y\rangle
\end{align}
where, using \eqref{eq:nicerelation},
\begin{align}
    \langle y|y\rangle = \frac{3}{\beta ((a\bar{r}_{\mathrm{eq}}+b)^2-\bar{x}_{\mathrm{eq}}^2)} = 1.
\end{align}
Thus for $|\phi\rangle = |y\rangle$,
\begin{align}
    \phi + \delta \bar{H}(\phi) =  y - y =0.
\end{align}
A rather tedious calculation, provided in Appendix \ref{subsec:Appendix_kernel_W}, shows that $|r\rangle$  and $|x\rangle$ are not null vectors.  This is indeed expected on physical grounds: $|y\rangle$ corresponds to the ``Goldstone boson" of the superfluid in this mean-field model, which is expected to be the only gapless mode.  

\subsection{Finite-$N$ corrections: diffusion of the order parameter}
\label{subsec:Diffusion_Goldstone_mode}
The Goldstone mode $\phi\propto y$ is an exact zero mode of the linearized mean-field operator $W$ in the $N\to\infty$ limit. Therefore, the deterministic PDE \eqref{eq:Nonlin_PDE} exerts no restoring force along this direction, and the microscopic noise self-averages away. At finite $N$ the cancellation is incomplete, and the collective phase diffuses slowly.  In this sense, as in Model A of Hohenberg-Halperin's classification \cite{HohenbergHalperin1977}, there is no true hydrodynamic description at finite $N$, since fluctuations can slowly cause the order parameter to wind around.  Still, we may compute the diffusion constant $D_\theta$ exactly, to leading order in $1/N$, as it vanishes at $N=\infty$.  This will also be a useful warm-up for our later discussion.

The natural collective coordinate is the global phase itself 
\begin{align}
    \bar\theta = \arg\big(\vec{\bar s}\,\big)
    = \arg\Big(\sum_j r_j\,\mathrm{e}^{\mathrm{i}\theta_j}\Big).
\end{align}
The aim of this section is to examine the long time dynamics of the Goldstone mode in equilibrium. All averages below are therefore taken in the stationary ordered state, $\langle\cdot\rangle_\mu$, with $\mu$ the Gibbs measure
\begin{align}
    d\mu = \frac{e^{-\beta H}d^{2N}\vec{x}}{Z}
\end{align}
restricted to the ordered phase. One subtlety deserves a remark. The phase $\bar\theta$ becomes singular near $\vec{\bar s}=0$, and the full finite-$N$ Gibbs measure gives the near-core region enough weight that $\langle\bar{s}^{-2}\rangle$ diverges logarithmically. We therefore do not attempt to characterize the unrestricted long-time winding statistics. In Appendix \ref{sec:Appendix_finite_N_phase_diffusion} we make the ordered regime precise by introducing a reflected basin $\bar{s}\geq s_*$ and prove the finite-$N$ Brownian scaling limit for that stationary process. The reflected and physical trajectories coincide exactly up to the first entrance into the excluded near-core region. The latter is expected to occur on an exponentially long timescale at large $N$, providing the metastable physical interpretation of the reflected calculation.

The evolution of $\bar\theta$ follows from It\^o's formula applied to \eqref{eq:langevin}. Writing $\mathcal{L}$ for the generator of the Langevin dynamics and $A:=\mathcal{L}\bar{\theta}$ as the phase drift, the phase increment decomposes exactly into an accumulated drift and a pure noise term,
\begin{align}
\label{eq:theta_Ito}
    \bar\theta(t)-\bar\theta(0) = \int_0^t A(X_s) ds + M_t,
\end{align}
where $X_s$ denotes the configuration of all $N$ rotors at time $s$. Using $|\nabla_i\bar\theta|^2 = (N\bar s)^{-2}$, we further estimate \begin{equation}
\label{eq:martingale_main_text}
     \left\langle M_t^2\right\rangle = 2\gamma T\Big\langle\sum_i|\nabla_i\bar\theta|^2\Big\rangle t
    = \frac{2\gamma T}{N}\Big\langle\frac{1}{\bar s^{2}}\Big\rangle t .
\end{equation} 
If the drift could be neglected, \eqref{eq:theta_Ito} would immediately give free diffusion of the phase with the ``bare" diffusion constant
\begin{align}
\label{eq:bare_phase_diffusion}
    D_{\theta}^{(0)} = \frac{\gamma T}{N\bar{s}_{\text{eq}}^2},
\end{align}
and this is indeed the answer one obtains by reading off the noise variance along the Goldstone direction. This is not the full answer at finite $N$, however, because the deterministic drift of $\bar\theta$ vanishes only when all rotors have exactly identical length. Otherwise it is a mean-zero force of size $\mathcal{O}(N^{-1/2})$, built from non-Goldstone fluctuations whose correlations are expected to decay on an $\mathcal{O}(1)$ microscopic timescale. Integrated over a time $t$, such a force accumulates a variance of order $t/N$, the same order as the direct noise. Physically, $\mathrm{U}(1)$ symmetry forbids any restoring force along the Goldstone direction, therefore kicks
delivered by the gapped modes are never relaxed away. A fast, gapped force acting on a flat
coordinate contributes to its long-time diffusion analogous to how molecular collisions drive the Brownian motion of a free particle. The gapped sector therefore renormalizes $D_\theta$ at leading order.

In Appendix \ref{sec:Appendix_finite_N_phase_diffusion} we resum the drift exactly using the spectral and martingale corrector methods underlying the Kipnis--Varadhan theory \cite{KipnisVaradhan1986}. We find that the long-time diffusion constant obeys the two-sided bound
\begin{align}
\label{eq:diffusion_renormalization_bound}
    \frac{\gamma T}{NQ_{\text{eq}}}\leq D_{\theta} \leq \frac{\gamma T}{N\bar{s}_{\text{eq}}^2},
\end{align}
up to $\mathcal{O}(N^{-2})$ corrections. Equivalently, we may write
\begin{align}
\label{eq:renormalized_phase_diffusion}
D_{\theta} = (1-\mathcal{C})\frac{\gamma T}{N\bar{s}_{\text{eq}}^2}, \qquad 0\leq \mathcal{C}\leq 1-\frac{\bar{s}_{\text{eq}}^2}{Q_{\text{eq}}} < 1,
\end{align}
where $\mathcal{C}$ is the time-integrated drift autocorrelation in units of the bare noise, see \eqref{eq:Appendix_definition_C}. Note that $Q_{\text{eq}} = \langle r^2\rangle \geq |\langle\vec{s}\rangle|^2 = \bar{s}_{\text{eq}}^2$ automatically, so the two bounds are always compatible. At large $T$ both $\bar{s}_{\text{eq}}^2$ and $Q_{\text{eq}}$ grow linearly in $T$, so both sides of \eqref{eq:diffusion_renormalization_bound} scale as $\mathcal{O}(1/N)$ with temperature-independent coefficients. We return to the physical significance of this in the next subsection.

Finally, we connect this to the linearized analysis where the Goldstone mode appears as $\phi_G \propto y$. The Goldstone mode corresponds to an infinitesimal rotation of the ordered equilibrium distribution. Under a rotation by $\delta\theta$ near the ordered equilibrium with $\vec{\bar{s}}_{\text{eq}}=(\bar{x}_{\text{eq}},0)$,
\begin{align}
    P_{\theta} = \frac{1}{Z}\text{exp}\left[-\beta \left((a\bar{r}_{\text{eq}}+b)r - \bar{x}_{\text{eq}}x\cos\delta \theta-\bar{x}_{\text{eq}}y\sin\delta \theta \right)\right]\approx P_{\text{eq}}\left(1+\beta \bar{x}_{\text{eq}} y\delta \theta \right)
\end{align}
Comparing with $P\approx P_{\rm eq}(1+\beta\phi)$ identifies the Goldstone mode $\phi_G = \bar x_{\rm eq}y\,\delta\theta$. In the chosen frame the transverse direction is $y$, so the only collective moment that shifts is
\begin{align}
    \bar{y}[P_{\theta}] = \bar{s}_{\text{eq}} \delta \theta 
\end{align}
The mode $\phi = y$ and the global phase $\bar{\theta}$ are therefore in the same direction and are related by the constant Jacobian $\bar{s}_{\text{eq}}$. Since $\bar{y}_{\text{eq}} =0$, we have that $d\bar{y} = \bar{s}_{\text{eq}} \delta \theta $. Additionally, the rotation parameter $\delta\theta$ is precisely the increment $d\bar\theta$. Therefore, 
\begin{align}
\label{eq:y_relation_goldstone}
    \langle d\bar{y}^2\rangle = \bar{s}_{\text{eq}}^2 \langle d\bar{\theta}^2 \rangle  = \frac{2T\gamma}{N}dt
\end{align}
One may verify this directly from $|\nabla_i \bar{y}|^2 = 1/N^2$. This consistency reflects the normalization $\langle y|y\rangle =1$ which follows exactly from equilibrium self consistency conditions. This states that $|y\rangle $ is the unit-normalized Goldstone direction in the $L^2(P_{\text{eq}})$ inner product. The drift along $\bar{y}$ does not vanish on every configuration at finite $N$. Its projection onto the Goldstone manifold vanishes identically by symmetry, so it produces no restoring force on the phase. However, its time-integrated fluctuations are precisely the gapped-mode contribution that generates the factor $1-\mathcal{C}$ in \eqref{eq:renormalized_phase_diffusion}. The relation \eqref{eq:y_relation_goldstone} is an instantaneous statement about the noise and reproduces only the bare coefficient $D_\theta^{(0)}$. It is not a second derivation of the long-time diffusion constant.

\subsection{MSR effective theory}
\label{subsec:MSR_formalism_soft_rotor}
In principle, we can solve the nonlinear equation \eqref{eq:Nonlin_PDE} far from equilibrium to determine the relaxational dynamics of the entropic-ordered superfluid.  However, we will be more interested in spatial dynamics of interacting superfluid clusters in the remainder of the paper.  In this setting, it can feel rather tedious to keep track of a high-dimensional equation -- as we will need one pair of $(x,y)$ coordinates for each lattice site.   A popular alternative in the physics literature is, therefore, to work with an MSR Lagrangian \cite{MartinSiggiaRose1973,Janssen1976,DeDominicis1976}.  The general form of an MSR effective theory is:
\begin{align}
    \mathcal{L} =\pi_{a} \dot{q}_a - \mathcal{H}
\end{align}
where $\pi_a$ is the noise field conjugate to the coordinate $q_a$ and $\mathcal{H}$ is the MSR Hamiltonian. Suppose the time evolution of the system is described by a deterministic PDE of the form
\begin{align}
\label{eq:Fokker_Planck_form}
    \partial_t |P\rangle = -\hat{W}|P\rangle
\end{align}
where $\hat{W}$ is a differential operator. Then the MSR Lagrangian can be constructed by identifying $\mathcal{H} = -i\hat{W}$ and replacing the partial derivatives in $\hat{W}$ with the conjugate noise field, $\partial_a \to i\pi_a$. The mean-field PDE \eqref{eq:Nonlin_PDE} is nonlinear in $P$ and so cannot be cast directly in this form. We therefore start from the full $N$-body Fokker-Planck equation on the microscopic configuration space, where the dynamics are linear, 
\begin{align}
    \partial_t \rho_N = \gamma \sum_i \nabla_i \cdot \left(\rho_N\nabla_iH\right)+\gamma T\sum_i \nabla_i^2\rho_N
\end{align}
where $H$ is the microscopic Hamiltonian, \eqref{eq:microscopic_overdamped_Hamiltonian}, and $\rho_N = \rho_N(x_1,y_1,...,x_N,y_N,t)$ is the full $N$-body probability density on the microscopic configuration space. 
By replacing $\nabla_i \to (i\pi_{x_i},i\pi_{y_i})$, we can immediately read off the full MSR Lagrangian
\begin{align}
\label{eq:micro_MSR}
    \mathcal{L} = \sum_i \bigg[\pi_{x_i}\dot{x}_i + \pi_{y_i}\dot{y}_i + iT\gamma \pi_{x_i}\left(\pi_{x_i}-i\beta \frac{\partial H}{\partial x_i}\right)+ iT\gamma \pi_{y_i}\left(\pi_{y_i}-i\beta \frac{\partial H}{\partial y_i}\right)\bigg]
\end{align}
In the spirit of an effective theory, we now keep only the slow mode. The linearized analysis of Section~\ref{subsec:linear_MF_PDE} identified the collective phase $\bar{\theta}$ as the unique gapless hydrodynamic mode, while all transverse directions are gapped and relax at a finite rate. Additionally, global $\mathrm{U}(1)$ symmetry forbids any dependence of $H$ on $\bar\theta$, so the Goldstone coordinate has no potential. Integrating out the gapped modes must therefore produce the most general MSR Lagrangian for a single coordinate with no restoring force, namely pure diffusion. The leading hydrodynamic MSR Lagrangian is then,
\begin{align}
     \mathcal{L}_{\text{slow}} = \bar{\pi}_{\theta}\dot{\bar{\theta}} +iD_{\theta}\bar{\pi}_{\theta}^2
\end{align}
If the dynamics were restricted rigidly to the Goldstone manifold, the direct microscopic noise would give the bare phase-diffusion coefficient,
\begin{align}
    D_{\theta}^{(0)} = \gamma T \sum_i \big|\nabla_i\bar{\theta}\big|^2= \frac{\gamma T}{N\bar{s}^2},
\end{align}
This is precisely the direct Brownian term appearing in Section \ref{subsec:Diffusion_Goldstone_mode}, with $\mathcal{C} =0$. However, the transformation from the microscopic Cartesian coordinates to the collective phase and transverse fluctuations is nonlinear. Consequently, integrating out the massive modes also incorporates their coupling to the phase. Although these modes relax on an $\mathcal{O}(1)$ timescale, their fluctuating deterministic force produces an $\mathcal{O}(N^{-1})$ contribution to the long-time diffusion coefficient. The complete coefficient is therefore fixed by matching the effective theory to the microscopic finite-$N$ calculation of Section \ref{subsec:Diffusion_Goldstone_mode}. From \eqref{eq:renormalized_phase_diffusion},
\begin{align}
\label{eq:MSR_matched_diffusion}
   D_{\theta}= \left(1-\mathcal{C}\right)\frac{\gamma T}{N\bar{s}^2}+o\left(\frac{1}{N}\right)
\end{align}
Thus, $D_{\theta}^{(0)}$ should be interpreted as the rigid-manifold or direct-noise contribution, while the factor $1-\mathcal{C}$ accounts for the renormalization generated by the finite-$N$ transverse dynamics. The coefficient $D_\theta$ appearing here describes smooth fluctuations of the Goldstone mode about the ordered state, and it should not be identified with a diffusion constant extracted from the unrestricted winding of $\bar\theta$ at finite $N$. The singularity of the unrestricted diffusion constant, discussed in Section \ref{subsec:Diffusion_Goldstone_mode}, is an artifact of polar coordinates. The noise acting on the Cartesian components of the order parameter is regular everywhere, as in \eqref{eq:y_relation_goldstone}, and the factor of $\bar{s}^{-2}$ in the angular noise is just the Jacobian of the change of variables. The relaxation rates derived in the following sections depend only on the combination $\bar{s}^2D_\theta$, in which this factor cancels. The effective theory is therefore controlled by typical configurations near $\bar{s}_{\text{eq}}>0$, while excursions of the order parameter to the core are exponentially rare in $N$ and are expected to give only non-perturbative corrections.

Lastly we comment on the scaling of $D_{\theta}$. Conventional intuition tells us that  increasing temperature should destroy order by making the collective phase diffuse arbitrarily quickly. That intuition would be correct if the rotor amplitude were bounded, in which case one would have $D_{\theta} \sim T/N$ and thus $\tau_{\theta} \sim N/T$. In the entropically ordered phase, however, the collective amplitude itself grows with temperature as $\bar{s}^2 \sim T$. Additionally, the collective second radial moment scales with temperature as $Q\sim T$. Since 
\begin{align}
    \frac{\gamma T}{NQ} \leq D_{\theta}\leq \frac{\gamma T}{N\bar{s}^2}
\end{align}
the upper and lower bound of $D_{\theta}$ scale as $\mathcal{O}(1/N)$. As a result, the increase in thermal noise is compensated by an increase in the stiffness and the diffusion constant remains 
\begin{align}
    D_{\theta} \sim \mathcal{O}\left(\frac{1}{N}\right)
\end{align}
Therefore, in the high temperature limit, the ordered phase still supports a well-defined hydrodynamic mode. Explicitly using the self consistent equilibrium values derived in Section \ref{subsec:phase_diagram_of_model}, the bound on the renormalization factor becomes
\begin{align}
\label{eq:bound_renormalized_D}
    0 \leq
\limsup_{T\rightarrow\infty} \mathcal{C}\leq\frac{2a}{9-4a}<1,
    \qquad
    1<a<\frac{3}{2}.
\end{align}
Thus, the massive modes can produce an order-one renormalization of the phase-diffusion coefficient, but they cannot change its $N^{-1}$ scaling. The collective phase therefore remains parametrically long lived, with a characteristic diffusion time scaling as
\begin{align}
    \tau_{\theta}
    \sim
    \mathcal{O}(N)
\end{align}
even in the high-temperature limit.

\section{Cluster lattice model}
\label{sec:Cluster_lattice}
We now consider a lattice of weakly coupled clusters. Each site $\mu$ contains $N$ soft rotors governed by the same intra-cluster Hamiltonian as in the single-cluster problem, while neighboring clusters are coupled by a weak inter-cluster interaction of strength $\epsilon$. We use $\mu,\nu$ to label lattice sites and $(i,j)$ to label rotors within each cluster. The full Hamiltonian is 
\begin{align}
    H = \sum_{\mu}\left(H_0^{\mu} + \sum_{\mu\sim \nu} H_{\text{int}}^{\mu\nu}\right)
\end{align}
where 
\begin{subequations}\begin{align}
    H_0^{\mu} &= \frac{1}{2N}\sum_{i\neq j \in \mu}\left(a|s_i^{\mu}||s_j^{\mu}|-\mathbf{s}_i^{\mu}\cdot \mathbf{s}_j^{\mu}\right)+ b\sum_{i\in \mu}|s_i^{\mu}| \\
    H_{\text{int}}^{\mu\nu} &= \frac{\epsilon}{2N}\sum_{i\in \mu}\sum_{j\in \nu}\left(a|s_i^{\mu}||s_j^{\nu}|-\mathbf{s}_i^{\mu}\cdot \mathbf{s}_j^{\nu}\right)
\end{align}\end{subequations}
Here $\epsilon$ controls the strength of the inter-cluster coupling. Notice that we can write the inter-cluster Hamiltonian exactly in terms of the collective variables on each site,
\begin{align} \label{eq:H_int_munu}
   H_{\text{int}}^{\mu\nu} = \frac{\epsilon N}{2}\left(a\bar{r}^{\mu}\bar{r}^{\nu}-\bar{s}^{\mu}\bar{s}^{\nu}\cos(\bar{\theta}^{\mu}-\bar{\theta}^{\nu})\right)
\end{align}

\subsection{Slow dynamics}
\label{subsec:slow_dynamics_cluster_lattice_soft_rotor}
As shown in Section \ref{subsec:linear_MF_PDE}, the only physical gapless mode of a single ordered cluster is the Goldstone mode associated with an infinitesimal rotation of the order parameter. All radial and amplitude fluctuations are therefore gapped and relax at a finite rate. On timescales long compared to this relaxation, these modes are frozen at their equilibrium values, so that each cluster behaves as a rigid collective rotor. The hydrodynamic limit is then governed solely by the collective phases $\bar{\theta}^{\mu}$ of each cluster, whose dynamics describe the slow relaxation of neighboring clusters toward alignment. Therefore, the radial term in the inter-cluster Hamiltonian is an additive constant and does not affect the long-time dynamics of the system. Additionally, each cluster has internal $\mathrm{U}(1)$ symmetry, so the effective intra-cluster Hamiltonian cannot depend on $\bar{\theta}^{\mu}$. 

Thus, when analyzing the dynamics on long time scales, in the large-$N$ limit we only need to consider the following effective potential for nearest neighbor interactions
\begin{align}
    H_{\text{eff}}^{\mu} = -\frac{\epsilon N}{2}\sum_{\mu\sim\nu} \bar{s}_{\mu}\bar{s}_{\nu}\cos(\bar{\theta}_{\mu}-\bar{\theta}_{\nu})
\end{align}
We can extend our results from the single cluster section by summing over all lattice sites $\mu$
\begin{align}
     \mathcal{L}_{\text{slow}} =  \sum_{\mu} \mathcal{L}_{\text{slow}}^{\mu}
\end{align}
where, since each cluster is identical, 
\begin{align}
    \mathcal{L}_{\text{slow}}^{\mu} = \bar{\pi}_{\theta}^{\mu}\dot{\bar{\theta}}^{\mu} + iD_{\theta}\bar{\pi}_{\theta}^{\mu}\left(\bar{\pi}_{\theta}^{\mu}-i\beta \frac{\partial H_{\text{eff}}}{\partial \bar{\theta}_{\mu}}\right)
\end{align}
Therefore the noise-free equation of motion for the collective phase on site $\mu$ is given by 
\begin{align}
    \dot{\bar{\theta}}^{\mu} = -\beta D_{\theta}\frac{\partial H_{\text{eff}}}{\partial \bar{\theta}_{\mu}}.
\end{align}
Furthermore, suppose we are at sufficiently long timescales such that each site is close to alignment. Then $\bar{\theta}_{\mu}-\bar{\theta}_{\nu}\ll1$, and so $\sin(\bar{\theta}_{\mu}-\bar{\theta}_{\nu}) \approx \bar{\theta}_{\mu}-\bar{\theta}_{\nu}$. Then the angular drift term for each site becomes
\begin{align}
    \frac{\partial H_{\text{eff}}}{\partial \bar{\theta}_{\mu}}\approx \epsilon N\bar{s}^2\sum_{\mu\sim\nu}(\bar{\theta}_{\mu}-\bar{\theta}_{\nu})
\end{align}
Note that the discrete Laplace operator is defined as 
\begin{align}
    \nabla^2 \phi(\mu) = \sum_{\nu:d(\mu,\nu)=1} [\phi(\nu) - \phi(\mu)]
\end{align}
Thus the equation of motion for each site $\mu$ can be written as 
\begin{align}
     \partial_t \bar{\theta}_{\mu} = (1-\mathcal{C})\gamma \epsilon\nabla^2\bar{\theta}_{\mu}
\end{align}
This is a purely diffusive mode with a diffusion constant \begin{equation}
    D_{\text{lat.}}= (1-\mathcal{C})\gamma\epsilon.
\end{equation}
The bound \eqref{eq:bound_renormalized_D} on $\mathcal{C}$ from Section \ref{subsec:MSR_formalism_soft_rotor} transfers, so $D_{\text{lat.}}$ remains $\mathcal{O}(\epsilon)$ with a temperature independent coefficient. This result is what is expected based on general ``hydrodynamic" grounds \cite{HohenbergHalperin1977,ChaikinLubensky1995}.  In the long-wavelength limit the free energy of a phase configuration is $\frac{\rho_s}{2}\int \mathrm{d}^dx\, (\nabla \theta)^2$, so the steady state is the Boltzmann weight 
\begin{align}
    P_{\rm eq}[\theta]\propto \exp\!\big(-\frac{\beta\rho_s}{2}\int \mathrm{d}^dx\,(\nabla\theta)^2\big),
\end{align}
and it is the vanishing cost of a uniform phase rotation that gives rise to diffusion.  Namely, the diffusion is not a consequence in this model of any genuine conservation law: the overdamped dynamics is $\mathrm{U}(1)$-covariant but does not conserve the charge that generates the symmetry. The reason that a ``hydrodynamic" mode still exists is simply that the steady state has a ``flat" direction to diffuse around.

\subsection{Connection to XY model}
\label{subsec:lattice_XY_comparison}
Before continuing on, let us stress that the ``hydrodynamics" of the entropic-ordered superfluid is identical to that which arises at low temperature in the ordinary XY model \eqref{eq:XY_model_Hamiltonian}.  From \eqref{eq:H_int_munu}, we have seen that the only relevant dynamics on long time scales is generated by 
\begin{align} \label{eq:Heffsec3}
     H_{\text{eff}} = -\frac{\epsilon N\bar{s}^2}{2}\sum_{\mu\sim\nu} \cos(\bar{\theta}_{\mu}-\bar{\theta}_{\nu})
\end{align}
In the ordinary XY model, the effective Boltzmann weight is set by the dimensionless ratio $\beta J$, as we saw in Section \ref{subsec:XY_model_review}. When $\beta J > (\beta J)_{\mathrm{c}}$, there is superfluidity. Equation \eqref{eq:Heffsec3} is identical to the XY model, with \begin{equation}
    \beta J \rightarrow \beta \epsilon N \bar s^2 .
\end{equation} 
Recall that  $\bar{s}_{\mu}^2 \sim T$. Explicitly, using the high-temperature asymptotics of the symmetry-broken branch in Section \ref{subsec:phase_diagram_of_model},
\begin{align}
    \beta\epsilon N\bar{s}^2 \to \epsilon N\,\frac{3-2a}{a-1}
\end{align}
as $T\to \infty$. The lattice of clusters therefore orders whenever this exceeds the critical coupling $(\beta J)_{\mathrm{c}}$ of the XY model on the chosen lattice, which for the simple cubic lattice is $(\beta J)_{\mathrm{c}}\approx 0.45$ \cite{Campostrini2001}. At fixed $\epsilon>0$ this is guaranteed by taking $N$ sufficiently large. Up to O(1) constants, therefore, as $T\rightarrow \infty$ we have an effective $\beta J \sim \epsilon N$.  We can clearly see that in the large $N$ limit, we are at an ``effective XY temperature" $T_{\mathrm{XY,eff}}\sim N^{-1}$, which is a more microscopic way to understand the emergence of robust superfluid ``hydrodynamics" in the entropic-ordered superfluid phase.

\section{Adding a conserved charge}
\label{sec:Superfluid_model}
So far, we have seen an entropic-ordered superfluid characterized by spontaneous symmetry breaking in equilibrium and overdamped diffusive relaxation of a corresponding Goldstone mode.  However, this Goldstone mode is not characteristic of ``true" superfluid hydrodynamics on the lattice because the MSR dissipative dynamics introduced previously was not based on an intrinsically Hamiltonian system.  In particular, in order to observe an honest superfluid sound mode, we will need to build an entropically-ordering Hamiltonian model.

At first glance, this is quite straightforward -- we just need to add a conjugate momentum to the $\theta_i$ degree of freedom (for completeness we will go ahead and add one for both components of $\mathbf{s}$).  However, we will see that this can render the thermodynamics of the system rather delicate (and interesting!).  In this section, we will revisit the statistical mechanics of models with a conserved U(1) charge.   Given that the U(1) symmetry above corresponds to the rotation of the rotors in the plane, we will call this corresponding conserved charge the angular momentum.

\subsection{Adding angular momentum}
\label{subsec:Introducing_SF_model}
For each soft rotor we write the Cartesian components as
$\vec{s}_i=(x_i,y_i)$, with conjugate momenta $p_{x_i}$ and
$p_{y_i}$. The canonical coordinates obey the following Poisson brackets 
\begin{align}
    \{x_i,p_{x_j}\} =  \{y_i,p_{y_j}\} = \delta_{ij}
\end{align}
while all other fundamental brackets vanish. The global U(1) symmetry corresponds to simultaneous rotations
of all rotors in the $(x,y)$ plane, and is generated by the total angular momentum
\begin{align}
    L = \sum_i \left(x_i p_{y_i}-y_i p_{x_i}\right).
\end{align}
We look for a Hamiltonian that is invariant under this symmetry, $\{L,H\}=0$. Because the momenta enter only through the kinetic term, which is already rotationally invariant, the conservation law constrains the soft-rotor potential alone: $V$ may depend on the rotors only through the rotationally invariant combinations $r_i$ and $\vec{s}_i\cdot\vec{s}_j$.
The natural family of such potentials in $d$ dimensions is
\begin{align}
\label{eq:general_family}
    V_{(n,m)} = \frac{1}{2N}\sum_{i\neq j} \left(ar_i^nr_j^n - r_i^{n-1}r_j^{n-1}(\vec{s}_i\cdot \vec{s}_j)\right) + b\sum_i r_i^m
\end{align}
Not every member of \eqref{eq:general_family} entropically orders. Because the momenta are Gaussian and decouple from the configurational sector, the ordering criterion only depends on $V_{(n,m)}$. In Appendix
\ref{sec:Appendix_allowed_models} we show that it admits only two solutions for $d\geq 2$. We find that if $m>2n$, the confining term $br^m$ suppresses the radial fluctuations needed to support the entropic ordering mechanism at high temperature. If $m<2n$, the self-consistent interaction can support an entropically ordered phase, but boundedness of the Hamiltonian and the ordering criterion together force $n=1$, hence $m=1$. The only remaining possibility is the marginal case $m=2n$, which again forces $n=1$ and hence $m=2$: see Table \ref{tab:allowed_exponents}. Within the family \eqref{eq:general_family}, the two admissible microscopic models are thus the linear potential
\begin{align}
\label{eq:linear_model}
    V_{(1,1)} = \frac{1}{2N}\sum_{i\neq j}
        \left(a r_ir_j - \vec{s}_i\cdot \vec{s}_j\right) + b\sum_i r_i ,
\end{align}
which is the overdamped soft rotor potential already studied in Section \ref{subsec:Introducing_soft_rotor_model}, and the marginal quadratic potential
\begin{align}
\label{eq:quadratic_model}
    V_{(1,2)} = \frac{1}{2N}\sum_{i\neq j}
        \left(a r_ir_j - \vec{s}_i\cdot \vec{s}_j\right) + b\sum_i r_i^2 ,
\end{align}
which supports an entropically ordered phase precisely when (see Appendix \ref{sec:Appendix_allowed_models} for details)
\begin{align}
\label{eq:parameter_constraints_quadratic_model}
    (1-2b)<a<a_c(b),\qquad 0< b< \frac{1}{2} .
\end{align}
In each case the lower bound is the condition that $H$ be bounded below and the upper bound the condition that alignment be entropically favorable.  $a_c(b)$ is the coupling at which the disordered state becomes linearly unstable, and is given parametrically in \eqref{eq:boundary_parametrization}. In Appendix \ref{subsec:Appendix_generalized_kernel_W}, we show that, for the case $d=2$, it is also the full phase boundary, with a unique ordered state throughout \eqref{eq:parameter_constraints_quadratic_model}.

One feature of \eqref{eq:quadratic_model} deserves emphasis before we go further. Every term in $H$ is homogeneous of degree two in the phase-space variables, so the rescaling $(\vec{s}_i,\vec{p}_i)\to\sqrt{T}\,(\vec{s}_i,\vec{p}_i)$ removes $\beta$ from the Boltzmann weight entirely. The wedge \eqref{eq:parameter_constraints_quadratic_model} is therefore the same for every $T>0$, and inside it the mean-field amplitudes $\bar{r}$, $\bar{s}$ and the second moment $\langle r^2\rangle$ are all exactly proportional to $\sqrt{T}$, $\sqrt{T}$ and $T$ respectively, not merely asymptotically as in the linear model. In other words, the quadratic model does not undergo a transition at a positive critical temperature: it is ordered at every $T>0$ in the wedge, and $T=0$ is the only endpoint of its phase diagram. The same scaling will later make its dimensionless dynamics temperature independent as well. Since we found no advantage to taking $d>2$, we specialize to $d=2$ in what follows, where $a_c(b)\approx \tfrac{3}{2}-4b$ at small $b$.

\begin{table}[t]
\centering
\renewcommand{\arraystretch}{1.6}
\begin{tabular}{|c|p{0.68\textwidth}|}
\hline
\textbf{Regime} & \textbf{Consequence for the ordered phase} \\
\hline
$m<2n$ 
& Entropic ordering requires $n=1$. \\
\hline
$m>2n$ 
& No high-temperature ordered phase. \\
\hline
$m=2n$ 
& Entropic ordering requires $n=1$ and sufficiently small $b$. \\
\hline
\end{tabular}
\caption{Summary of the high-temperature ordering criteria for integer exponents in $d\geq 2$. The viable choices are $(n,m)=(1,1)$ and the marginal case $(n,m)=(1,2)$, with the latter requiring sufficiently small $b$.}
\label{tab:allowed_exponents}
\end{table}

Since $L$ is conserved, equilibrium states organize into sectors of fixed total angular momentum.  For a macroscopic system in such a sector, a subsystem may exchange angular momentum with the remainder, so by equivalence of ensembles \cite{Touchette2015, Ruelle1999} we will pass, for now, to a generalized canonical ensemble in which $L$ is controlled by a conjugate field $\Omega$, exactly as particle number is controlled by a chemical potential in the grand canonical ensemble. The value of $\Omega$ is fixed by the desired angular momentum density. Completing the square for the Hamiltonian in the frame co-rotating at angular velocity $\Omega$, 
\begin{align}
\label{eq:rotating_frame_H}
    H-\Omega L = \frac{1}{2}\sum_i\left(\left(p_{x_i}+\Omega y_i\right)^2 + \left(p_{y_i}-\Omega x_i\right)^2\right) -\frac{\Omega^2}{2}\sum_i r_i^2 + V
\end{align}
The first term is the rotating-frame kinetic energy, and the second is the centrifugal potential. Since the momentum shift is a translation at fixed position, the Gaussian momentum integrals are independent of $\Omega$ and of the configuration. Hence, 
\begin{align}
    Z(\Omega) = \left(\frac{2\pi}{\beta}\right)^N \int \prod_i dx_i dy_i e^{-\beta \left(V - \frac{\Omega^2}{2}\sum_i r_i^2\right)}
\end{align}
All of the $\Omega$ dependence of the thermodynamics therefore resides in the single centrifugal term added to the potential. Since the centrifugal term is quadratic, the general family of potentials \eqref{eq:general_family} is closed under the shift only when $m=2$. In that case, rotation merely renormalizes a coupling rather than changing the form of the potential: letting $\kappa := b-\frac{1}{2}\Omega^2$, 
\begin{align}
\label{eq:rotating_quad_potential}
    V_{(1,2)}^{\Omega} := V_{(1,2)} - \frac{\Omega^2}{2}\sum_i r_i^2 = \frac{1}{2N}\sum_{i\neq j}(ar_ir_j - \vec{s}_i\cdot \vec{s}_j) + \kappa \sum_i r_i^2,
\end{align}
so that the rotating theory at coupling $b$ is the static theory at coupling $\kappa$. For $V_{(1,1)}$, by contrast, the centrifugal term stands alone and the resulting potential is unbounded below for any $\Omega>0$. The linear model therefore admits at best a metastable configuration in a rotating ensemble, and we adopt \eqref{eq:quadratic_model} for the remainder of the paper. 
\subsubsection{Mean-field model}
\label{subsubsec:SF_mean_field_model}
As in the static case of Section~\ref{subsec:Introducing_soft_rotor_model}, the mean-field description replaces the pair interaction by the self-consistent fields $(\bar{r},\bar{x})$, with the ordered direction again taken along the $x$ axis. The single-rotor Hamiltonian is
\begin{align}
    H = \frac{p_x^2}{2}+\frac{p_y^2}{2} + V_{\text{mf}}, \qquad V_{\text{mf}} = a\bar{r}\,r-\bar{x}\,x + br^2.
\end{align}
In the rotating ensemble, the completing-the-square manipulation of \eqref{eq:rotating_frame_H} applies verbatim at the single-rotor level. In terms of the shifted momenta,
\begin{align}
    P_x = p_x + \Omega y, \qquad P_y = p_y-\Omega x, \label{eq:Pxpx}
\end{align}
the rotating-frame Hamiltonian is
\begin{align}
\label{eq:rotating_mf_potential}
    H -\Omega\ell = \frac{P_x^2}{2}+\frac{P_y^2}{2} + V^{\Omega}, \qquad V^{\Omega} = a\bar{r}\,r - \bar{x}\,x + \kappa r^2,
\end{align}
so the rotating mean-field theory at coupling $b$ is the static mean-field theory at coupling $\kappa$, in accordance with \eqref{eq:rotating_quad_potential}. 
The single-rotor partition function in the rotating ensemble is\footnote{Note that there is no Jacobian factor since the coordinate transformation \eqref{eq:Pxpx} has unit Jacobian.}
\begin{align}
    z(\Omega) = \int d^2P\, d^2x\;\, e^{-\beta\left(H-\Omega\ell\right)}.
\end{align}
 In the shifted variables, the angular momentum density decomposes as
\begin{align}
    \ell = xp_y - yp_x = xP_y -yP_x + \Omega \left(x^2+y^2\right).
\end{align}
The first term is odd in the shifted momenta, so its expectation vanishes identically. The angular momentum density is therefore carried entirely by the configurational second moment. At fixed mean fields $(\bar{r},\bar{x})$, 
\begin{align}
\label{eq:ell_equals_OmegaQ}
    \langle \ell\rangle_{\Omega} = \frac{1}{\beta}\frac{\partial \log z}{\partial \Omega} =  \Omega\,  \langle r^2\rangle_{\Omega} =: \Omega Q(\Omega)
\end{align}
where $Q$ plays the role of an effective moment of inertia.
\subsubsection{Critical angular velocity }
\label{subsubsec:critical_angular_velocities}
Because \eqref{eq:rotating_quad_potential} is an exact rewriting, every constraint derived at $\Omega=0$ transfers to the rotating ensemble under $b\to\kappa$. 
In \eqref{eq:parameter_constraints_quadratic_model}, we showed that the Hamiltonian is bounded below if and only if 
\begin{align}
    a\geq 1-2b, \quad \text{and}\quad  b\geq 0
\end{align}
In the rotating ensemble, upon transferring $b\to \kappa$, the rotating Hamiltonian is therefore bounded below if and only if 
\begin{align}
    2\kappa + a-1\geq 0, \quad\text{and}\quad \kappa \geq 0
\end{align}
Equivalently, the system possesses two critical angular velocities,
\begin{align}
\label{eq:critical_velocities}
     \quad \Omega_1 = \sqrt{2b+a-1},\quad \Omega_2 = \sqrt{2b}
\end{align}
Which threshold is reached first as $\Omega$ is increased from zero is determined by the sign of $a-1$: for $a<1$ one has $\Omega_1<\Omega_2$, while for $a>1$ the order is reversed. The two thresholds signal the softening of qualitatively different modes. To see this, note that the two boundedness conditions probe the potential along different directions in configuration space. Evaluating \eqref{eq:rotating_quad_potential} on a homogeneous aligned configuration $\vec{s}_i=\vec{s}$ gives
\begin{align}
    V_{(1,2)}^{\Omega}\big|_{\vec{s}_i=\vec{s}}
    = \frac{N}{2}\left(2\kappa+a-1\right)r^2  + \mathcal{O}\left(1\right)
\end{align}
so that $2\kappa+a-1>0$ is stability against a coherent expansion in which every rotor participates equally. By contrast, $\kappa$ is the stiffness seen by a single rotor. Since each pair interaction carries a factor $1/N$, the cost of displacing one rotor to radius $r$ while the remainder stay in a typical configuration is, at mean-field level,
\begin{align}
\label{eq:single_rotor_cost}
    \mathrm{\Delta} V = \left(a\bar{r} - \bar{x}\right) r + \kappa r^2,
\end{align}
and $\kappa>0$ is stability against the escape of individual rotors.

For $a<1$ the collective stiffness vanishes first, at $\Omega=\Omega_1$, while individual rotors remain harmonically confined there since
\begin{align}
   2\kappa_c + a-1 =0 \implies  \kappa_c = \frac{1}{2}(1-a)>0
\end{align}
Thus, the instability is shared uniformly across the population and as $\Omega \to \Omega_1$, the system swells coherently, and the equilibrium radial distribution remains described by a homogeneous saddle point whose radius diverges at the threshold. For $a>1$ the situation is reversed. The collective mode remains stiff up to $\Omega_2$ and beyond, 
\begin{align}
    \kappa_c = 0 \implies 2\kappa_c + a-1 = a-1 >0
\end{align}
so a uniform expansion is always penalized. Thus for the $a>1$ regime, it is the harmonic confinement of individual rotors that is lost at $\Omega=\Omega_2$. Because the collective mode remains stiff, the system's angular momentum and the radius of the cluster do not diverge at this threshold. By \eqref{eq:ell_equals_OmegaQ}, the angular momentum density approaches
\begin{align}
    \langle \ell\rangle_{\Omega}  \to \langle \ell\rangle_{\Omega_2} = \Omega_2 Q(\Omega_2), \quad Q(\Omega_2) = \frac{9-4a}{3\beta(a-1)}<\infty
\end{align}
where the value of $Q(\Omega_2)$ follows from the self-consistent moments of the residual linear potential \eqref{eq:single_rotor_cost} at $\kappa=0$. Since the theory diverges for $\Omega >\Omega_2$, it appears as though the system has a finite maximum angular momentum density. However, since the harmonic confinement of individual rotors is lost, the system can carry excess angular momentum $\ell > \ell(\Omega_2)$ by transferring a vanishing fraction of rotors to large radius, where a rotor at radius $r$ carries angular momentum $\Omega r^2$ at only linear energy cost. The radial distribution accordingly splits into a confined bulk and a macroscopically separated large-$r$ component which we denote as the ``condensate", rather than expanding homogeneously. The two regimes thus exhibit sharply different critical behavior and must
be analyzed separately.  Since $a_c$ is decreasing and rotation only lowers the coupling, $\kappa\leq b$;  a cluster that is ordered at rest, $a<a_c(b)$, satisfies $a<a_c(\kappa)$ for every $0\leq\Omega<\Omega_c$, where $\Omega_c:=\min(\Omega_1,\Omega_2)$ is the first threshold reached. We therefore separate the ordered wedge into two sectors,
\begin{align}
\label{eq:sector_split}
    &a\leq 1: \begin{cases}
        0<b\leq b_*, & (1-2b)<a\leq 1\\
        b_*< b <\frac{1}{2}, & (1-2b)<a< a_c(b)
    \end{cases}\\[1em]
    &a>1: 0<b<b_*, \quad 1< a < a_c(b)
\end{align}
where $b_*\approx 0.142$ is the coupling at which the phase boundary crosses $a=1$, i.e. $a_c(b_*)=1$.
In Section~\ref{subsec:homogeneous_saddle} we treat the homogeneous regime $a<1$ by a saddle-point analysis, and in Section~\ref{subsec:Condensate} we construct the condensate that governs the regime $a>1$.

Finally, we briefly explain why the marginal case $a=1$, where the two critical points coincide, belongs to the homogeneous regime. Although both stiffnesses vanish simultaneously at $\Omega_1=\Omega_2$, the two directions in configuration space remain sharply distinguished at the threshold. Along the uniform expansion the potential is exactly quadratic at $a=1$,
\begin{align}
    \left.V_{(1,2)}^{\Omega}\right|_{\vec{s}_i=s} = N\kappa r^2 + \mathcal{O}(1)
\end{align}
and therefore, as $N\to \infty$, it becomes an exactly flat direction when $\kappa\to 0$. A single rotor displaced against the bulk, by contrast, remains linearly confined even at $\kappa=0$,
\begin{align}
    \left.\mathrm{\Delta} V\right|_{\kappa =0} = \left(a\bar{r} - \bar{x}\right)r,
\end{align}
and since $\bar{r}> |\bar{x}|$ at any finite temperature, the coefficient $\bar{r}-\bar{x}$ remains strictly positive at $a=1$. Individual escape therefore remains exponentially suppressed.

\subsection{Homogeneous cloud when $a\le 1$}
\label{subsec:homogeneous_saddle}
In this section we determine the critical behavior of the angular momentum density in the homogeneous sector $a\leq 1$ of \eqref{eq:sector_split}, as the angular velocity approaches the collective threshold of \eqref{eq:critical_velocities},
\begin{align}
    \Omega_c := \Omega_1 = \sqrt{2b+a-1}.
\end{align}
By \eqref{eq:ell_equals_OmegaQ}, the angular momentum density is carried entirely by the effective moment of inertia $Q(\Omega)$, so it suffices to determine the critical scaling of $Q$. We do so by a saddle-point analysis of the rotating mean-field potential \eqref{eq:rotating_mf_potential}, treating $a<1$ first and returning to the marginal case $a=1$ at the end of the section.
We first consider the angular sector. At fixed $r$, the conditional expectation of the alignment is
\begin{align}
    \langle \cos\theta\rangle_r = \frac{I_1(z)}{I_0(z)} = 1- \frac{1}{2z} + \mathcal{O}(z^{-2}), \qquad z=\beta \bar{x}r,
\end{align}
where we have used the large-argument asymptotics of the modified Bessel functions, valid in the ordered phase near the threshold where the distribution is dominated by large $r$. The angular self-consistency condition is therefore
\begin{align}
\label{eq:xbar_selfconsistency}
    \bar{x} = \langle r\cos\theta\rangle = \bar{r} - \frac{1}{2\beta \bar{x}}.
\end{align}
Integrating out the angle in \eqref{eq:rotating_mf_potential}, the radial density is
\begin{align}\label{eq:radial_density_reduced}
   P(r) \propto r\, I_0(\beta \bar{x}r)\, e^{-\beta (\kappa r^2 + a\bar{r}r)} \sim \sqrt{r}\; e^{-\beta \left(\kappa r^2 - (\bar{x}-a\bar{r})r\right)}
\end{align}
where we used asymptotics at large $r$ for the last step.
For $a<1$ near the threshold the linear coefficient is positive so the exponent is minimized at the finite radius
\begin{align}
\label{eq:homogeneous_r_saddle}
    r_* = \frac{\bar{x}-a\bar{r}}{2\kappa}.
\end{align}
In order to expose the scaling limit, we rescale $r=r_*\xi$ so that all moments factorize as $\langle r^n\rangle = r_*^n \langle \xi^n\rangle$. Let us define, 
\begin{align}
    J_{\rho}(m) = \int_0^{\infty} d\xi\, \xi^{\rho} e^{-m(\xi-1)^2}, \quad m:= \beta \kappa r_*^2
\end{align}
where $m$ controls the sharpness of the radial peak. The collective variables are then
\begin{align}
\label{eq:homogeneous_collective_vars}
    \bar{r} = r_* \frac{J_{3/2}(m)}{J_{1/2}(m)}, \quad Q = r_*^2 \frac{J_{5/2}(m)}{J_{1/2}(m)}
\end{align}
and the large-$m$ expansion of the second moment gives
\begin{align}
\label{eq:Q_large_m}
    Q = \frac{m}{\beta \kappa} + \frac{1}{\beta \kappa} +\mathcal{O}\left(m^{-1}\right)
\end{align}
The critical behavior of $Q$ is therefore fixed by the scaling of $m$ as $\Omega\to \Omega_c$. Collecting the self-consistency relations,
\begin{align}
    \bar{r} = r_*A(m), \qquad A(m):= \frac{J_{3/2}(m)}{J_{1/2}(m)},
\end{align}
along with \eqref{eq:xbar_selfconsistency} and \eqref{eq:homogeneous_r_saddle}, and solving for $\kappa$ in terms of $m$ yields
\begin{align}
    \kappa(m) &= \frac{1}{4}\left(A(m) (1-2a) -\frac{1}{4m} + \sqrt{\left(A(m)-\frac{1-2a}{4m}\right)^2 + \frac{a(1-a)}{4m^2}}\right) \notag \\
    &\approx 
     \frac{1-a}{2} + \frac{1-a^2}{32m^2}+ \mathcal{O}(m^{-3}). \label{eq:kappa_expansion}
\end{align}
The leading term reproduces the critical coupling $\kappa_c = \frac{1}{2}(1-a)$ identified in Section~\ref{subsubsec:critical_angular_velocities}, confirming that the saddle becomes critical precisely at the collective threshold. On the other hand, directly from the definitions of $\kappa$ and $\Omega_c$,
\begin{align}
    \kappa - \kappa_c = \frac{1}{2}\left(\Omega_c^2-\Omega^2\right).
\end{align}
Matching against \eqref{eq:kappa_expansion} gives the critical scaling of the peak sharpness,
\begin{align}
    m = \frac{1}{4}\sqrt{\frac{1-a^2}{\Omega_c^2-\Omega^2}} + \mathcal{O}(1)
\end{align}
so that by \eqref{eq:Q_large_m}, the effective moment of inertia diverges as 
\begin{align}
Q = \frac{1}{2\beta}\sqrt{\frac{1+a}{1-a}} (\Omega_c^2-\Omega^2)^{-1/2} + \mathcal{O}(1) 
\end{align}
By \eqref{eq:ell_equals_OmegaQ}, the angular momentum density inherits this divergence as well: $\ell \sim \Omega (\Omega_c^2-\Omega^2)^{-1/2}$.
Arbitrarily large angular momentum density is therefore attained within the homogeneous saddle itself, in accordance with the collective mechanism identified in Section~\ref{subsubsec:critical_angular_velocities}: the system swells coherently, the effective moment of inertia diverging while every rotor remains typical.

Finally, the marginal case $a=1$ requires separate treatment within this regime. In this case, $\bar{x}-\bar{r} \approx 0$, making the saddle \eqref{eq:homogeneous_r_saddle} degenerate. Integrating by parts against the Gibbs measure yields the equipartition identity 
\begin{align}
    \langle \vec{s}\cdot \nabla V^{\Omega}\rangle = \frac{2}{\beta}
\end{align}
At the self-consistent saddle,
\begin{align}
    2\kappa Q + \bar{r}^2-\bar{x}^2 = \frac{2}{\beta}
\end{align}
Solving the angular self-consistency equation \eqref{eq:xbar_selfconsistency} for $\bar{x}$,
\begin{align}
    \bar{x}^2 = \frac{\bar{r}^2}{2}-\frac{1}{2\beta} + \frac{\bar{r}}{2}\sqrt{\bar{r}^2-\frac{2}{\beta}}
\end{align}
Therefore, in terms of the dimensionless variable $\bar{r} = \bar{r}_0\beta^{-1/2}$,
\begin{align}
    \bar{r}^2-\bar{x}^2 = \frac{1}{2\beta}\left(\bar{r}_0^2 + 1-\bar{r}_0^2\sqrt{1-\frac{2}{\bar{r}_0^2}}\right)
\end{align}
In the ordered phase near the critical point, $\bar{r}_0$ is large, hence
\begin{align}
    \bar{r}^2 - \bar{x}^2 \approx \frac{1}{\beta}
\end{align}
Therefore, using the equipartition identity, we have that
\begin{align}
    Q \approx \frac{1}{2\kappa \beta}
\end{align}
Since $\kappa_c =0$, 
\begin{align}
    \kappa = \frac{1}{2}(\Omega_c^2-\Omega^2)
\end{align}
It follows that 
\begin{align}
    Q \approx \frac{1}{\beta (\Omega_c^2-\Omega^2)}
\end{align}
Therefore, the effective moment of inertia and, consequently, the angular momentum density diverge more strongly in the marginal case $a=1$, but by the same mechanism as the $a<1$ regime.

\subsection{Condensate of angular momentum when $a>1$}
\label{subsec:Condensate}
In this section, we restrict the constant parameters $(a,b)$ to the sector
\begin{align}
    0<b<b_*, \quad 1<a<a_c(b)
\end{align}
of \eqref{eq:sector_split}. As shown in Section \ref{subsec:Introducing_SF_model}, the critical angular velocity in this regime is set by the centrifugal instability,
\begin{align}
    \Omega_c = \sqrt{2b}
\end{align}
Unlike in the $a\leq 1$ regime, as $\Omega \to \Omega_c$ in the fixed-$\Omega$ ensemble the angular momentum density does not diverge. Instead, it approaches the finite limiting value derived in Section \ref{subsubsec:critical_angular_velocities},
\begin{align}
    \ell_{\mathrm{c}} = \Omega_c Q_{\mathrm{max}}, \qquad Q_{\mathrm{max}} = \frac{9-4a}{3\beta(a-1)}
\end{align}
where $Q_{\mathrm{max}}$ is the moment of inertia of the homogeneous saddle at the threshold. The fixed-$\Omega$ ensemble therefore terminates at $\ell_{\mathrm{c}}$, and ensembles with $\ell>\ell_{\mathrm{c}}$ must be described by a different equilibrium configuration.

The nature of this configuration is suggested by the instability itself. At $\Omega=\Omega_c$, the centrifugal potential exactly cancels the harmonic confinement of an individual rotor since the effective single-particle radial potential is $\kappa r^2$, which becomes flat at the threshold. A single rotor can therefore be displaced to an arbitrarily large radius at no confinement cost, paying only the linear interaction energy $(a\bar{r}-\bar{x})r$. If it is displaced to a macroscopic radius $R\sim\sqrt{N}$, it contributes an extensive amount $\Omega_c R^2 \sim N$ to the total angular momentum, while its interaction energy is only $\mathcal{O}(\sqrt{N})$, sub-extensive in density. This suggests that for $\ell>\ell_{\mathrm{c}}$ the equilibrium is inhomogeneous. In other words, it consists of a bulk of rotors at the critical homogeneous state supplying $Q_{\mathrm{max}}$, together with a ``condensate'' of one or more rotors at macroscopic radius carrying the entire excess. We now verify this picture variationally and show that the condensate consists of exactly one rotor. 

Suppose the excess moment of inertia is carried by $m$ rotors displaced to macroscopic radii $r_i = \sqrt{u_i}$, so that
\begin{align}
    \sum_{i=1}^m u_i = N\mathrm{\Delta} Q, \qquad \mathrm{\Delta} Q := Q - Q_{\mathrm{max}}
\end{align}
Since every carrier sits at macroscopic radius, its statistical weight is deep in the asymptotic regime of the Bessel function, $\log I_0(\beta\bar{x}r)\approx \beta\bar{x}r$, so the angular fluctuations of the carriers are frozen along the ordered direction and their free-energy cost reduces to the mean-field interaction energy,
\begin{align}
\label{eq:condensate_cost}
    F_{\mathrm{c}} = \left(a\bar{r}-\bar{x}\right)\sum_{i=1}^m \sqrt{u_i} -\frac{1}{\beta} h(m)
\end{align}
where $a\bar{r}-\bar{x}>0$ in the ordered phase, and $h(m)$ is the mixing entropy,
\begin{align}
    h(m) = -m\log\frac{m}{N}-\left(N-m\right)\log \left(1-\frac{m}{N}\right)
\end{align}
Because the square root is concave, the cost at fixed total excess is minimized by concentrating the excess in a single rotor,
\begin{align}
    \sum_{i=1}^m \sqrt{u_i} \;\geq\; \sqrt{\sum_{i=1}^m u_i} = \sqrt{N\mathrm{\Delta} Q}
\end{align}
with equality if and only if only one $u_i$ is nonzero. The mixing entropy gained by splitting the excess among $m$ carriers can never compensate for this cost, even if $m$ grows with $N$. In the entropically most favorable case of equal sharing, 
\begin{align}
    \sum_i \sqrt{u_i} = \sqrt{mN\mathrm{\Delta} Q}
\end{align}
an energetic excess of order $\sqrt{mN}$ over the single-carrier cost, while the mixing entropy is 
\begin{align}
    h(m)\approx m\log\left(\frac{N}{m}\right)
\end{align}
Since $\sqrt{mN} \gg m\log(N/m)$ for every subextensive $m\ll N$, all such configurations are suppressed. More generally, any configuration with $m$ carriers at radii $r_i\geq r_0$ pays, by concavity, an excess of at least order $(m-1)r_0$ which dominates $h(m)\leq m\log N$ whenever $r_0\gg \log N$. If the condensate contains additional rotors at microscopic radius, then any such ``condensate" rotors with $u_i \sim \mathcal{O}(1)$ are statistically indistinguishable from a bulk rotor, so such configurations are relabelings of the same macrostate rather than distinct ones. This excludes the remaining case where $m = \mathcal{O}(N)$. The only genuine multiplicity is the choice of which rotor carries the excess, contributing an entropy density $\log N/ N\to 0$. We remark that the same conclusion holds without invoking the asymptotic regime: the exact single-carrier cost $a\bar{r}r - \beta^{-1}\log I_0(\beta\bar{x}r)$ is strictly increasing and strictly concave in $r$, since $\log I_0$ is convex, so the minimizing configuration is a vertex of the constraint simplex for any $N$. The condensate is therefore a single rotor at radius
\begin{align}
    R = \sqrt{N\mathrm{\Delta} Q}
\end{align}
This inhomogeneous configuration is self-consistent. The condensate rotor is aligned with the ordered direction, so the collective moments of the full population are
\begin{align}
    \bar{r} = \left(1-\frac{1}{N}\right)\bar{r}_b + \frac{R}{N} = \bar{r}_b + \mathcal{O}\left(N^{-1/2}\right), \qquad
    \bar{x} = \bar{x}_b + \mathcal{O}\left(N^{-1/2}\right)
\end{align}
The condensate shifts the collective moments only sub-extensively, so the bulk self-consistency conditions are unmodified in the thermodynamic limit. Therefore, the bulk remains at the critical homogeneous saddle, with moments $\bar{r}_b,\bar{x}_b$ and moment of inertia $Q_{\mathrm{max}}$. 

It remains to determine the equilibrium moment of inertia at fixed $\ell$. The free energy density consists of the rotational terms together with the bulk contribution (see Appendix \ref{sec:Appendix_free_energy_density} for the homogeneous branch),
\begin{align}
\label{eq:f_supercritical}
    f(\ell) = bQ + \frac{\ell^2}{2Q} + f_b + \mathcal{O}\left(N^{-1/2}\right)
\end{align}
where $f_b$ is the free energy density of the bulk at the critical homogeneous saddle, independent of $\ell$, and the $\mathcal{O}(N^{-1/2})$ term is the condensate cost \eqref{eq:condensate_cost}. Here $f_b$ denotes only the interaction and entropic contributions of the critical saddle. The confinement term $bQ$, with $Q = Q_{\text{max}}+\mathrm{\Delta} Q$, is displayed explicitly in the first term so no bulk contribution is double-counted. Since the condensate supplies excess moment of inertia at vanishing cost in density, $Q$ is free to adjust above $Q_{\mathrm{max}}$, and minimizing over $Q$ gives
\begin{align}
    Q_*(\ell) = \frac{|\ell|}{\sqrt{2b}} = \frac{|\ell|}{\Omega_c}
\end{align}
which satisfies $Q_*\geq Q_{\mathrm{max}}$ precisely when $|\ell|\geq \ell_{\mathrm{c}}$. The fixed-$\ell$ free energy density on the supercritical branch is therefore linear,
\begin{align}
    f(\ell) = \sqrt{2b}\,|\ell| + f_b, \qquad |\ell|\geq \ell_{\mathrm{c}}
\end{align}
The slope of this branch is the angular velocity, $\partial f/\partial\ell = \Omega_c$. Thus, every ensemble with $\ell>\ell_{\mathrm{c}}$ rotates at exactly the critical angular velocity, and additional angular momentum is absorbed entirely by the radius of the condensate rotor at fixed $\Omega$.

\section{Sound mode in a superfluid cluster lattice}
\label{sec:dynamics_of_propagating_SF_model}
In this section, we generalize Section \ref{sec:Cluster_lattice} to incorporate conservation of angular momentum, giving us a genuine superfluid hydrodynamics characterized by a conserved charge and a Goldstone boson. 
\subsection{Kinetic theory in a single cluster}
\label{subsec:Kinetic_theory_of_SF_model}

In Section \ref{subsec:Introducing_SF_model}, we identified the most natural Hamiltonian model with momenta which admits an entropic-ordered superfluid phase: 
\begin{align}
    H = \frac{1}{2}\sum_i \left(p_{x_i}^2+p_{y_i}^2\right) +  \frac{1}{2N}\sum_{i\neq j} \left(ar_ir_j -\vec{s}_i\cdot \vec{s}_j\right) + b\sum_i r_i^2
\end{align}
Following the method outlined in Section \ref{subsec:dynamics_of_soft_rotor_model}, we now formulate the mean-field theory in terms of the effective single-particle probability distribution
\begin{align}
    P(x,y,p_x,p_y,t) = \frac{1}{N}\sum_{i=1}^ N \delta (x-x_i(t))\delta(y-y_i(t))\delta(p_x-p_{x_i}(t))\delta(p_y-p_{y_i}(t))
\end{align}
In the mean-field limit, each rotor is described by the following mean-field Hamiltonian which depends on the distribution $P(x,y,p_x,p_y,t)$ of all rotors:
\begin{align}
    \bar{H}(x,y,p_x,p_y,t) = \frac{p_x^2}{2}+\frac{p_y^2}{2} + \left(a\bar{r}[P]+br\right)r -\bar{x}[P]x-\bar{y}[P]y.
\end{align}
\subsubsection{Constructing the effective kinetic equation}
\label{subsubsec:constructing_effective_kinetic_eq}
We now construct the deterministic mean-field PDE for the soft rotor model with momentum. In Section \ref{subsec:dynamics_of_soft_rotor_model} we derived the mean-field PDE from the microscopic Langevin equations. Here, however, the physically correct noise must conserve angular momentum exactly, as we do not want to explicitly break the U(1) symmetry generated by $L$ that the superfluid spontaneously breaks. Angular-momentum conserving noise is necessarily multiplicative in the $(p_x,p_y,x,y)$ coordinates, which makes its construction at the level of It\^o SDEs subtle.\footnote{This Fokker--Planck equation is realized microscopically by Langevin dynamics with multiplicative noise, schematically
$dX_a = \{X_a,\bar{H}\}\,dt - \beta\Gamma_{ab}\partial_b\bar{H}\,dt + \partial_b\Gamma_{ab}\,dt + B_{ab}\,dW_b$ with $(BB^{\mathsf{T}})_{ab} = 2\Gamma_{ab}$, where the It\^o drift correction $\partial_b\Gamma_{ab}$ is required to reproduce the divergence form of \eqref{eq:nonlinear_FPE_momenta}. We will not need this explicit construction in what follows.} It is therefore much cleaner to work directly at the level of the Fokker-Planck equation, where each physical requirement can be imposed transparently. We take as our starting point the nonlinear Fokker-Planck equation
\begin{align}
\label{eq:nonlinear_FPE_momenta}
    \partial_t P = \{\bar{H}[P],P\} + \partial_a \left(\Gamma_{ab} \left(\partial_b + \beta \partial_b \bar{H}[P]\right)P\right)
\end{align}
where the indices $a,b$ run over the phase space coordinates $\{x,y,p_x,p_y\}$. The non-dissipative term is the mean-field limit of the microscopic Liouville evolution and is fixed unambiguously. The dissipative term is the most general form consistent with the fluctuation-dissipation theorem. By construction, it annihilates any Boltzmann weight of the mean-field Hamiltonian. It remains to constrain the mobility matrix $\Gamma$. Recall that the microscopic dynamics conserves the total angular momentum: \begin{equation}
    \{H,L_{\text{tot.}}\}=0
\end{equation}
where
\begin{align}
     L_{\text{tot.}} = \sum_i x_ip_{y_i}-y_ip_{x_i}
\end{align}
so the kinetic equation must conserve angular momentum exactly. In the mean-field description, the single-particle angular momentum is $L = xp_y-yp_x$. We demand that the dissipative sector conserve $L$ for every realization of noise; hence we need $\Gamma_{ab}=\Gamma_{ba}$ (noise variance must be symmetric) and 
\begin{align}
    \Gamma_{ab}\partial_bL =0.
\end{align}
Additionally, to avoid explicitly breaking $\mathrm{U}(1)$ symmetry, we require $\Gamma$ to be $\mathrm{U}(1)$-covariant. The mobility matrix satisfying all of the above constraints is found to be
\begin{align}
    \Gamma = Q^TGQ
\end{align}
where in the basis $\{x,y,p_x,p_y\}$,
\begin{align}
\label{eq:mobility_matrix}
     G = \begin{pmatrix}
               \gamma_1 & \gamma_5 & \gamma_6 \\
        \gamma_5 & \gamma_2 &\gamma_4\\
        \gamma_6 & \gamma_4 & \gamma_3
    \end{pmatrix}, \quad Q = \begin{pmatrix}
        p_x & p_y & x & y\\
    x&y&-p_x&-p_y\\
        -y&x&-p_y&p_x
    \end{pmatrix}
\end{align}
Additionally, $\Gamma$ must be positive semidefinite. An equivalent condition is that the damping matrix $G$ be positive semidefinite.  In principle, $G$ can itself depend on the phase space coordinates. In particular, if $G$ is positive definite, then for generic phase space coordinates away from the origin, the kernel of $\Gamma$ is one dimensional,
\begin{align}
\label{eq:kernel_of_Gamma}
    \ker \Gamma = \text{span}\{\nabla L\}
\end{align}
Finally, we demand that the dissipative sector be consistent with detailed balance. Since the microscopic Hamiltonian is time-reversal invariant, the mobility matrix must satisfy the generalized detailed balance/time-reversal symmetry condition \cite{Onsager1931a,Onsager1931b,Casimir1945}
\begin{align}
    \Gamma(\varepsilon z) = \varepsilon \Gamma (z)\varepsilon, \quad \varepsilon=\text{diag}(1,1,-1,-1)
\end{align}
which guarantees that \eqref{eq:nonlinear_FPE_momenta} arises from coupling the system to a time-reversal invariant bath, rather than merely admitting the Boltzmann weight as a stationary state. Under time reversal the frame vectors of $Q$ transform with definite parity, $w_1(\varepsilon z)=-\varepsilon w_1(z)$ and $w_{2,3}(\varepsilon z) = +\varepsilon w_{2,3}(z)$, so for constant damping coefficients the condition forces the couplings between the time-reversal odd channel $w_1$ and the even channels $w_2,w_3$ to vanish\footnote{More generally, the entries of $G$ may be arbitrary functions of the U(1) invariants $(r^2, p^2, s, L)$, where $s = xp_x + yp_y$. Detailed balance 
then requires $G_{11}$, $G_{22}$, $G_{33}$, and $G_{23}$ to be even and $G_{12}$, 
$G_{13}$ to be odd under $(s,L)\to(-s,-L)$, so that the off-diagonal couplings 
to $w_1$ need not vanish but must be proportional to odd invariants, e.g. 
$G_{12} = s\, g_{12}(r^2,p^2,s^2,L^2)$. Positive semidefiniteness must then be 
verified pointwise on phase space. Since such couplings would require a bath 
with rather complex structure, we restrict to constant $G$ throughout.}, leaving
\begin{align}
\label{eq:G_time_reversal}
    G = \begin{pmatrix}
        \gamma_1 & 0&0\\
        0 & \gamma_2 & \gamma_4 \\
        0&\gamma_4 & \gamma_3
    \end{pmatrix}, \qquad \gamma_1> 0, \quad \gamma_2> 0,\quad \gamma_3> 0, \quad \gamma_2\gamma_3 > \gamma_4^2
\end{align}
For the discussion that follows, we restrict $G$ to be positive definite such that \eqref{eq:kernel_of_Gamma} holds.

Next, consider the action of the transport term on the angular momentum. For a general distribution $P$,
\begin{align}
    \{\bar{H}[P],L\} = \bar{x}[P]\, y - \bar{y}[P]\, x
\end{align}
so the single-particle angular momentum is not conserved pointwise. However, the angular momentum moment is conserved by the self-consistent mean-field dynamics, since
\begin{align}
    \frac{\mathrm{d}}{\mathrm{d}t}\langle L\rangle = \big\langle \{\bar{H}[P],L\}\big\rangle = \bar{x}[P]\,\bar{y}[P]-\bar{y}[P]\,\bar{x}[P] =0
\end{align}
Simply put, this captures the fact that angular momentum can be exchanged between particles within the single cluster. Together with $\Gamma_{ab}\partial_bL=0$, this establishes that \eqref{eq:nonlinear_FPE_momenta} conserves the total angular momentum per particle exactly: the dissipative sector by construction, and the transport term by self-consistency of the mean-field moments.

We now turn to stationary solutions of \eqref{eq:nonlinear_FPE_momenta}. The equilibrium distribution
\begin{align}
\label{eq:KT_SF_equilibrium}
    P_{\mathrm{eq}} = \frac{e^{-\beta \bar{H}_{\mathrm{eq}}}}{Z}, \quad \text{where} \quad
    \bar{H}_{\mathrm{eq}} =  \frac{p_x^2}{2}+\frac{p_y^2}{2} + \left(a\bar{r}_{\mathrm{eq}}+br\right)r -\bar{x}_{\mathrm{eq}}x-\bar{y}_{\mathrm{eq}}y
\end{align}
is a stationary solution where the equilibrium moments are determined self-consistently following Section \ref{subsec:dynamics_of_soft_rotor_model}. Indeed, the dissipative term annihilates $P_{\mathrm{eq}}$ by the fluctuation-dissipation form, while $\partial_a P_{\mathrm{eq}} = -\beta P_{\mathrm{eq}}\partial_a \bar{H}_{\mathrm{eq}}$ implies $\{\bar{H}_{\mathrm{eq}},P_{\mathrm{eq}}\}=0$. The conservation of angular momentum promotes $P_{\mathrm{eq}}$ to a one-parameter family of rotating equilibria,
\begin{align}
    P_{\mathrm{eq}}^{\Omega} = \frac{e^{-\beta(\bar{H}_{\mathrm{eq}}-\Omega L)}}{Z}
\end{align}
where the self-consistent moments entering $\bar{H}_{\mathrm{eq}}$ are now computed with respect to $P_{\mathrm{eq}}^{\Omega}$. Since $\Gamma_{ab}\partial_bL=0$, the dissipative term of \eqref{eq:nonlinear_FPE_momenta} is invariant under $\bar{H}\to\bar{H}-\Omega L$, and therefore annihilates the entire family $P_{\mathrm{eq}}^{\Omega}$, not only the $\Omega=0$ member. The transport term, on the other hand, does not vanish on $P_{\mathrm{eq}}^{\Omega}$,
\begin{align}
    \{\bar{H}_{\mathrm{eq}},P_{\mathrm{eq}}^{\Omega}\} = \beta\Omega\, P_{\mathrm{eq}}^{\Omega}\{\bar{H}_{\mathrm{eq}},L\}
\end{align}
However, this non-stationarity is trivial: it merely generates a rigid rotation of the distribution at angular velocity $\Omega$, i.e. the condensate globally precesses but the state is otherwise unchanged. In particular, $P_{\mathrm{eq}}^{\Omega}$ is genuinely stationary in the frame rotating with angular velocity $\Omega$. Denoting the mean-field Hamiltonian in the co-rotating frame as
\begin{align}
    \bar{H}^{\Omega}[P] = \bar{H}[P]-\Omega L
\end{align}
we have $\partial_a P_{\mathrm{eq}}^{\Omega} = -\beta P_{\mathrm{eq}}^{\Omega}\partial_a\bar{H}_{\mathrm{eq}}^{\Omega}$, so that
\begin{align}
    \{\bar{H}_{\mathrm{eq}}^{\Omega}, P_{\mathrm{eq}}^{\Omega}\} =0
\end{align}
In the rotating frame, the kinetic equation \eqref{eq:nonlinear_FPE_momenta} becomes
\begin{align}
\label{eq:rotating_FPE_momenta}
    \partial_t P = \{\bar{H}^{\Omega}[P],P\} + \partial_a\left(\Gamma_{ab}\left(\partial_b +\beta \partial_b\bar{H}^{\Omega}[P]\right)P\right)
\end{align}
where the dissipative term is unchanged by the invariance noted above, and $P_{\mathrm{eq}}^{\Omega}$ is a stationary solution. 

\subsubsection{Zero modes of the linearized theory}
\label{subsubsec:zero_modes_with_momenta}
We now linearize about this rotating equilibrium. Following Section \ref{subsec:linear_MF_PDE}, suppose that the system is perturbed by a small $\phi$ from the stationary distribution such that to linear order in $\phi$, 
\begin{align}
    P = e^{-\beta (\bar{H}_{\mathrm{eq}}^{\Omega}-\phi)} \approx e^{-\beta \bar{H}_{\mathrm{eq}}^{\Omega}} \left(1+\beta \phi\right)
\end{align}
Here, we again impose the normalization condition $\langle \phi|1\rangle =0$. Since the mean-field Hamiltonian depends on the distribution only through the same moments as in Section \ref{subsec:linear_MF_PDE}, its linear perturbation under a shifted distribution is unchanged. This remains true in both the lab and rotating frames. Therefore, to linear order in $\phi$ we have 
\begin{align}
    \bar{r}[P]&= \bar{r}_{\mathrm{eq}} + \langle \phi|r\rangle,\quad 
    \bar{x}[P]= \bar{x}_{\mathrm{eq}} + \langle \phi|x\rangle ,\quad 
    \bar{y}[P]= \bar{y}_{\mathrm{eq}} + \langle \phi|y\rangle
\end{align}
where it is understood that the equilibrium moments and the weighted $L^2$ norm are taken with respect to the rotating stationary distribution $P_{\mathrm{eq}}^{\Omega}$. Additionally, 
\begin{align}
    \bar{H}^{\Omega}[P] = \bar{H}_{\mathrm{eq}}^{\Omega}+\delta \bar{H}(\phi)
\end{align}
where 
\begin{align}
     \delta \bar{H}(\phi)=ar \langle \phi|r\rangle  - x\langle \phi|x\rangle  - y\langle \phi|y\rangle
\end{align}
Expanding the mean-field PDE to linear order in the perturbation $\phi$, we have 
\begin{align}
    \partial_t \phi = \{\bar{H}_{\mathrm{eq}}^{\Omega},\phi + \delta \bar{H}(\phi)\} +\frac{1}{P_{\mathrm{eq}}^{\Omega}}\partial_a\left(\Gamma_{ab}P_{\mathrm{eq}}^{\Omega}\partial_b\left(\phi+\delta \bar{H}(\phi)\right)\right)
\end{align}
By introducing a test function $\Psi$ and integrating by parts, we can express the linearized PDE in terms of the inner-product, defined analogously to \eqref{eq:innerproduct} but now using $P^\Omega_{\mathrm{eq}}$:
\begin{align}
    \langle \Psi|\partial_t\phi\rangle = -\langle \Psi|W|\phi\rangle
\end{align}
where 
\begin{align}
    \langle \Psi|W|\phi\rangle = -\langle \Psi |\{\bar{H}_{\mathrm{eq}}^{\Omega},\phi+\delta \bar{H}(\phi)\}\rangle +  \left\langle \partial_a \Psi|\Gamma_{ab}\partial_b\left(\phi+\delta \bar{H}(\phi)\right)\right\rangle 
\end{align}
Let us define 
\begin{align}
    u[\phi] = \phi + \delta \bar{H}(\phi)
\end{align}
Then any zero mode $\phi$ must satisfy both conditions simultaneously
\begin{align}
\label{eq:zero_mode_conditions}
    \{\bar{H}_{\mathrm{eq}}^{\Omega},u[\phi]\}=0 \quad \text{and} \quad \Gamma_{ab}\partial_bu[\phi]=0
\end{align}
We will work in the ordered frame where $\bar{x}_{\mathrm{eq}}=\bar{s}_{\mathrm{eq}}$ and $\bar{y}_{\mathrm{eq}}=0$. We can immediately identify any mode, $\phi$, which satisfies 
\begin{align}
    u[\phi] = \text{const.}
\end{align}
as a zero mode. Indeed, one can readily check that $u[\phi]=\text{const.}$ trivially satisfies the zero mode conditions \eqref{eq:zero_mode_conditions}. As in Section \ref{subsec:linear_MF_PDE}, since $\delta\bar{H}$ has range $\text{span}\{r,x,y\}$, the condition
\begin{align}
    \phi + \delta \bar{H}(\phi) = \text{const.}
\end{align}
confines $\phi$ to the finite-dimensional space $\text{span}\{1,y,r,x\}$. The constant mode is the unphysical normalization direction, while for $\phi = y$ the transverse reflection $(y,p_x)\to(-y,-p_x)$, which preserves $P_{\mathrm{eq}}^{\Omega}$, gives $\langle y|r\rangle = \langle y|x\rangle = 0$, and the self-consistency conditions again fix the Goldstone normalization $\langle y|y\rangle = 1$ (see \eqref{eq:app_ward}), so that $u[y] = 0$. Therefore,
\begin{align}
    \text{span}\{1,y\}\subset \ker W
\end{align}
In Appendix \ref{subsubsec:Appendix_exact_kernel_SF} we show that, exactly as in the overdamped model, no solution of $u[\phi]=\text{const.}$ has a component along $r$ or $x$ anywhere in the ordered phase, so that in fact
\begin{align}
\label{eq:u_const_iff_span}
    \phi + \delta \bar{H}(\phi) = \text{const.} \iff \phi\in \text{span}\{1,y\}.
\end{align}
Let us now look for additional zero modes outside of $\text{span}\{1,y\}$. We first consider the dissipative sector.  Recall that at generic points of phase space (i.e. for $z\neq (0,0,0,0)$), the null space of $\Gamma$ is one dimensional and spanned by $\nabla L$,
\begin{align}
    \ker \Gamma = \text{span}\{\nabla L\}
\end{align}
This means that any zero mode of the dissipative sector must satisfy
\begin{align}
\label{eq:dissipative_zero_condition}
    \nabla u[\phi] = \alpha \nabla L
\end{align}
where $\alpha$ is a priori an arbitrary function on phase space. However, for $\alpha \nabla L$ to be a gradient, $\alpha$ must be constant along the level sets of $L$. Therefore the zero modes of the dissipative sector are exactly the modes of the form
\begin{align}
\label{eq:dissipative_zero_modes}
    u[\phi] = F(L)
\end{align}
for some function $F$. It remains to check which of these modes also satisfy the transport condition in \eqref{eq:zero_mode_conditions}. For a mode $\phi_0$ of the form \eqref{eq:dissipative_zero_modes}, the transport term gives
\begin{align}
    \{\bar{H}_{\mathrm{eq}}^{\Omega}, u[\phi_0]\} = F'(L)\{\bar{H}_{\mathrm{eq}}^{\Omega},L\} = \bar{x}_{\mathrm{eq}}F'(L) y
\end{align}
This vanishes against arbitrary test functions only if $F'(L) = 0$, which returns us to $u[\phi]=\text{const}$. We conclude that there are no additional exact zero modes,
\begin{align}
    \ker W = \text{span}\{1,y\}
\end{align}
However, if $F'(L) = \text{const.}$, then the transport term maps a zero mode of the dissipative sector into $\ker W$. In other words, a zero mode $\phi$ of the dissipative sector which satisfies
\begin{align}
    u[\phi] = \alpha L, \quad \alpha \in \mathbb{R}
\end{align}
is mapped into $\ker W$ by the transport term and is therefore an exact zero mode of $W^2$. Let us define the mode $\phi_L$ as 
\begin{align}
    u[\phi_L] = L
\end{align}
For $\Omega = 0$ this is simply $\phi_L = L$, but in the rotating ensemble $\phi_L \neq L$, and we give its explicit form at the end of this section. 

The mode $\phi_L$ is the generalized zero mode associated with the conserved charge $L$. Physically, angular momentum is conjugate to the Goldstone angle $\bar{\theta}$, so a charge fluctuation causes the Goldstone mode to drift,
\begin{align}
    \dot{\bar{\theta}} \propto \delta L
\end{align}
An angular momentum perturbation therefore does not relax directly to zero; instead, its time evolution produces a rotation of the order parameter. We can see this explicitly in the linearized dynamics. Consider a perturbation $\phi = \epsilon \phi_L$. The dissipative term annihilates $u[\phi_L]=L$, so the initial evolution is generated by the transport term alone,
\begin{align}
    \partial_t \phi \big|_{t=0} = \epsilon\, \bar{x}_{\mathrm{eq}}\, y
\end{align}
Since $y$ is the Goldstone mode associated with the broken $\mathrm{U}(1)$ symmetry, the perturbation acquires a component along the Goldstone direction at a rate proportional to the angular momentum fluctuation. In operator language,
\begin{align}
    W\phi_L \propto y
\end{align}
and since the Goldstone mode is an exact zero mode,
\begin{align}
    W^2\phi_L = 0
\end{align}
Thus $\phi_L$ is a generalized zero mode: it maps into the Goldstone rotation mode, which then maps to zero. Furthermore, in Appendix \ref{subsec:Appendix_generalized_kernel_W}, we show that the generalized kernel
\begin{align}
    \text{gen}\ker W := \bigcup_{n\geq 1} \ker W^n 
\end{align}
stabilizes at $n=2$,
\begin{align}
    \text{gen}\ker W=\ker W^2 = \text{span}\{1,y,\phi_L\}
\end{align}
Thus, the only generalized zero modes of $W$ are associated with the Goldstone mode and angular momentum. Additionally, in Appendix \ref{subsec:Appendix_Spectral_Gap_momenta}, we outline the proof, given in full in the companion paper \cite{companion}, that the hydrodynamic modes are the only generalized zero modes of $W$ and that all other modes are gapped.

Lastly, we give the explicit form of $\phi_L$ and explain why $\phi_L \neq L$ in the rotating ensemble. Since $W$ acts on a perturbation $\phi$ only through $u[\phi]$, the relevant object is
\begin{align}
      u[L]=L+\delta \bar H(L)
\end{align}
Recall from Section \ref{subsubsec:SF_mean_field_model} that in the rotating ensemble the mean angular momentum at fixed position is set by the rigid rotation profile, $\langle L \rangle = \Omega \langle r^2\rangle$. Consequently, by \eqref{eq:deltaH_centered},
\begin{align}
    \delta \bar{H}(L) = \Omega ar\langle r^2|\tilde{r}\rangle -\Omega x\langle r^2|\tilde{x}\rangle
\end{align}
This vanishes at $\Omega =0$, but for $\Omega \neq 0$ it is a non-zero function in $\text{span}\{r,x\}$. As a result, $L$ itself fails the dissipative zero mode condition: since neither $\nabla r$ nor $\nabla x$ is annihilated by $\Gamma$,
\begin{align}
    \Gamma_{ab}\partial_b u[L] = \Gamma_{ab}\partial_b \delta\bar{H}(L) \neq 0
\end{align}
The mode $\phi_L$ is precisely the correction which removes this self-consistent dressing, retaining only the generalized zero mode content of $L$. Since $u$ maps $\text{span}\{r,x\}$ to itself, the correction lies in $\text{span}\{r,x\}$ and we find
\begin{align}
    \phi_L = L - arC_r +xC_x
\end{align}
where the constants are fixed to be 
\begin{subequations}\begin{align}
   C_r&= \frac{(1-\langle \tilde{x}|\tilde{x}\rangle)\langle L|\tilde{r}\rangle+\langle \tilde{r}|\tilde{x}\rangle\langle L|\tilde{x}\rangle}{(1+a\langle \tilde{r}|\tilde{r}\rangle)(1-\langle \tilde{x}|\tilde{x}\rangle)+a\langle \tilde{r}|\tilde{x}\rangle^2} \\
   C_x &=  \frac{(1+a\langle \tilde{r}|\tilde{r}\rangle)\langle L|\tilde{x}\rangle-a\langle \tilde{r}|\tilde{x}\rangle\langle L|\tilde{r}\rangle}{(1+a\langle \tilde{r}|\tilde{r}\rangle)(1-\langle \tilde{x}|\tilde{x}\rangle)+a\langle \tilde{r}|\tilde{x}\rangle^2}
\end{align}\end{subequations}
Using $\langle L|\tilde{r}\rangle = \Omega\langle r^2|\tilde{r}\rangle$ and $\langle L|\tilde{x}\rangle = \Omega\langle r^2|\tilde{x}\rangle$, both constants are manifestly $\mathcal{O}(\Omega)$, so that $\phi_L = L + \mathcal{O}(\Omega)$, recovering $\phi_L = L$ smoothly in the non-rotating limit.

\subsection{Single Cluster Analysis}
\label{subsec:single_cluster_analysis_SF}
We now construct the MSR effective theory for a single cluster, following the strategy of Section \ref{subsec:MSR_formalism_soft_rotor}. As in the overdamped case, the mean-field kinetic equation \eqref{eq:rotating_FPE_momenta} is nonlinear in $P$ and cannot be cast directly in MSR form. We therefore start from the full $N$-body Fokker-Planck equation on the microscopic phase space, where the dynamics is linear. The construction of the MSR Lagrangian then proceeds exactly as in Section \ref{subsec:MSR_formalism_soft_rotor}, with the isotropic noise kernel $\gamma T$ replaced by the mobility matrix $\Gamma$, and we can jump directly to the slow MSR Lagrangian.

In Section \ref{subsec:Kinetic_theory_of_SF_model}, we showed that the only hydrodynamic degrees of freedom are the collective phase and the total angular momentum, $(\bar{\theta},L)$, with conjugate response fields $(\bar{\pi}_{\theta},\bar{\sigma}_{\theta})$. Three structural features fix the form of the slow MSR Lagrangian. First, rotational invariance of the microscopic Hamiltonian forces $\partial H/\partial \bar{\theta}=0$. Second, total angular momentum is exactly conserved, i.e. $\{H,L\}=0$. Lastly, by construction, the mobility matrix $\Gamma$ conserves angular momentum exactly since $\Gamma\nabla L =0$. Consequently, the dissipative sector is independent of $\bar{\sigma}_{\theta}$. The slow MSR Lagrangian is therefore
\begin{align}
\label{eq:slow_MSR_single_cluster_momenta_form}
    \mathcal{L}_{\text{slow}} = \bar{\pi}_{\theta}\dot{\bar{\theta}}+\bar{\sigma}_{\theta} \dot{L} -\bar{\pi}_{\theta}\left\langle \frac{\partial H}{\partial L} \right\rangle + iD_{\theta}\bar{\pi}_{\theta}^2
\end{align}
where the averages are taken over the gapped modes at fixed $(\bar{\theta},L)$. Additionally, we expand about the non-rotating saddle $\Omega=0$, retaining $L$ as a slow fluctuating field. The coefficient $D_{\theta}$ is the phase mobility. Exactly as in the overdamped model of Section \ref{subsec:MSR_formalism_soft_rotor}, it describes smooth fluctuations of the phase about the ordered state, and it has two contributions. If the dynamics were restricted rigidly to the Goldstone manifold, the direct microscopic noise projected onto the phase would give the bare coefficient
\begin{align}
     D_{\theta}^{(0)} := \langle M^{\theta\theta}\rangle, \quad M^{\theta\theta} = \sum_i \frac{\partial \bar{\theta}}{\partial a_i}\Gamma_{a_i b_i}\frac{\partial \bar{\theta}}{\partial b_i},
\end{align}
where $a_i,b_i \in \{x_i,y_i,p_{x_i},p_{y_i}\}$ and $M^{\theta\theta}$ is the projected mobility matrix onto the Goldstone direction. However, the map from the microscopic phase-space coordinates to the collective phase and the transverse fluctuations is nonlinear, and the transverse coordinates that enter $\Gamma \nabla \bar{\theta}$ are precisely the fluctuating fast variables. Their elimination therefore couples back to the phase, and generates an $\mathcal{O}(N^{-1})$ correction to the mobility, by the same mechanism that produced the factor $(1-\mathcal{C})$ in Section \ref{subsec:Diffusion_Goldstone_mode}. In the overdamped model we could compute that correction exactly because the dynamics is reversible. Here the generator contains the Hamiltonian flow, which is antisymmetric, and the corrector method of Appendix \ref{sec:Appendix_finite_N_phase_diffusion} does not transfer verbatim. We therefore write
\begin{align}
    D_{\theta} = (1-\mathcal{C}_H)D_{\theta}^{(0)} + o\left(\frac{1}{N}\right)
\end{align}
where $\mathcal{C}_H<1$ is an $\mathcal{O}(1)$, non-universal renormalization factor that we do not compute, and $D_{\theta}^{(0)}$ should be interpreted, analogous to the overdamped model, as the rigid-manifold or direct-noise contribution. Three things can nevertheless be said about $\mathcal{C}_H$ without computing it. First, since the fast-mode-induced drift of $\bar{\theta}$ is $\mathcal{O}(N^{-1/2})$ and its correlation time is expected to be $\mathcal{O}(1)$, its contribution to the long-time mobility is $\mathcal{O}(N^{-1})$, the same order as $D_{\theta}^{(0)}$ itself. Therefore, it renormalizes the coefficient of $1/N$ by an order-one factor, which is what $1-\mathcal{C}_H$ represents, but it cannot generate a contribution of lower order in $1/N$. The $N^{-1}$ scaling of $D_{\theta}$ is therefore robust, exactly as in the overdamped model. Second, at $\Omega=0$ every term in the Hamiltonian is quadratic, and the mobility \eqref{eq:mobility_matrix} is quadratic in the phase-space coordinates, so under the rescaling $(x,y,p_x,p_y)\to \sqrt{T}(x,y,p_x,p_y)$ the full kinetic dynamics is independent of $T$. Consequently $\mathcal{C}_{H}$ is a function of $(a,b,G)$ only and is exactly temperature independent. Third, the fast modes cannot cancel the leading coefficient. The global-rotation component of the microscopic noise gives the strict lower bound $D_\theta\geq\gamma_{\mathrm{rot}}/N$ with $\gamma_{\mathrm{rot}}:=\gamma_3 -\gamma_4^2/\gamma_2>0$ (Appendix \ref{subsec:Appendix_rotation_bound}), which is what guarantees $\mathcal{C}_H<1$.

Integrating out the fast modes at the level of the rigid-manifold projection gives (see Appendix \ref{sec:Appendix_single_cluster_SF})
\begin{align}
\label{eq:single_cluster_coefficients_momenta}
   D_{\theta}^{(0)} = \frac{\gamma_3 Q}{N\bar{s}^2}+\frac{\gamma_1+\gamma_2-\gamma_3}{N\beta \bar{s}^2}, \qquad\left\langle \frac{\partial H}{\partial L} \right\rangle  = \frac{L}{NQ}
\end{align}
In what follows we take the simplification $\gamma_1+\gamma_2 =\gamma_3 =\gamma>0$, so that $D_{\theta}^{(0)} = \gamma Q/(N\bar{s}^2)$. The slow MSR Lagrangian is therefore
\begin{align}
\label{eq:slow_MSR_single_cluster_momenta}
    \mathcal{L}_{\text{slow}} = \bar{\pi}_{\theta}\dot{\bar{\theta}}+\bar{\sigma}_{\theta} \dot{L} -\bar{\pi}_{\theta}\frac{L}{NQ} + iD_{\theta}\bar{\pi}_{\theta}^2, \qquad D_{\theta} = (1-\mathcal{C}_H)\frac{\gamma Q}{N\bar{s}^2}
\end{align}
Since $\bar{s}^2\propto T$ and $Q\propto T$ in the ordered region of the quadratic model, $D_{\theta}$ is $\mathcal{O}(N^{-1})$ with a temperature-independent coefficient, whichever value $\mathcal{C}_H$ takes. Note that the total angular momentum of a rigidly rotating cluster is
\begin{align}
   L = \left\langle \sum_i x_ip_{y_i}-y_ip_{x_i}\right\rangle = N\Omega \langle r^2\rangle\implies   \Omega = \frac{L}{NQ}
\end{align}
The equations of motion for a single cluster are then $\dot L = 0$ and 
\begin{align}
     \dot{\bar{\theta}} = \frac{L}{NQ} + \sqrt{2D_\theta}\xi = \Omega+ \sqrt{2D_\theta}\xi
\end{align}
where $\xi$ is Gaussian white noise.
 In contrast with the overdamped model, where the phase was purely diffusive, the phase now has a deterministic drift set by the conserved charge. This is the structure which will produce a propagating sound mode once clusters are coupled on the lattice. The mobility $D_{\theta}$ enters only the noise and, through fluctuation-dissipation, the damping of that mode.  The non-dissipative part of the dynamics is fixed by $\langle\partial H/\partial L\rangle$ alone and is unaffected by $\mathcal{C}_H$.

\subsection{Cluster lattice model with sound modes}
\label{subsec:cluster_lattice_SF_model}
By extending the MSR Lagrangian to include momentum and dissipation that respects conservation of the $\mathrm{U}(1)$ generator, we promote the purely diffusive relaxation of the phase to a propagating sound mode. Just like the previous lattice model without momentum, we consider a lattice where each lattice site is a cluster with $N$ rotors. Each cluster follows the same internal dynamics as in the previous section, but it is now also weakly coupled to its nearest neighbors. Let $\mu,\nu$ denote the site and $i$ denote the rotors within each site. Then the full Hamiltonian is 
\begin{align}
    H = \sum_{\mu} H^{\mu},\qquad  H^{\mu}:=H_0^{\mu} + \sum_{\mu\sim \nu} H_{\mathrm{int}}^{\mu\nu}
\end{align}
where
\begin{align}
    H_0^{\mu} &= \frac{1}{2}\sum_{i\in \mu} \left((p_{x_i}^{\mu})^2+ (p_{y_i}^{\mu})^2\right)+ \frac{1}{2N}\sum_{i\neq j \in \mu}\left(a|s_i^{\mu}||s_j^{\mu}|-\mathbf{s}_i^{\mu}\cdot \mathbf{s}_j^{\mu}\right)+ b\sum_{i\in \mu}|s_i^{\mu}|^2 \\
    H_{\text{int}}^{\mu\nu} &= \frac{\epsilon}{2N}\sum_{i\in \mu}\sum_{j\in \nu}\left(a|s_i^{\mu}||s_j^{\nu}|-\mathbf{s}_i^{\mu}\cdot \mathbf{s}_j^{\nu}\right)
\end{align}
From the analysis of Section \ref{subsec:Kinetic_theory_of_SF_model}, we know that the collective radial modes are fast and thus do not contribute to the analysis of the slow modes. Therefore, we use the results from the single cluster analysis and extend them to the lattice model by summing over all sites,
\begin{align}
    \mathcal{L} = \sum_{\mu}\mathcal{L}_{\mu}
\end{align}
where each $\mathcal{L}_{\mu}$ has the same form as the single cluster MSR Lagrangian,
\begin{align}
\mathcal{L}_{\mu}&=\bar{\pi}_{\theta}^{\mu}\dot{\bar{\theta}}^{\mu} + \bar{\sigma}_{\theta}^{\mu} \dot{L}^{\mu}  -\bar{\pi}_{\theta}^{\mu} \frac{\partial H}{\partial L^{\mu}} + \bar{\sigma}_{\theta}^{\mu} \frac{\partial H}{\partial \bar{\theta}^{\mu}}
    +i D_{\theta}\bar{\pi}_{\theta}^{\mu}\left(\bar{\pi}_{\theta}^{\mu} -i\beta \partial_{\bar{\theta}^{\mu}}H\right)
\end{align}
It remains to integrate out the fast variables. Since the amplitude sector is gapped and all clusters have identical parameters, each cluster's amplitude relaxes to the same equilibrium value, $\bar{s}_{\mu}=\bar{s}$. Therefore, to leading order in the $\Omega=0$ expansion, the diffusion coefficient $D_{\theta}$ is unchanged from \eqref{eq:slow_MSR_single_cluster_momenta}. 
Additionally, the angular inter-cluster interaction term is identical to that in Section \ref{sec:Cluster_lattice}, and the remaining average carries over from the single cluster analysis,
\begin{align}
     \left\langle \frac{\partial H}{\partial \bar{\theta}^{\mu}}\right\rangle   =  \epsilon N \bar{s}^2\sum_{\mu\sim\nu}\sin(\bar{\theta}^{\mu}-\bar{\theta}^{\nu}), \qquad \left\langle \frac{\partial H}{\partial L^{\mu}}\right\rangle&= \frac{L^{\mu}}{NQ_{\mu}}
\end{align}
The effective MSR Lagrangian for the slow modes in the lattice model is then,
\begin{align} \label{eq:Leff_cluster_sound}
    \mathcal{L}&=\sum_{\mu} \left(\bar{\pi}_{\theta}^{\mu}\dot{\bar{\theta}}^{\mu} + \bar{\sigma}_{\theta}^{\mu} \dot{L}^{\mu} - \frac{\bar{\pi}_{\theta}^{\mu}L^{\mu}}{NQ_{\mu}}+ \epsilon N\bar{s}^2\left(\bar{\sigma}_{\theta}^{\mu}+D_{\theta} \beta \bar{\pi}_{\theta}^{\mu}\right)\sum_{\mu\sim\nu}\sin(\bar{\theta}_{\mu}-\bar{\theta}_{\nu}) +iD_{\theta}(\bar{\pi}_{\theta}^{\mu})^2\right)
\end{align}
Following Section \ref{sec:Cluster_lattice}, we now assume the clusters are close to alignment, $\bar{\theta}^{\mu}-\bar{\theta}^{\nu}\ll 1$. In this regime, the inter-cluster term is replaced by a discrete Laplacian and the equations of motion are, neglecting noise:
\begin{subequations}
\begin{align}
\dot{\bar{\theta}}_{\mu} &= \frac{L^{\mu}}{NQ} +\beta N\bar{s}^2D_{\theta} \epsilon\nabla^2\bar{\theta}_{\mu} \\
    \dot{L}^{\mu} &= \epsilon N\bar{s}^2\nabla^2\bar{\theta}_{\mu}
\end{align}
\end{subequations}
Note that summing the second equation over sites gives $\sum_{\mu}\dot{L}^{\mu}=0$. Thus, the inter-cluster coupling transports angular momentum between neighboring clusters while conserving the total, which is precisely the conserved-charge structure required for a propagating sound mode. The quasinormal modes have dispersion relation
\begin{align}
   \omega_{\pm}(k)= -\frac{i\beta N\bar{s}^2D_{\theta} \epsilon}{2}k^2\pm \sqrt{\frac{\epsilon \bar{s}^2 k^2}{Q}-(\beta N\bar{s}^2D_{\theta} \epsilon)^2 \frac{k^4}{4}}
\end{align}
At long wavelengths, $k\to 0$, this reduces to the form of a hydrodynamic sound mode
\begin{align}
    \omega_{\pm}(k) = \pm c_s k - i \Gamma_{\theta} k^2 + \mathcal{O}(k^3)
\end{align}
where the propagation speed and damping are given by 
\begin{align}
    c_s = \sqrt{\frac{\epsilon \bar{s}^2}{Q}},\quad \text{and}\quad \Gamma_{\theta} = \frac{\beta N\bar{s}^2D_{\theta} \epsilon}{2} = \frac{1}{2}\beta \gamma  Q\epsilon (1-\mathcal{C}_H)
\end{align}
Since $\bar{s}^2 \sim T$ and $Q\sim T$, the propagation speed remains finite in the high temperature limit. It follows that $\beta Q \sim \mathcal{O}(1)$, and from the discussion in Section \ref{subsec:single_cluster_analysis_SF} the renormalization $1-\mathcal{C}_H$ is an $\mathcal{O}(1)$ constant. Therefore the damping term also remains finite as $T\to\infty$. As in the overdamped model, the growth of thermal noise is compensated by the growth of the collective amplitude. Moreover, the speed of sound is $\mathcal{O}(1)$ while the phase noise is $D_{\theta}=\mathcal{O}(1/N)$. In the large cluster limit $(N\to\infty)$ the noise vanishes but the sound survives. Thus, the $\mathcal{O}(1/N)$ diffusive phase dynamics of the overdamped model has been promoted to an $\mathcal{O}(1)$ propagating mode.

\section{Conclusion}
\label{sec:Conclusion}
In this paper, we described minimal classical models for entropic-ordered superfluids based on ``soft rotor" XY models.  These models invert the usual phase diagram of a lattice superfluid:  they are symmetric at low temperature and spontaneously break the U(1) symmetry at high temperature.  

Unlike the models introduced in \cite{Han2026}, the models introduced in this paper involve only continuous variables, so it was also natural to study the classical dynamics of the resulting lattice models, including the hydrodynamic limit.  We found that these models in thermal equilibrium are consistent with the fluctuation-dissipation theorem in a non-trivial way -- the entropic order causes the order parameter of superfluidity to blow up as $T\rightarrow \infty$ fast enough to ensure that the hydrodynamic modes are long-lived even at a fixed $k$.\footnote{In another model of entropic-ordered superfluidity, we can't rule out the possibility that the limits $T\rightarrow \infty$ and $k\rightarrow 0$ would not commute, e.g. if the hydrodynamic modes had decay rates $\omega \sim -\mathrm{i}Tk^2$.}   We used a large-$N$ limit in order to make this limit analytically tractable, although we expect that our general conclusions are valid down to $N=1$ as well.  

Our construction illustrates a simple new family of entropic-ordered models on the lattice, distinct from those of \cite{Han2026}, all of which relied on discrete boson degrees of freedom on the lattice.   Moving forward, it might be interesting to compare and contrast how these entropically-ordered systems maintain order in very noisy high-temperature environments with active matter systems, which can also realize ordered phases in unconventional ways \cite{Huang2023GTR,Marchetti2013,TonerTu1995,TonerTu1998,Vicsek1995,Fruchart2021}.  A minimal example of this may be to compare entropic-ordered superfluid dynamics with the dynamics of phase synchronization in the Kuramoto model \cite{Kuramoto1984}.

\section*{Acknowledgements}
 We acknowledge ChatGPT 5.5/5.6 Pro for help with proofreading and checking the calculations.  This work was supported by the National Science Foundation under CAREER Grant DMR-2145544.

\begin{appendix}

\makeatletter


\let\@sectioncntformat\@seccntformat

\let\@hangfrom@section\@hang@from


\renewcommand\section{\@ifstar{\AndySectionStar}{\@startsection{section}{1}{\z@}%

  {-3.5ex \@plus -1ex \@minus -.2ex}%

  {2.3ex \@plus .2ex}%

  {\normalfont\Large\bfseries}}}

\renewcommand\subsection{\@ifstar{\AndySubsectionStar}{\@startsection{subsection}{2}{\z@}%

  {-3.25ex\@plus -1ex \@minus -.2ex}%

  {1.5ex \@plus .2ex}%

  {\normalfont\large\bfseries}}}

\renewcommand\subsubsection{\@ifstar{\AndySubsubsectionStar}{\@startsection{subsubsection}{3}{\z@}%

  {-3.25ex\@plus -1ex \@minus -.2ex}%

  {1.5ex \@plus .2ex}%

  {\normalfont\normalsize\bfseries}}}

\makeatother

\renewcommand{\thesubsection}{\thesection.\arabic{subsection}}
\renewcommand{\theequation}{\thesection.\arabic{equation}}
\renewcommand{\thesubsubsection}{\thesubsection.\arabic{subsubsection}}
\makeatletter
\renewcommand\section{\@ifstar{\AndySectionStar}{\@startsection {section}{1}{-\parindent}%
  {-3.5ex \@plus -1ex \@minus -.2ex}%
  {2.3ex \@plus .2ex}%
  {\normalfont\Large\bfseries}}}
\renewcommand\subsection{\@ifstar{\AndySubsectionStar}{\@startsection{subsection}{2}{-\parindent}%
  {-3.25ex\@plus -1ex \@minus -.2ex}%
  {1.5ex \@plus .2ex}%
  {\normalfont\large\bfseries}}}
\renewcommand\subsubsection{\@ifstar{\AndySubsubsectionStar}{\@startsection{subsubsection}{3}{-\parindent}%
  {-3.25ex\@plus -1ex \@minus -.2ex}%
  {1.5ex \@plus .2ex}%
  {\normalfont\normalsize\bfseries}}}
\makeatother

\section{Thermodynamics of the overdamped soft-rotor model}
\label{sec:Appendix_free_energy_overdamped}
In this appendix we derive the constrained free-energy density \eqref{eq:free_energy_density_overdamped}. The free energy at fixed order parameter is defined through the constrained partition function,
\begin{align}
    f(\bar s) = -\frac{1}{\beta N}\log Z(\bar s),
    \qquad
    Z(\bar s) = \int\prod_i r_i\,\mathrm{d}r_i\,\mathrm{d}\theta_i\;
    \mathrm{e}^{-\beta H}\;
    \delta^{(2)}\!\Big(N\vec{\bar s} - \sum_i\vec s_i\Big),
\end{align}
with $H$ the microscopic Hamiltonian \eqref{eq:microscopic_overdamped_Hamiltonian}. By the global $\mathrm{U}(1)$ symmetry we may take $\vec{\bar s} = (\bar s, 0)$ without loss of generality. It is convenient to introduce the mean length $\bar r$ as a second collective
variable via,
\begin{align}
   N \int d\bar{r}\,\delta\left(N\bar{r}-\sum_i r_i\right) = 1
\end{align}
since the Hamiltonian is then quadratic in the collective variables,
\begin{align}
    H = \frac{N}{2}\left(a\bar r^2 - \bar s^2 + 2b\bar r\right) - \frac{a-1}{2N}\sum_i r_i^2 .
\end{align}
The last term is $O(1)$ in the thermodynamic limit so it contributes $O(1/N)$ to the free energy density and is therefore dropped in what follows. We exponentiate the $\delta$-functions with Lagrange multipliers,
\begin{align}
&\delta \left(N\bar{s}-\sum_i r_i\cos\theta_i\right)=\int \frac{d\lambda}{2\pi} e^{-i\lambda\left(N\bar{s}-\sum_i r_i\cos\theta_i\right)} \\[1em]
&\delta \left(N\bar{r}-\sum_i r_i\right)=\int \frac{d\rho}{2\pi} e^{-i\rho\left(N\bar{r}-\sum_i r_i\right)}
\end{align}
The multiplier conjugate to the transverse component $\sum_i r_i\sin\theta_i$ vanishes at the saddle point by the $y\to-y$ symmetry of the frame choice, and is omitted. The integrand then factorizes over sites, and to leading exponential order in $N$
\begin{align}
    Z(\bar s) \propto \int\mathrm{d}\bar r\,\mathrm{d}\lambda\,\mathrm{d}\rho\;
    \mathrm{e}^{N\Phi},
    \qquad
    \Phi = -\frac{\beta}{2}\big(a\bar r^2 - \bar s^2 + 2b\bar r\big)
    - \mathrm{i}\rho\bar r - \mathrm{i}\lambda\bar s + \log z_1,
\end{align}
where the neglected prefactors contribute $O(\log N/N)$ to $f(\bar{s})$, and the single-site partition function is
\begin{align}
    z_1 = \int_0^\infty\!\mathrm{d}r\, r\int_0^{2\pi}\!\mathrm{d}\theta\;
    \mathrm{e}^{\,\mathrm{i}\rho r + \mathrm{i}\lambda r\cos\theta}.
\end{align}
At the saddle point the multipliers move onto the imaginary axis, as usual in steepest descent. Stationarity of $\Phi$ with respect to $\bar r$ gives 
\begin{align}
    i\rho = -\beta A, \quad A:= a\bar{r}+b
\end{align}
Additionally, we parametrize 
\begin{align}
    h := \frac{i\lambda}{\beta}, \qquad h \in \mathbb{R}
\end{align}
The remaining saddle conditions over the Lagrange multipliers are then the statements
\begin{align}
    \bar r = \langle r\rangle_1, \qquad \bar s = \langle r\cos\theta\rangle_1,
\end{align}
where $\langle\cdot\rangle_1$ denotes the average with respect to the single-site weight. Therefore, the collective variables must be reproduced by the effective
one-body problem in the fields $(A, h)$, which is the expected mean-field structure. On the saddle, the angular integral gives a Bessel function and the radial integral is elementary,
\begin{align}
    z_1 = 2\pi\int_0^\infty\mathrm{d}r\, r\,\mathrm{e}^{-\beta A r}\,
    I_0(\beta h r)
    = \frac{2\pi A}{\beta^2\left(A^2 - h^2\right)^{3/2}},
\end{align}
Differentiating $\log z_1$ with respect to the fields evaluates the one-body averages and puts the saddle conditions in the closed form,
\begin{align}
    \bar{r} = \frac{3A}{\beta(A^2-h^2)}-\frac{1}{\beta A}, \qquad \bar{s} = \frac{3h}{\beta(A^2-h^2)}
\end{align}
Substituting the saddle values back into $\Phi$, the terms reorganize to
\begin{align}
    \Phi_* = \frac{\beta}{2}\big(a\bar r^2 + \bar s^2\big) - \beta h\bar s
    + \log\!\left[\frac{2\pi A}{\beta^2(A^2-h^2)^{3/2}}\right],
\end{align}
In the large $N$ limit, the integral is dominated by its saddle point. Therefore, the constrained partition function is 
\begin{align}
    Z(\bar{s}) \propto e^{N\Phi_*}
\end{align}
The free energy density is,
\begin{align}
\label{eq:free_energy_density_overdamped}
f(\bar{s}) = -\frac{1}{2}a\bar{r}^2 - \frac{1}{2}\bar{s}^2 +\bar{s}h -\frac{1}{\beta}\log \left[\frac{2\pi A}{\beta^2(A^2-h^2)^{3/2}}\right]
\end{align}
We now establish which branch globally minimizes the free energy. By the envelope theorem, 
\begin{align}
    \frac{d f}{d\bar{s}} = h-\bar{s}
\end{align}
Thus, at a stationary point of the unconstrained free energy, $h=\bar{s}$, and the saddle equations reduce precisely to the self-consistency equations derived in Section \ref{subsec:phase_diagram_of_model}. The solutions found previously are therefore stationary points of the thermodynamic free energy. To determine their stability, we expand about the disordered state. Global $\mathrm{U}(1)$ symmetry implies that the free energy is even in $\bar{s}$,
\begin{align}
    f(\bar{s}) = f(0) + \frac{\alpha_2}{2}\bar{s}^2 + \frac{\alpha_4}{4}\bar{s}^4 +  \mathcal{O}(\bar{s}^6)
\end{align}
Correspondingly,
\begin{align}
    h(\bar{s}) &= h_1\bar{s} + h_3\bar{s}^3 +\mathcal{O}(\bar{s}^5)\\
    A(\bar{s})&= A_0 + A_2\bar{s}^2 + \mathcal{O}(\bar{s}^4)
\end{align}
so that
\begin{align}
    \alpha_2=h_1-1, \qquad  \alpha_4=h_3, 
\end{align}
Expanding the saddle equations and matching powers of $\bar{s}$ gives
\begin{align} 
A_0 &= \frac{1}{2}\left(b+\sqrt{b^2+\frac{8a}{\beta}}\right)\\[1em]
\alpha_2 &= \frac{A_0^2\beta}{3}-1 \\[1em]
\alpha_4 &= -\frac{A_0^4\beta^3 (A_0^2\beta -4a)}{27(2a+A_0^2\beta)} 
\end{align}
The quadratic coefficient changes sign at
\begin{align}
    \beta_c = \frac{3}{A_0^2}= \frac{(3-2a)^2}{3b^2},
\end{align}
in agreement with the point at which the symmetry-broken self-consistent branch appears, derived in Section \ref{subsec:phase_diagram_of_model}. For $1<a<3/2$, we have $\alpha_2>0$ for $\beta>\beta_c$, so the disordered solution is locally stable. At the critical point,
\begin{align}
    \alpha_2=0, \qquad \alpha_4>0,
\end{align}
while for $\beta<\beta_c$, the quadratic coefficient becomes negative, so the disordered solution becomes unstable and the minimum moves continuously to $\bar{s}\neq0$. Since the self-consistency equations exhaust the stationary mean-field branches, there are no additional extrema of the free energy to consider. Therefore, the system undergoes a continuous phase transition at $T=T_c$, from the disordered phase with $\bar{s}=0$ for $T<T_c$ to the symmetry-broken phase with $\bar{s}\neq0$ for $T>T_c$.
\section{Spectral properties of the overdamped model}
\label{sec:Appendix_spectral_properties_of_overdamped_model}
\subsection{Kernel of $W$}
\label{subsec:Appendix_kernel_W}
In this subsection, we show
\begin{align}
    \ker W = \text{span}\{1,y\}
\end{align}
Two facts from Section \ref{subsec:linear_MF_PDE} reduce the problem considerably. First, both candidate directions are exact zero modes,
\begin{align}
    \text{span}\{1,y\}\subseteq\ker W
\end{align}
Second, every zero mode must satisfy
\begin{align}
\label{eq:app_zero_mode_condition}
     \phi + \delta \bar{H}(\phi)=\text{const.}
\end{align}
which confines the kernel to the four-dimensional space $\text{span}\{1,y,r,x\}$. It therefore remains to rule out zero modes with components along $r$ and $x$.

As discussed in Section \ref{subsec:linear_MF_PDE}, the component of $\phi$ along the constant function only changes the normalization of the distribution and is unphysical. It is convenient to remove the equilibrium means,
\begin{align}
    \tilde{r} = r-\bar{r}_{\text{eq}}, \qquad \tilde{x} = x-\bar{x}_{\text{eq}}
\end{align}
which span the same space as $r,x$ modulo the constant zero mode, and to replace $\delta \bar{H}$ by its centered version,
\begin{align}
\label{eq:centered_K}
    K\phi := a\tilde{r}\langle \phi|\tilde{r}\rangle -\tilde{x}\langle \phi|\tilde{x}\rangle -y\langle \phi|y\rangle 
\end{align}
For mean zero $\phi$, the centered map $K$ differs from $\delta \bar{H}$ only by a constant function, so the centering does not change the zero mode condition. Moreover, $K\phi$ itself has zero mean, so taking the mean of \eqref{eq:app_zero_mode_condition} fixes the constant to zero. In centered variables the zero-mode condition therefore becomes
\begin{align}
    \phi + \delta \bar{H}(\phi)=\text{const.} \iff \phi + K\phi = 0
\end{align}
and it suffices to show that $\phi + K\phi = 0$ has no nonzero solution in $\text{span}\{\tilde{r},\tilde{x}\}$.

Restricted to this two-dimensional subspace, the condition is a $2\times 2$ linear system. Writing $\phi = c_1 \tilde{r} + c_2 \tilde{x}$, with $c_1, c_2 \in \mathbb{R}$, we collect the basis functions and coefficients into $v = (\tilde{r}\quad \tilde{x})$, and $c^T = (c_1\quad c_2)$, and introduce the Gram matrix of the basis together with the diagonal matrix of the couplings appearing in \eqref{eq:centered_K},
\begin{align}
\label{eq:matrices_ker_overdamped}
    M := \begin{pmatrix}
        \langle \tilde{r}| \tilde{r} \rangle & \langle \tilde{r}| \tilde{x} \rangle \\
        \langle \tilde{r}|\tilde{x} \rangle & \langle \tilde{x}| \tilde{x} \rangle
    \end{pmatrix}, \quad D :=\begin{pmatrix}
        a & 0 \\
        0 & -1
    \end{pmatrix}
\end{align}
In this notation $K\phi = vDMc$, and the zero mode condition takes the matrix form
\begin{align}
   Bc =0,\qquad  B:= \mathbb{I} + DM
\end{align}
There are additional zero modes if and only if $\det B =0$, so it suffices to show that $\det B \neq 0$ throughout the ordered region. Since the inner product carries a factor of $\beta$, the entries of $M$ are $\beta$ times the equilibrium covariances of the single-rotor variables, which follow from derivatives of the single-rotor partition function evaluated at the ordered saddle point detailed in Appendix \ref{sec:Appendix_free_energy_overdamped}, with $h = \bar{x}$ and $A = a\bar{r} + b$, 
\begin{align}
   \langle \tilde{r}|\tilde{r}\rangle &= \beta \text{Var}(r) = \frac{1}{\beta}\frac{\partial^2 \log z_1}{\partial A^2},\\
    \langle \tilde{x}|\tilde{x}\rangle&= \beta \text{Var}(x) = \frac{1}{\beta}\frac{\partial^2 \log z_1}{\partial \bar{x}^2}, \\
     \langle \tilde{r}|\tilde{x}\rangle&=\beta \text{Cov}(x,r) = -\frac{1}{\beta}\frac{\partial^2 \log z_1}{\partial A\partial \bar{x}}
\end{align}
Using the self-consistent stationary identity \eqref{eq:nicerelation},
\begin{align}
\label{eq:nice_relation_appendix}
    A^2 - \bar{x}^2 = \frac{3}{\beta}
\end{align}
and introducing the dimensionless combination $q := \beta A^2$, the entries simplify to
\begin{align}
    \langle \tilde{r}|\tilde{r}\rangle = \frac{2q}{3}-1-\frac{1}{q}, \quad  \langle \tilde{x}|\tilde{x}\rangle = \frac{2q}{3}-1, \quad \langle \tilde{r}|\tilde{x}\rangle = \frac{2\beta A\bar{x}}{3}
\end{align}
The determinant of $B$ is then
\begin{align}
\label{eq:determinant_of_B}
    \det B = \frac{2}{3q}\left(q -3\right)\left(q(a-1)+a\right)
\end{align}
In the ordered phase, $\bar{x}\neq 0$, so \eqref{eq:nice_relation_appendix} gives $q>3$. Additionally, throughout the ordered region, $a>1$. Both factors in \eqref{eq:determinant_of_B} are therefore positive,
\begin{align}
    \det(B)> 0
    \qquad
    \text{for all }
    1<a<\frac32,\quad
    0<b<\sqrt{\frac{T}{3}}(3-2a),\quad T>0.
\end{align}
Hence $B$ is invertible, there are no additional zero modes, so 
\begin{align}
    \ker W = \text{span}\{1,y\}
\end{align}
\subsection{Spectrum of $B$}
\label{subsubsec:Appendix_spectrum_B}

The kernel computation only required $\det B\neq 0$. The spectral gap argument in Appendix \ref{subsec:Appendix_spectral_gap_overdamped} will additionally require that both eigenvalues of $B$ are real and positive, which we will establish here. Since $B$ is not symmetric, positivity of its determinant alone does not imply positivity of its spectrum. However, $M$ is symmetric as well as strictly positive definite:
\begin{align}
    \det M = \frac{\beta A^2 + 3}{3\beta A^2} >0 , \qquad M^T=M
\end{align}
Therefore, there exists a unique symmetric positive definite square root $M^{1/2}$ with inverse $M^{-1/2}$. Conjugating $B$ by this square root produces a symmetric matrix,
\begin{align}
\label{eq:similar_matrices}
    M^{1/2}B M^{-1/2} = M^{1/2} \left(\mathbb{I}+DM\right)M^{-1/2} = \mathbb{I} + M^{1/2}DM^{1/2} =: S
\end{align}
Thus, $B$ is similar to the symmetric matrix $S$. Therefore, the two share the same characteristic polynomial and hence the same eigenvalues, and because $S$ is symmetric, those eigenvalues $\lambda_\pm$ are real, with 
\begin{align}
   \det B= \lambda_+ \lambda_{-} , \qquad \text{tr}(B) = \lambda_{+}+\lambda_{-} 
\end{align}
The trace evaluates to
\begin{align}
    \text{tr}(B) = \frac{2(a-1)q^2 + 3(3-a)q -3a}{3q}
\end{align}
which is an increasing function of $q$,
\begin{align}
    \frac{d}{dq}\text{tr}(B) = \frac{2}{3}(a-1) + \frac{a}{q^2} >0
\end{align}
and using $q > 3$,
\begin{align}
    \text{tr}(B)> 1+ \frac{2a}{3}>0
\end{align}
Real eigenvalues with a positive sum and a positive product are individually positive,
\begin{align}
    \lambda_+,\lambda_{-} >0
\end{align}

\subsection{Spectral gap of W}
\label{subsec:Appendix_spectral_gap_overdamped}
We now show that on the orthogonal complement of the kernel,
\begin{equation}
    V := \left(\operatorname{span}\{1, y\}\right)^{\perp}
\end{equation}
the spectrum of $W$ is bounded away from zero, so that every mode other than the Goldstone mode relaxes at a finite rate. Throughout we work with the centered map $K$ of \eqref{eq:centered_K}. Since we take all $\phi\perp 1$, $K\phi$ differs from $\delta \bar{H}(\phi)$ by only a constant function. Therefore, for all $f,g\in V$,
\begin{align}
    \langle f| W|g\rangle = \frac{\gamma}{\beta}\langle \nabla f| \nabla g\rangle + \frac{\gamma }{\beta}\langle \nabla f|\nabla \delta \bar{H}(g)\rangle = \frac{\gamma}{\beta}\langle \nabla f| \nabla g\rangle + \frac{\gamma }{\beta}\langle \nabla f| \nabla Kg \rangle 
\end{align}
A direct bound on the quadratic form of $W$ fails because $W$ is not self-adjoint in the $L^2(P_{\text{eq}})$ norm. This means its quadratic form can in fact be negative for some perturbations orthogonal to the kernel, even though the spectrum of $W$ is strictly positive there. However, this does not mean that $W$ does not have a spectral gap on $V$. For a non-self-adjoint operator, positivity of the spectrum guarantees eventual exponential decay but permits transient growth of the $L^2(P_{\text{eq}})$ norm, so no bound of the form 
\begin{align}
    \langle \phi| W|\phi \rangle \geq \lambda \|\phi\|^2, \quad \lambda >0
\end{align}
can hold. The resolution is to measure perturbations with an equivalent modified norm for which $W$ is self-adjoint. This section will proceed as follows. First, we will construct a modified norm where $W$ is self-adjoint and show that this norm is equivalent to the equilibrium norm. Next, we establish a spectral gap for $W$ in the modified norm. Then, we convert back to the $L^2(P_{\text{eq}})$ norm via norm equivalence and show that all modes other than the Goldstone mode are gapped and relax at a finite rate.

Let us define the modified norm as 
\begin{align}
    (f| f) := \langle Uf|f\rangle, \qquad U:= \mathbb{I} + K
\end{align}
In this inner product $W$ is self-adjoint,
\begin{align}
    (f|W|g) = \langle Uf| Wg\rangle =\frac{\gamma}{\beta}\langle \nabla Uf | \nabla Ug\rangle  = (g| W|f) 
\end{align}
We will now show that the modified norm is equivalent to the $L^2(P_{\text{eq}})$ norm. Let $f\in V$ and decompose $f = f_{\parallel} + f_{\perp}$ such that $f_{\parallel}\in \text{span}\{\tilde{r},\tilde{x}\}$ and $f_{\perp} \in (\text{span}\{\tilde{r},\tilde{x}\})^{\perp}\cap V$. Two facts make this decomposition nice in this norm. First, $K f_{\perp} = 0$, so $U f_{\perp} = f_{\perp}$ and
\begin{equation}
\label{eq:modified_split}
    (f|f) = \lVert f_{\perp} \rVert^2 + \langle U f_{\parallel}| f_{\parallel} \rangle
\end{equation} 
Second, writing $f_{\parallel} = c_1 \tilde{r} + c_2 \tilde{x}$, with $c_1,c_2\in \mathbb{R}$, the fact that $M$ \eqref{eq:matrices_ker_overdamped} is the Gram matrix of $(\tilde{r}, \tilde{x})$ gives
\begin{align}
\label{eq:f_parallel_spec_W_overdamped}
    \|f_{\parallel}\|^2 = c^TMc
\end{align}
By \eqref{eq:modified_split}, the orthogonal component $f_{\perp}$ is unchanged in the modified norm, so the two norms are equivalent if there exist positive $\alpha_1, \alpha_2\in \mathbb{R}$, such that for all $f\in V$,
\begin{align}
    \alpha_1\|f_{\parallel}\|^2 \leq \langle Uf_{\parallel}|f_{\parallel}\rangle \leq \alpha_2 \|f_{\parallel}\|^2
\end{align}
From $U f_{\parallel} = vBc$ and the Gram matrix,
\begin{align}
    \langle U f_{\parallel}|f_{\parallel}\rangle = (Bc)^TMc = c^T(M+MDM)c
\end{align}
Since $M$ is positive definite (Appendix \ref{subsubsec:Appendix_spectrum_B}), we may change variables to 
\begin{align}
    \tilde{c} := M^{1/2}c, \qquad c = M^{-1/2}\tilde{c}
\end{align}
Then, 
\begin{align}
    \|f_{\parallel}\|^2 = |\tilde{c}|^2, \qquad  \langle Uf_{\parallel}|f_{\parallel}\rangle =  \tilde{c}^TS\tilde{c}
\end{align}
with the same symmetric matrix $S$ defined in \eqref{eq:similar_matrices}. Its Rayleigh quotient yields the following bound,
\begin{align}
    \lambda_{\min}(S) |\tilde{c}|^2 \leq \tilde{c}^TS\tilde{c}\leq   \lambda_{\max}(S) |\tilde{c}|^2
\end{align}
In section \ref{subsubsec:Appendix_spectrum_B}, we established that $B$ and $S$ have equivalent eigenvalues which are real and strictly positive. Let us therefore define the following positive constants, 
\begin{align}
    \tau_1 = \min\{1,\lambda_{\min}(B)\}, \qquad   \tau_2 = \max\{1,\lambda_{\max}(B)\}
\end{align}
Combining all the above steps establishes norm equivalence (up to O(1) prefactor): 
\begin{align}
\label{eq:norm_equivalence_overdamped}
    \tau_1 \|f\|^2 \leq (f|f)\leq \tau_2 \|f\|^2
\end{align}

Next, we show that the $L^2(P_{\mathrm{eq}})$ norm satisfies a Poincar\'e inequality. If $P =  e^{-\mathcal{V}(x)}$ then $P$ is said to be a log-concave measure if $\mathcal{V}(x)$ is convex. Then Bobkov's theorem  says that any log-concave measure on $\mathbb{R}^n$ satisfies a Poincar\'e inequality \cite{Bobkov1999,BrascampLieb1976}. In our case $\mathcal{V} = \beta \bar{H}_{\mathrm{eq}}$ with
\begin{align}
\bar{H}_{\mathrm{eq}} = Ar -\bar{x}_{\mathrm{eq}}x
\end{align}
which is convex since $r= \sqrt{x^2+y^2}$ is convex, $A>0$, and the remaining term is linear. Hence, there exists $C_P>0$ such that 
\begin{align}
    \|f-\langle f|1\rangle\|^2 \leq C_{P}\|\nabla f\|^2
\end{align}
Every $f \in V$ has zero mean, and since $Kf$ has zero mean as well, so does $Uf$. Therefore,
\begin{align}
    \|Uf\|^2 \leq C_P\|\nabla Uf\|^2
\end{align}
We can now establish the spectral gap of $W$ on $V$ in the modified norm. A direct evaluation together with the Poincar\'e inequality yields
\begin{align}
    (f|W|f) = \langle Uf|Wf\rangle  = \frac{\gamma}{\beta}\|\nabla Uf\|^2 \geq \frac{\gamma}{\beta C_P}\|Uf\|^2
\end{align}
where
\begin{align}
    \|Uf\|^2 = \|f_{\perp}\|^2 + \tilde{c}^TS^2\tilde{c}
\end{align}
and since $S$ is positive definite, $S^2 \geq \lambda_{\min}S$, so
\begin{align}
    \|Uf\|^2 \geq \|f_{\perp}\|^2 + \lambda_{\min}\langle Uf_{\parallel}|f_{\parallel}\rangle \geq \tau_1 (f|f)
\end{align}
This gives us the spectral gap in the modified norm,
\begin{align}
    (f|W|f) \geq \lambda (f|f) ,\qquad \lambda := \frac{\gamma \tau_1}{\beta C_P}>0
\end{align}
The gap converts directly into exponential decay. For any $f \in V$ evolving under the linearized dynamics,
\begin{align}
\label{eq:decay_rate_modified_norm}
    \frac{d}{dt}(f|f) = 2(f|\partial_tf) = -2(f|W|f) \leq -2\lambda (f|f)
\end{align}
and integrating this differential inequality (Gronwall's lemma) gives
\begin{align}
    (f(t)|f(t))\leq e^{-2\lambda t}(f(0)|f(0))
\end{align}
Transferring back to $L^2(P_{\mathrm{eq}})$ with the norm equivalence \eqref{eq:norm_equivalence_overdamped}, 
\begin{align}
    \|f(t)\|^2 \leq \frac{1}{\tau_1} (f(t)|f(t))\leq \frac{e^{-2\lambda t}}{\tau_1}(f(0)|f(0))\le \frac{\tau_2}{\tau_1}e^{-2\lambda t}\|f(0)\|^2
\end{align}
Therefore every $f \in V$ decays at a finite rate,
\begin{align}
    \|f(t)\|\leq \sqrt{\frac{\tau_2}{\tau_1}}e^{-\lambda t}\|f(0)\|, \quad \lambda >0 
\end{align}
In summary, the zero modes of $W$ correspond to perturbations that do not relax even at the zero wavevector, i.e. they are gapless. We have demonstrated that there are only two zero modes and only one of the zero modes is physical. Additionally, every other mode is gapped, so it decays at a finite rate even at long wavelengths. We have therefore shown that there is only one physical hydrodynamic mode which is the Goldstone mode arising from broken $\mathrm{U}(1)$ symmetry.

\section{Finite-$N$ phase diffusion}
\label{sec:Appendix_finite_N_phase_diffusion}
The purpose of this appendix is to give a complete finite-$N$ derivation of the Goldstone-phase diffusion for the stationary ordered process discussed in Section \ref{subsec:Diffusion_Goldstone_mode}. Because the phase becomes singular near $\bar{s}=0$, we first formulate the problem on a reflected ordered basin that excludes a small neighborhood of the core while preserving stationarity and detailed balance. At fixed $N$ and fixed cutoff, we derive the phase variance and diffusion constant, obtain its variational representation, and use the Kipnis--Varadhan martingale approximation
\cite{KipnisVaradhan1986,KomorowskiLandimOlla2012} to establish the Brownian long-time scaling limit. We then compare the reflected and physical processes up to the first entrance into the excluded near-core region and discuss the corresponding metastable interpretation. Finally, we use the topology of the $\mathrm{U}(1)$ orbit to derive the upper and lower bounds quoted in the main text, establishing $D_\theta^{(s_*)}=\Theta(N^{-1})$ for every fixed admissible cutoff and a uniform high-temperature bound on the relative reduction of the bare phase-diffusion coefficient by the drift term.

The configuration is $X= (\vec{s}_1, ..., \vec{s}_N) \in (\mathbb{R}^{2})^N$, with each rotor $\vec{s}_i =(x_i,y_i)$, evolving by the overdamped Langevin dynamics
\begin{align}
\label{eq:app_microscopic_langevin}
    d\vec{s}_i = -\gamma \nabla_i H dt + \sqrt{2\gamma T} d\vec{W}_i,
\end{align}
where $H$ is the overdamped microscopic Hamiltonian \eqref{eq:microscopic_overdamped_Hamiltonian} and $\vec{W}_i$ are independent standard Brownian motions. The generator of the microscopic overdamped Langevin dynamics is
\begin{align}
\label{eq:app_generator}
    \mathcal{L} =\gamma T\sum_i \nabla_i^2-\gamma\sum_i \nabla_i H\cdot\nabla_i
\end{align}
So for any smooth $f$, the process $f(X_t)$ satisfies It\^o's formula,
\begin{align}
\label{eq:app_Ito_formula}
    f(X_t)-f(X_0) = \int_0^t (\mathcal{L}f)(X_s)ds  + M_t^f, \quad dM_t^f = \sqrt{2\gamma T} \sum_i \nabla_i f(X_t)\cdot d\vec{W}_i(t)
\end{align}
where $M^f$ is a martingale, i.e. a process with no systematic drift, whose quadratic variation is 
\begin{align}
\label{eq:app_martingale_variation}
    [M^f]_t = 2\gamma T\int_0^t \sum_i|\nabla_i f(X_s)|^2 ds
\end{align}
Recall that the purpose of this appendix is to characterize the long-time wandering of the Goldstone mode, the collective phase $\bar{\theta}$, due to equilibrium fluctuations at finite $N$. In the $N\to\infty$ mean-field theory, the Goldstone mode is the unique physical zero mode of the linearized dynamics about the stationary ordered state. The finite-$N$ analogue is therefore obtained by sampling the initial microscopic configuration from the stationary ordered Gibbs measure,
\begin{align}
    d\mu = \frac{e^{-\beta H}d^{2N}X}{Z},\qquad Z = \int d^{2N}X\,e^{-\beta H}
\end{align}
which is stationary for \eqref{eq:app_microscopic_langevin}. Sampling from the Gibbs measure guarantees the correct equilibrium distribution of microscopic configurations, but this alone is not sufficient to ensure that the dynamics of $\bar{\theta}$ reflects equilibrium fluctuations. Reversibility of the finite-$N$ dynamics with respect to the Gibbs measure ensures detailed balance and hence the absence of stationary probability currents or other irreversible dynamical biases. Physically, reversibility means that a movie of the stationary process played backwards is statistically indistinguishable from the movie played forwards. Without reversibility, one could preserve the same Gibbs stationary measure while introducing non-equilibrium currents that alter the long-time dynamics of $\bar{\theta}$. Thus detailed balance is needed to interpret the observed finite-$N$ motion of the zero mode as arising purely from equilibrium fluctuations.

The dynamics are reversible if and only if the generator is symmetric on $L^2(\mu)$. Indeed, integrating by parts with respect to the Gibbs measure over all of $(\mathbb{R}^2)^N$ (the boundary term at infinity vanishes because $e^{-\beta H}$ decays exponentially) yields,
\begin{align}
\label{eq:app_dirichlet_form_full}
\left\langle f\mathcal{L}g \right\rangle_{\mu} = -\int d\mu\, \left(\gamma T\sum_i\nabla_i f\cdot \nabla_i g\right) = \left\langle g\mathcal{L}f \right\rangle_{\mu}
\end{align}
Therefore, $\mathcal{L}$ is self-adjoint on $L^2(\mu)$ which implies the Langevin dynamics satisfy detailed balance, i.e. $\mu$ is reversible. Every step of the corrector method below rests on the symmetric bilinear form in \eqref{eq:app_dirichlet_form_full} (the ``Dirichlet form''), and the main difficulty we have to address is how to keep it true once we restrict attention to the ordered region.

Indeed, the phase, $\bar{\theta} = \arg(\vec{\bar{s}})$, and the phase drift, defined as 
\begin{align}
\label{eq:app_phase_and_drift}
   A :=\mathcal{L}\bar{\theta} = \frac{\gamma (a\bar{r}+b)}{N\bar{s}\bar{r}}\sum_i (r_i-\bar{r})\sin(\theta_i-\bar{\theta})
\end{align}
are only defined on the set $\{\vec{\bar{s}}\neq 0\}$. The expression for $A$ follows because $\bar\theta$ is harmonic in each $(x_i,y_i)$, so the Laplacian term of $\mathcal{L}$ vanishes, and because the definition of $\bar\theta$ implies
\begin{align}
\label{eq:phase_identity}
\sum_i r_i \sin(\theta_i-\bar{\theta}) =0 .
\end{align}
Let $\rho_N(\vec{\bar{s}})$ be the probability density of the order parameter $\vec{\bar{s}}$ under $\mu$, 
\begin{align}
    \rho_N(\vec{\bar{s}}) = \int d\mu(X)\delta^{(2)}\left(\vec{\bar{s}}-\frac{1}{N}\sum_i \vec{s}_i\right)
\end{align}
For $N\geq 2$, this is a continuous, strictly positive function of $\vec{\bar{s}}$, including at $\vec{\bar{s}}=0$. Indeed, the constraint surface $\{\sum_i \vec{s}_i=0\}$ is a $(2N-2)$-dimensional submanifold on which $e^{-\beta H}>0$, so $\rho_N(0)$ is a finite positive number. In the ordered phase, it is exponentially small in $N$, but it is non-zero. Therefore, since $\rho_N$ is continuous and $\rho_N(0)>0$, there exists $\delta >0$ such that $\rho_N(\vec{\bar{s}})>\frac{1}{2}\rho_N(0)$ for $\bar{s}<\delta$. Consequently, in polar coordinates $\vec{\bar{s}} = (r\cos\phi, r\sin\phi)$, 
\begin{align}
\label{eq:app_core_divergence}
    \left\langle \frac{1}{\bar{s}^2}\right\rangle_{\mu} = \int \frac{\rho_N(\vec{\bar{s}})}{|\vec{\bar{s}}|^2}d^2\bar{s} \geq \pi \rho_N(0)\int_0^{\delta} \frac{dr}{r} = \infty 
\end{align}
The divergence is logarithmic and arises from an arbitrarily small neighborhood of the core $\{\vec{\bar{s}}=0\}$. Since the direct noise variance of the phase is 
\begin{align}
    D_{\theta}^{(0)} = \frac{\gamma T}{N}\left\langle \frac{1}{\bar{s}^2}\right\rangle_{\mu}
\end{align}
(see \eqref{eq:app_martingale_variation}), that variance is infinite under the unrestricted Gibbs measure. The same difficulty also arises in the variational formula \eqref{eq:variational_diffusion_constant_app} below, since the phase winds by $2\pi$ around $\vec{\bar{s}}=0$ while any single-valued corrector $u$ does not, so the Dirichlet energy of $\bar{\theta}-u$ also diverges logarithmically at the core for every admissible $u$.

Physically, configurations with very small $\bar{s}$ correspond to near-core excursions during which the phase can undergo large, rapid changes. Under the unrestricted finite-$N$ Gibbs measure, the angular noise coefficient is not square-integrable, as reflected in the divergence of $\langle \bar{s}^{-2}\rangle_\mu$. We do not attempt to characterize the asymptotic winding statistics of the unrestricted process. Instead, our goal is to isolate the smooth Goldstone diffusion within the metastable ordered regime, before the process enters a small-amplitude neighborhood of the core. The difficulty lies in restricting to that regime while preserving stationarity and reversibility of the Gibbs measure. If we condition the Gibbs measure on $\{\bar{s}>\delta\}$ while keeping the dynamics unchanged, then the conditioned measure is no longer stationary (trajectories cross the surface $\bar{s}=\delta$). Moreover, integrating by parts over the truncated domain produces a surface term which breaks the symmetry \eqref{eq:app_dirichlet_form_full} and consequently violates detailed balance making the dynamics on the conditioned Gibbs measure non-reversible. In other words, one cannot change the measure without also changing the dynamics.

Therefore, we define an ordered basin which reflects trajectories back into the relevant region. In the following subsection, we show that this formulation preserves stationarity and reversibility and that the reflection produces no direct boundary contribution to the phase.

\subsection{The ordered basin}
\label{subsec:Appendix_ordered_basin}
The divergence found above comes entirely from configurations in which the order parameter approaches the core $\bar{s}=0$. We therefore exclude a small neighborhood of the core by fixing
\begin{align}
    0<s_*<\bar{s}_{\text{eq}},
\end{align}
such that as $N\to \infty$, $\bar{s}_{\text{eq}} -s_* = \mathcal{O}(1)$. Define the ordered basin and its boundary,
\begin{align}
    \mathcal{B}:= \{X \in (\mathbb{R}^2)^N: \bar{s}(X)\geq s_*\}, \quad \partial \mathcal{B}= \{\bar{s} = s_*\}
\end{align}
For definiteness one may take $s_*=\bar{s}_{\mathrm{eq}}/2$. We keep $s_*$ fixed throughout the finite-$N$ calculation. At the end of this appendix, we discuss the dependence of the static bounds on the choice of fixed positive cutoff and show that the precise choice is unimportant at the order considered.

Before defining the dynamics on $\mathcal{B}$, we record a few simple geometric properties that will be used repeatedly. Since
\begin{align}
    \nabla_i\bar{s}=\frac{\hat{\bar{s}}}{N},\qquad\sum_i|\nabla_i\bar{s}|^2=\frac{1}{N}
\end{align}
the boundary is smooth, and the unit normal pointing into the ordered region is
\begin{align}
    \nu:=\frac{\nabla\bar{s}}{|\nabla\bar{s}|}=\sqrt{N}\nabla\bar{s},\qquad\nu_i=\frac{\hat{\bar{s}}}{\sqrt{N}}
\end{align}
Likewise, away from the core the phase gradient is smooth and satisfies
\begin{align}
    \nabla_i\bar{\theta}=\frac{\hat{\bar{s}}^\perp}{N\bar{s}}, \qquad \hat{\bar{s}}^{\perp} := (-\sin\bar{\theta},\cos\bar{\theta}),\qquad\sum_i|\nabla_i\bar{\theta}|^2=\frac{1}{N\bar{s}^2}\leq\frac{1}{Ns_*^2},
\end{align}
Although $\bar{\theta}$ is not globally single-valued on $\mathcal B$, its gradient is well-defined, and along every continuous trajectory in $\mathcal B$ the phase admits a continuous lift.

The basin is connected. Indeed, any configuration $X\in\mathcal B$ can be continuously deformed, at fixed order parameter $\vec{\bar{s}}$, into the aligned configuration in which every rotor equals $\vec{\bar{s}}$. The aligned configurations with $|\vec{\bar{s}}|\geq s_*$ form a connected annulus. The basin is also invariant under global $\mathrm{U}(1)$ rotations and under reflection of every rotor about the ordered axis. The latter symmetry preserves both $H$ and $\bar{s}$ and maps $A\to -A$.

\subsubsection{Reflected dynamics}

As described above, simply conditioning the Gibbs measure on $\mathcal B$ is not enough, because the original dynamics would still cross the boundary. We now modify the dynamics, rather than the measure, so that the process never leaves $\mathcal{B}$, by placing a reflecting wall at $\bar{s}=s_*$. Physically, the system evolves exactly as before in the interior of the ordered region, while a trajectory that reaches the wall receives only the minimal inward push required to prevent it from entering the excluded near-core region.

The Skorokhod construction is the canonical way to obtain a Markov process whose reversible measure is a given Gibbs measure restricted to a domain \cite{Skorokhod1961,Tanaka1979,LionsSznitman1984}. Therefore, let $X_t^{(s_*)}$ be the solution of the Skorokhod problem, 
\begin{align}
\label{eq:app_reflected_SDE}
    dX_t = -\gamma \nabla H(X_t)dt + \sqrt{2\gamma T}dW_t + \nu(X_t)d\ell_t, \quad X_t \in \mathcal{B} \,\,\forall t
\end{align}
where $\ell_t$ is a continuous non-decreasing process, the ``boundary local time'', that increases only at times when $X_t \in \partial\mathcal{B}$. The process follows the original Langevin dynamics in the interior, and whenever it touches the boundary wall, it receives the minimal inward push, along $\nu$, needed to keep it inside $\mathcal{B}$. Therefore, $\ell_t$ simply keeps count of how much total push has been delivered up to time $t$. Existence and pathwise uniqueness hold because the domain has a smooth boundary, the noise is additive and non-degenerate, and the drift is bounded on bounded sets.\footnote{The drift $-\gamma\nabla_i H$ is discontinuous at $r_i=0$ because of the $|\vec{s}_i|$ terms in $H$. This is a property of the original model and is unrelated to the wall. It does not affect anything since $\nabla_i H$ is bounded near $r_i=0$ (only its direction jumps), and a bounded drift with additive non-degenerate noise gives a well-posed SDE. Moreover $\{r_i=0\}$ has codimension two in the $(x_i,y_i)$-plane and is never visited by the process.\cite{Veretennikov1981}} Then for a smooth function $f$ on $\mathcal{B}$, It\^o's formula acquires one extra term, 
\begin{align}
\label{eq:app_reflected_Ito}
    f(X_t^{(s_*)})-f(X_0^{(s_*)}) = \int_0^t (\mathcal{L}f)(X_s^{(s_*)})ds + M_t^f + \int_0^{t}\left(\nu \cdot \nabla f\right)(X_s^{(s_*)})d\ell_s
\end{align}
with $\mathcal{L}$ the same differential operator \eqref{eq:app_generator} and $M^f$ the same martingale with quadratic variation \eqref{eq:app_martingale_variation}. The last term records the kicks received at the wall, and it is absent precisely for functions with vanishing normal derivative on the wall. This is the usual Neumann boundary condition for a reflecting wall. We denote the corresponding domain by
\begin{align}
    \mathcal{D}_N := \{f\in C^2(\mathcal{B}): \partial_{\nu}f = \nu \cdot \nabla f=0\,\,\text{on}\,\,\partial\mathcal{B}\}
\end{align}
The generator of the reflected process is $\mathcal{L}$ acting on the closure of $\mathcal{D}_N$. Equivalently, the reflected semigroup 
\begin{align}
    P_tf(x) := \mathbb{E}[f(X_t)|X_0=x]
\end{align}
satisfies $\partial_t P_t f = \mathcal{L}P_tf$ with the boundary condition $\partial_{\nu}P_tf=0$. From now on we drop the superscript on $X_t^{(s_*)}$; until Section \ref{subsubsec:Appendix_physical_process} every statement refers to the reflected process.

\subsubsection{Reversibility on the basin}
\label{subsubsec:Appendix_reversibility_basin}
Define the restricted Gibbs measure,
\begin{align}
\label{eq:app_restricted_gibbs}
    d\mu_{s_*} := \frac{e^{-\beta H}\mathds{1}_{\mathcal{B}}\,d^{2N}X}{Z_{s_*}}, \qquad Z_{s_*} := \int_{\mathcal{B}} d^{2N}X\, e^{-\beta H}
\end{align}
where $\mathds{1}_{\mathcal{B}}$ is the indicator function of the basin. The crucial point is that reflection preserves the same equilibrium structure as the unrestricted Langevin dynamics. For $f,g\in\mathcal D_N$, integration by parts gives
\begin{align}
\label{eq:app_surface_term}
     \left\langle f\mathcal{L}g\right\rangle_{\mu_{s_*}}  = \frac{\gamma T}{Z_{s_*}} \oint_{\partial \mathcal{B}} f e^{-\beta H}\left(n\cdot \nabla g\right)d\sigma -\frac{\gamma T}{Z_{s_*}}\int_{\mathcal{B}} e^{-\beta H}\nabla f\cdot \nabla g 
\end{align}
where $n=-\nu$ is the outward unit normal. The Neumann condition makes the surface term vanish. The remaining expression is symmetric under $f\leftrightarrow g$, so for all $f,g \in \mathcal{D}_N$, 
\begin{align}
\label{eq:app_dirichlet_form_basin}
    \left\langle f\mathcal{L}g\right\rangle_{\mu_{s_*}} = -\gamma T\left\langle \sum_i \nabla_i f\cdot \nabla_i g\right\rangle_{\mu_{s_*}}= \left\langle g\mathcal{L}f\right\rangle_{\mu_{s_*}} 
\end{align}
Thus $\mu_{s_*}$ is stationary and the reflected dynamics are reversible. The surface term in \eqref{eq:app_surface_term} is exactly the term that prevents the conditioned-measure dynamics described above from being reversible. Reflecting the dynamics enforces the Neumann condition, and the Neumann condition is exactly what removes the surface term. The reflected process is, by construction, the process whose Dirichlet form is the restriction of \eqref{eq:app_dirichlet_form_full} to $\mathcal{B}$. Two consequences are used repeatedly. First, if the process starts in the stationary state $\mu_{s_*}$, then for every window $[0,t]$, the reversed trajectory
\begin{align}
    \tilde{X}_u := X_{t-u}, 
\end{align}
has the same law, as a process, as $(X_u)_{u\in[0,t]}$. Second, since $\mathcal{B}$ is connected, the only functions annihilated by $\mathcal{L}$ are the constants. Indeed if $\mathcal{L}f = 0$ then
\begin{align}
    0 = \langle f\mathcal{L}f\rangle_{\mu_{s_*}} = -\gamma T\left\langle\sum_i|\nabla_i f|^2\right\rangle_{\mu_{s_*}}
\end{align}
so $\nabla f = 0$ and $f$ is constant. Hence, the reflected process is ergodic.

\subsubsection{Absence of a direct phase contribution from reflection}
\label{subsubsec:Appendix_phase_blind}
The wall pushes the configuration along $\nabla\bar{s}$. Since this direction does not rotate the order parameter, the reflected local-time term makes no direct contribution to the phase. Indeed,
\begin{align}
\label{eq:app_phase_blind}
    \nu\cdot \nabla\bar{\theta} = \sqrt{N}\sum_i \nabla_i \bar{s}\cdot \nabla_i\bar{\theta} = \frac{1}{\sqrt{N}\bar{s}}\,\hat{\bar{s}}\cdot \hat{\bar{s}}^{\perp} = 0
\end{align}
everywhere on $\mathcal{B}$. As noted above, the phase admits a continuous lift along every trajectory in $\mathcal{B}$, while its gradient is globally well-defined. Applying the reflected It\^o formula locally to this lifted phase, the boundary term is proportional to $\nu\cdot\nabla\bar{\theta}$ and therefore vanishes identically by \eqref{eq:app_phase_blind}. Thus the reflected wall changes the radial motion of the order parameter but never directly rotates it. Consequently, the phase obeys the same It\^o decomposition as in the unrestricted dynamics,
\begin{align}
\label{eq:app_Ito_direct_theta}
    \bar{\theta}(t)-\bar{\theta}(0)=\int_0^t A(X_s)\,ds+M_t^\theta, \qquad \mathbb{E}\left[(M_t^{\theta})^2\right] = \frac{2\gamma T}{N}t\left\langle \frac{1}{\bar{s}^2}\right\rangle_{\mu_{s_*}}
\end{align}
The only difference is that now $\bar{s}\geq s_*$, so every quantity appearing in this equation is finite:
\begin{align}
    \left\langle\frac{1}{\bar{s}^2}\right\rangle_{\mu_{s_*}} \leq\frac{1}{s_*^2},
\end{align}
and $A\in L^2(\mu_{s_*})$. Moreover, transverse-reflection symmetry
preserves $\mu_{s_*}$ and sends $A\to-A$, so
\begin{align}
    \langle A\rangle_{\mu_{s_*}}=0.
\end{align}
Thus reflection removes the singular near-core region without adding any direct boundary contribution to the phase dynamics. The reflected and physical trajectories are compared separately below, where they are shown to coincide exactly up to the first wall hit.

\subsubsection{Relation to the physical dynamics}
\label{subsubsec:Appendix_physical_process}
The reflected process is an auxiliary equilibrium process, but it agrees exactly with the original dynamics until the latter first attempts to enter the excluded core region. To see this, drive the original process \eqref{eq:app_microscopic_langevin} and the reflected process \eqref{eq:app_reflected_SDE} by the same Brownian motion from the same initial condition in $\mathcal{B}$. They coincide pathwise up to the first hitting time of the wall, 
\begin{align}
    \tau_{s_*} := \inf\{t\geq 0: \bar{s}(X_t)\leq s_*\}
\end{align}
since before that time the local time $\ell_t$ vanishes. Hence, for any functional of the path on $[0,t]$, the expectation restricted to the event $\{t<\tau_{s_*}\}$ is the same for both processes. The reflected calculation therefore describes the physical dynamics on timescales before the first entrance time into the near-core region. This motivates the metastable window
\begin{align}
\label{eq:app_window}
    \tau_{\mathrm{rel}}
    \ll
    t
    \ll
    \tau_{s_*}(N),
\end{align}
where $\tau_{\mathrm{rel}}$ denotes the relaxation time of the non-Goldstone modes. The fixed-$N$ invariance principle derived below does not require an $N$-uniform estimate on $\tau_{\mathrm{rel}}$. Establishing a joint large-$N$ limit for the physical process on this window would additionally require uniform mixing and exit-probability estimates, which we do not prove here. To estimate the width of the window, recall that Appendix \ref{sec:Appendix_free_energy_overdamped} computes the constrained free energy per rotor $f(\bar{s})$ at fixed order parameter. In the ordered phase it has an unstable maximum at $\bar{s}=0$ and its minimum at $\bar{s}_{\text{eq}}$, so 
\begin{align}
    \Delta f(s_*) := f(s_*)-f(\bar{s}_{\text{eq}})>0
\end{align}
The marginal density of $\bar{s}$ under $\mu$ is $e^{-\beta N f(\bar{s})}$ up to sub-exponential factors, so the equilibrium weight of configurations near the wall is exponentially small, $\sim e^{-\beta N \mathrm{\Delta} f(s_*)}$. This suggests, by the usual Kramers picture \cite{Kramers1940}, an exponentially large characteristic time for entering the near-core region,
\begin{align}
    \tau_{s_*}(N) \sim e^{\beta N \mathrm{\Delta} f(s_*)}
\end{align}
up to sub-exponential prefactors. We use this only as the expected metastable scaling. The exact statement needed below is the pathwise agreement of the physical and reflected processes up to the first entrance time $\tau_{s_*}$. All results below are derived at fixed $s_*$ and fixed $N$, and all expectations in the remainder of this appendix are with respect to the stationary reflected process, i.e. we sample $X_0\sim\mu_{s_*}$ and then evolve.

\subsection{Spectral measure of the drift}
We will now introduce the operator $\mathcal{S}$. Since the reflected dynamics are reversible, 
\begin{align}
    \langle f \mathcal{L}f\rangle_{\mu_{s_*}} = -\gamma T \left\langle \sum_i |\nabla_i f|^2\right\rangle_{\mu_{s_*}} \leq 0,
\end{align}
the spectrum of $\mathcal{L}$ is non-positive. We therefore introduce the positive self-adjoint operator
\begin{align}
    \mathcal{S} := -\mathcal{L}\geq 0
\end{align}
We may then apply the spectral theorem to $\mathcal{S}$ which states that there exists a unique projection-valued measure $E_{\lambda}$ such that
\begin{align}
    \mathcal{S} = \int_0^{\infty} \lambda dE_{\lambda}
\end{align}
More generally, for any bounded, measurable $\phi$, 
\begin{align}
    \phi(\mathcal{S}) = \int_0^{\infty}\phi(\lambda)dE_{\lambda}
\end{align}
(Heuristically: $E_{\lambda}$ projects onto the relaxation modes of the dynamics with rate at most $\lambda$, and $\phi(\mathcal{S})$ applies $\phi$ to each rate.) Define the function 
\begin{align}
F(\lambda) := \|E_{\lambda}A\|^2_{L^2(\mu_{s_*})}, \quad A:=\mathcal{L}\bar{\theta}
\end{align}
Since $E_{\lambda}$ projects onto everything in the spectrum of $\mathcal{S}$ at or below $\lambda$, $F(\lambda)$ is non-decreasing and ranges from $0$ to $\|A\|^2$, which is finite (see Section \ref{subsubsec:Appendix_phase_blind}). Since any non-decreasing, right continuous function defines a positive measure, we define the following spectral measure
\begin{align}
 \nu_A([\lambda_1,\lambda_2]) = F(\lambda_2)-F(\lambda_1^-) = \|(E_{\lambda_2}-E_{\lambda_1^-})A\|^2
\end{align}
Then, for bounded measurable $\phi$,
\begin{align}
\langle A,\phi(\mathcal{S})A\rangle_{\mu_{s_*}} = \int_0^{\infty} \phi(\lambda)\nu_A(d\lambda)
\end{align}
In words: $\nu_A$ tells us how much of the drift $A$ lives on modes relaxing at rate $\lambda$. Crucially, $A$ is a mean-zero observable in $L^2(\mu_{s_*})$, and since $\mathcal{B}$ is connected the kernel of $\mathcal{S}$ consists of the constants alone, so $E_0 A = \langle A\rangle_{\mu_{s_*}} = 0$. Therefore, $\nu_A(\{0\})=0$. This removes the trivial divergence that can appear when $A$ has mass at $\lambda =0$. 

Since the generator $\mathcal{L}$ defines the Markov semigroup acting on observables by
\begin{align}
    (e^{t\mathcal{L}}f)(x) = \mathbb{E}[f(X_t)|X_0=x]
\end{align}
we can use the spectral representation to define the correlation function. For $t>s$,
\begin{align}
\label{eq:correlation_function_A}
    \langle A(X_t)A(X_s)\rangle_{\mu_{s_*}} = \langle A e^{-(t-s)\mathcal{S}}A\rangle_{\mu_{s_*}} = \int_0^{\infty} e^{-(t-s)\lambda}d\nu_A(\lambda) =: C(t-s)
\end{align}

\subsection{Variance of phase}
We will now proceed with the calculation. We first expand the It\^o formula applied directly to the global phase,
\begin{align}
\langle \mathrm{\Delta} \bar{\theta}(t)^2\rangle_{\mu_{s_*}} =   \left\langle \left(M_t^{\theta}\right)^2 \right\rangle_{\mu_{s_*}} + 2\left\langle M_t^{\theta}\int_0^t A(X_s)ds \right\rangle_{\mu_{s_*}}+ \left\langle \left(\int_0^t A(X_s)ds\right)^2 \right\rangle_{\mu_{s_*}}
\end{align}
For the martingale term, by direct calculation we obtain
\begin{align}
\left\langle (M_t^{\theta})^2 \right\rangle_{\mu_{s_*}}  = \frac{2\gamma T}{N}\left\langle \frac{1}{\bar{s}^2} \right\rangle_{\mu_{s_*}}t, \qquad \nabla_i \bar{\theta} = \frac{1}{N\bar{s}}(-\sin\bar{\theta},\cos\bar{\theta})
\end{align}
which is finite on the basin. For the cross term, we rearrange \eqref{eq:app_Ito_direct_theta}, 
\begin{align}
M_t^{\theta} = \mathrm{\Delta}\bar{\theta}(t) - \int_0^tA(X_s)ds
\end{align}
Then, 
\begin{align}
\left\langle M_t^{\theta}\int_0^tA(X_s)ds \right\rangle = \int_0^t \left\langle \mathrm{\Delta}\bar{\theta}(t)A(X_s) \right\rangle_{\mu_{s_*}} ds - \left\langle \left(\int_0^t A(X_s)ds\right)^2 \right\rangle
\end{align}
Notice that we can decompose the following expectation value as
\begin{align}
\left\langle \mathrm{\Delta} \bar{\theta}(t)A(X_s) \right\rangle_{\mu_{s_*}} = \left\langle \left(\bar{\theta}(X_t)-\bar{\theta}(X_s)\right)A(X_s) \right\rangle_{\mu_{s_*}} + \left\langle \left(\bar{\theta}(X_s)-\bar{\theta}(X_0)\right)A(X_s) \right\rangle_{\mu_{s_*}}
\end{align}
Since the dynamics are reversible, 
\begin{align}
    \left\langle \left(\bar{\theta}(X_s)-\bar{\theta}(X_0)\right)A(X_s) \right\rangle_{\mu_{s_*}} = -\left\langle \left(\bar{\theta}(X_s)-\bar{\theta}(X_0)\right)A(X_0) \right\rangle_{\mu_{s_*}}
\end{align}
By the martingale property of $M_t^{\theta}$, only the phase drift of $\Delta \bar{\theta}(t)$ contributes to the expectation. Hence, 
\begin{align}
\label{eq:theta_A_cross_term}
\left\langle \mathrm{\Delta} \bar{\theta}(t)A(X_s) \right\rangle_{\mu_{s_*}}= \int_0^{t-s}C(u)du -\int_0^s C(u)du 
\end{align}
This does not vanish for a given $s$ (except at $s=t/2$), but its integral over $s\in[0,t]$ does. The change of variables $s\mapsto t-s$ maps the first term onto the second. Hence,
\begin{align}
\label{eq:theta_A_cross_term_integrated}
\int_0^t \left\langle \mathrm{\Delta} \bar{\theta}(t)A(X_s) \right\rangle_{\mu_{s_*}}ds = 0
\end{align}
which is the statement we actually use. Then, 
\begin{align}
\left\langle (\bar{\theta}(t)-\bar{\theta}(0))^2\right\rangle_{\mu_{s_*}} = \frac{2\gamma T}{N}\left\langle \frac{1}{\bar{s}^2}\right\rangle_{\mu_{s_*}}t - \left\langle \left(\int_0^t A(X_s)ds\right)^2\right\rangle_{\mu_{s_*}}
\end{align}
For the remaining term,
\begin{align}
\left\langle \left(\int_0^t A(X_s)ds\right)^2 \right\rangle_{\mu_{s_*}} = \int_0^t\int_0^t C(|s-s'|)dsds' = 2\int_0^t (t-u)C(u)du
\end{align}
Using the spectral representation of the correlator \eqref{eq:correlation_function_A}, and integrating over $u$,
\begin{align}
     2\int_0^t (t-u)C(u)du = 2\int_0^{\infty} \frac{1}{\lambda^2}\left(\lambda t -1+e^{-\lambda t}\right)d\nu_A(\lambda )
\end{align}
Since $\nu_A$ is a positive measure, this integral is strictly positive for all $t> 0$, and is zero only when $t=0$. Hence, 
\begin{align}
\label{eq:app_variance_direct}
\left\langle \mathrm{\Delta} \bar{\theta}(t)^2 \right\rangle_{\mu_{s_*}} = \frac{2\gamma T}{N}\left\langle \frac{1}{\bar{s}^2} \right\rangle_{\mu_{s_*}}t-2\int_0^{\infty} \frac{1}{\lambda^2}\left(\lambda t -1+e^{-\lambda t}\right)d\nu_A(\lambda )
\end{align}
Note that the drift always reduces the variance below the bare martingale value. Physically, this means in equilibrium, the gapped modes can only slow the phase down.
\subsubsection{Diffusion constant and variational representation}
We now calculate the diffusion constant directly from the variance of the phase derived in the preceding section. Dividing \eqref{eq:app_variance_direct} by $2t$, define 
\begin{align}
    k_t(\lambda):=\frac{\lambda t-1+e^{-\lambda t}}{t\lambda^2}
\end{align}
For every $\lambda>0$, $k_t(\lambda)$ increases monotonically to $1/\lambda$ as $t\to\infty$. Since $\nu_A(\{0\})=0$ and the phase variance is non-negative,
\begin{align}
    \int_0^\infty k_t(\lambda)\,\nu_A(d\lambda)\leq\frac{\gamma T}{N}\left\langle\frac{1}{\bar{s}^2}\right\rangle_{\mu_{s_*}} 
\end{align}
Monotone convergence therefore gives
\begin{align}
    \int_0^\infty\frac{1}{\lambda}\,\nu_A(d\lambda)\leq\frac{\gamma T}{N}
    \left\langle\frac{1}{\bar{s}^2}\right\rangle_{\mu_{s_*}}<\infty
\end{align}
and hence
\begin{align}
\label{eq:diffusion_constant_app_1}
    D_{\theta}^{(s_*)} = \lim_{t\to \infty} \frac{\langle \Delta \bar{\theta}(t)^2\rangle_{\mu_{s_*}}}{2t} = \frac{\gamma T}{N}\left\langle \frac{1}{\bar{s}^2} \right\rangle_{\mu_{s_*}} - \int \frac{1}{\lambda}d\nu_A(\lambda) \geq 0
\end{align}
We now derive the variational representation of the diffusion constant. Let us define the functional
\begin{align}
    J[u]:= \gamma T \left\langle \sum_i |\nabla_i(\bar{\theta}-u)|^2\right\rangle_{\mu_{s_*}} 
\end{align}
By expanding the square, this can equivalently be represented by 
\begin{align}
    J[u] = D_{\theta}^{(0)} + \langle u\mathcal{S}u\rangle_{\mu_{s_*}} + 2\langle Au\rangle_{\mu_{s_*}}
\end{align}
Using that $A\in H_{-1}$, i.e. 
\begin{align}
    \|\mathcal{S}^{-\frac{1}{2}}A\|^2 = \int_0^{\infty}\frac{1}{\lambda}d\nu_A(\lambda)<\infty
\end{align}
we can express $\langle Au\rangle_{\mu_{s_*}} = \langle\mathcal S^{-1/2}A,\mathcal S^{1/2}u\rangle_{\mu_{s_*}}$. Then, after completing the square we obtain
\begin{align}
J[u]=D_\theta^{(0)}-\|\mathcal S^{-1/2}A\|^2+\left\|\mathcal S^{1/2}u+\mathcal{S}^{-1/2}A\right\|^2
\end{align}
Notice that the first two terms are exactly the diffusion constant derived in \eqref{eq:diffusion_constant_app_1}. The last term is non-negative, and since this holds for all admissible $u$, we have that 
\begin{align}
    \inf_u J[u]\geq D_{\theta}^{(s_*)}
\end{align}
Formally, equality would occur if $u = -\mathcal{S}^{-1}A$. However, that requires the stronger condition that $S^{-1}A\in L^2(\mu_{s_*})$, whereas we have established only $\mathcal{S}^{-\frac{1}{2}}A \in L^2(\mu_{s_*})$. Therefore, we use a spectral cutoff to show the reverse inequality. Define
\begin{align}
\label{eq:app_spectral_functions}
    g_n(\lambda) = \frac{\mathds{1}_{\{1/n\leq\lambda\leq n\}}}{\lambda}, \quad u_n = -g_n(\mathcal{S})A
\end{align}
Then, 
\begin{align}
\left\|\mathcal S^{1/2}u_n+\mathcal S^{-1/2}A
\right\|^2=\int_{\lambda<1/n\;\text{or}\;\lambda>n}
\frac1\lambda\,\nu_A(d\lambda)
\end{align}
Since $\|\mathcal{S}^{-1/2}A\|<\infty$, the tail outside $[\frac{1}{n},n]$ goes to zero,
\begin{align}
    \left\|\mathcal S^{1/2}u_n+\mathcal S^{-1/2}A\right\|^2  \to 0
\end{align} 
Therefore, for each $n$ we have
\begin{align}
    \inf_u J[u] \leq J[u_n] \to D_{\theta}^{(s_*)}
\end{align}
This proves the reverse inequality and we obtain the variational representation of the diffusion constant,
\begin{align}
\label{eq:variational_diffusion_constant_app}
    D_{\theta}^{(s_*)} = \gamma T \inf_u \left\langle \sum_i |\nabla_i(\bar{\theta}-u)|^2\right\rangle_{\mu_{s_*}}
\end{align}

\subsection{Distribution of phase}
The preceding calculation determines the asymptotic growth of the phase variance and hence identifies the diffusion constant $D_{\theta}$. This alone, however, does not determine the long-time probability law of $\bar{\theta}(t)$. In particular, linear growth of the variance does not imply that the rescaled phase fluctuations are Gaussian, nor that the phase converges to Brownian motion on diffusive time scales. Establishing such a distributional statement requires control of the full stochastic process, rather than only its second moment. We therefore now use the Kipnis--Varadhan \cite{KipnisVaradhan1986} martingale approximation to show that the phase increment is, up to a remainder negligible on the $\sqrt{t}$ scale, an effective martingale whose quadratic variation converges to $2D_{\theta}^{(s_*)}t$. This will yield the Brownian scaling limit of the Goldstone phase.

The version of the Kipnis--Varadhan theorem relevant here applies to a centered $L^2$ additive functional of a stationary, reversible, ergodic Markov process satisfying the $H_{-1}$ condition. All of these hypotheses have already been established above for the reflected process and the observable $A$. Therefore, the Kipnis--Varadhan theorem \cite{KipnisVaradhan1986} gives a martingale additive functional $M_t^A$ and a remainder $R_t$ such that
\begin{align}
    \int_0^t A(X_s)\,ds= M_t^A + R_t,
\end{align}
where the remainder is negligible on the diffusive scale,
\begin{align}
    \frac{\mathbb{E}[R_t^2]}{t}\longrightarrow 0 
\end{align}
Moreover, the remainder is negligible at the level of the functional central limit theorem, so for every fixed $t_{\max}<\infty$,
\begin{align}
\label{eq:app_KV_maximal_remainder}
    \frac{1}{\sqrt{\tau}} \sup_{0\leq t\leq t_{\max}} |R_{\tau t}| \xrightarrow[\tau\to\infty]{P} 0
\end{align}
Thus the remainder is negligible uniformly on compact intervals of the rescaled time variable. We now identify the martingale $M_t^A$ using the same spectral approximation that appeared in the variational argument above. Recall that
\begin{align}
    \left\|\mathcal S^{1/2}u_n+\mathcal S^{-1/2}A\right\|\longrightarrow 0
\end{align}
where $u_n$ is defined in \eqref{eq:app_spectral_functions}. Define the Dirichlet energy seminorm by 
\begin{align}
    \|u\|_1^2 := \langle u,\mathcal{S}u\rangle_{\mu_{s_*}}=\gamma T\left\langle\sum_i|\nabla_i u|^2\right\rangle_{\mu_{s_*}}
\end{align}
After quotienting by additive constants this becomes a norm. In the previous section, we showed that $\mathcal{S}^{\frac{1}{2}}u_n \to -\mathcal{S}^{-\frac{1}{2}}A$ in $L^2(\mu_{s_*})$. Therefore, as $n,m\to \infty$,
\begin{align}
    \|u_n-u_m\|_1 = \|\mathcal{S}^{\frac{1}{2}}(u_n-u_m)\| \to 0
\end{align}
Thus $u_n$ is Cauchy in this energy norm and we denote its limit in the corresponding completed energy space, modulo additive constants, by $\chi$. By construction, $\chi$ is defined through
\begin{align}
    \mathcal{S}^{1/2}\chi =-\mathcal{S}^{-1/2}A 
\end{align}
Here $\chi$ need not exist as an $L^2$ solution of the Poisson equation, since only its energy-space gradient is required. For each $u_n$, let $M_t^{u_n}$ denote the martingale in It\^o's formula. By the It\^o isometry and stationarity,
\begin{align}
    \mathbb{E}[(M_t^{u_n}-M_t^{u_m})^2] = 2t\|u_n-u_m\|_1^2 \longrightarrow 0
\end{align}
Thus $M_t^{u_n}$ converges in mean square to a square-integrable limit, which we denote by $M_t^\chi$. The Kipnis--Varadhan martingale associated with the additive functional of $A$ is
\begin{align}
    M_t^A=-M_t^\chi
\end{align}
Consequently,
\begin{align}
    \Delta \bar{\theta}(t) = M_t^{\mathrm{eff}} + R_t,\qquad M_t^{\mathrm{eff}}:=M_t^\theta - M_t^{\chi}
\end{align}
Thus the phase differs, on the diffusive scale, from a martingale by a negligible remainder. It remains only to identify the limiting quadratic variation of $M_t^{\mathrm{eff}}$. Since $M_t^\chi$ is the $L^2$ limit of the martingales $M_t^{u_n}$, its stochastic integrand is $\nabla\chi$ in the energy-space sense. Therefore,
\begin{align}
    [M^{\mathrm{eff}}]_t =2\gamma T\int_0^t\sum_i\left|\nabla_i\bar{\theta} -\nabla_i\chi\right|^2(X_s)\,ds 
\end{align}
The integrand is an $L^1(\mu_{s_*})$ observable of the stationary ergodic process. Hence the ergodic theorem gives
\begin{align}
    \frac{[M^{\mathrm{eff}}]_t}{t}\longrightarrow2\gamma T \left\langle\sum_i\left|
    \nabla_i\bar{\theta}-\nabla_i\chi\right|^2\right\rangle_{\mu_{s_*}}
\end{align}
almost surely, and therefore also in probability. Since $\chi$ is precisely the energy-space limit of the minimizing sequence constructed above, the variational representation \eqref{eq:variational_diffusion_constant_app} gives
\begin{align}
    \gamma T\left\langle\sum_i\left|\nabla_i\bar{\theta}-\nabla_i\chi\right|^2\right\rangle_{\mu_{s_*}}=D_\theta^{(s_*)}
\end{align}
We therefore obtain the required convergence of the quadratic variation,
\begin{align}
    \frac{[M^{\mathrm{eff}}]_t}{t}\longrightarrow2D_\theta^{(s_*)}
\end{align}
almost surely, and hence also in probability. The martingale functional central limit theorem \cite{KomorowskiLandimOlla2012,KipnisVaradhan1986} then gives, for every fixed $t_{\max}<\infty$,
\begin{align}
    \frac{M^{\mathrm{eff}}_{\tau t}}{\sqrt{\tau}}\Rightarrow\sqrt{2D_\theta^{(s_*)}}\,B_t,\qquad\tau\to\infty
\end{align}
in $C([0,t_{\max}])$ and here $B_t$ is a standard Brownian motion. Together with the Kipnis--Varadhan remainder estimate \eqref{eq:app_KV_maximal_remainder}, this yields
\begin{align}
    \frac{\bar{\theta}(\tau t)-\bar{\theta}(0)}{\sqrt{\tau}}\Rightarrow\sqrt{2D_\theta^{(s_*)}}\,B_t,\qquad\tau\to\infty
\end{align}
with convergence in $C([0,t_{\max}])$. Hence, at fixed $N$ and fixed $s_*$, the Goldstone phase of the stationary reflected process has a Brownian long-time scaling limit with diffusion constant $D_\theta^{(s_*)}$. Its relation to the physical large-$N$ time window is the separate metastable comparison discussed in Section \ref{subsubsec:Appendix_physical_process}.

\subsection{Establishing bounds}
So far we have shown that the Goldstone phase is asymptotically Brownian with diffusion constant, 
\begin{align}
\label{eq:Appendix_definition_C}
    D_{\theta}^{(s_*)} = (1-\mathcal{C})\frac{\gamma T}{N}\left\langle \frac{1}{\bar{s}^2}\right\rangle_{\mu_{s_*}}, \quad \mathcal{C} := \frac{N}{\gamma T}\left\langle \frac{1}{\bar{s}^2}\right\rangle_{\mu_{s_*}}^{-1}\int_0^{\infty}\frac{1}{\lambda}d\nu_A(\lambda)
\end{align}
with $0\leq \mathcal{C}\leq 1$. We now show that the correction can never consume the leading term. 

The lower bound ultimately follows from the topology of the Goldstone mode. For any configuration in the ordered basin, a simultaneous $2\pi$ rotation of all rotors traces out a closed $\mathrm{U}(1)$ orbit that remains entirely inside $\mathcal{B}$. Along this orbit, $\bar{\theta}$ winds once, whereas any admissible single-valued function $f$ returns to its original value. Consequently, the difference $\bar{\theta}-f$ must retain a nonzero gradient in the global rotation direction, independent of the choice of $f$. Since the variational formula minimizes precisely the Dirichlet energy of $\bar{\theta}-f$, this winding gives a nonzero lower bound on $D_\theta$. To make this statement quantitative, introduce the infinitesimal rotation vector field
\begin{align}
R_i = \hat{z}\times \vec{s}_i = (-y_i,x_i)
\end{align}
We have the following exact identity
\begin{align}
\sum_i R_i \cdot \nabla_i \bar{\theta}= \frac{\sin\bar{\theta}}{N\bar{s}}\sum_i y_i + \frac{\cos\bar{\theta}}{N\bar{s}}\sum_i x_i  = 1
\end{align}
Additionally, for an arbitrary phase space function $f$, 
\begin{align}
\frac{\partial f}{\partial \theta_i} &= \frac{\partial f}{\partial x_i}\frac{\partial x_i}{\partial \theta_i} + \frac{\partial f}{\partial y_i}\frac{\partial y_i}{\partial \theta_i} 
= -y_i\frac{\partial f}{\partial x_i} + x_i\frac{\partial f}{\partial y_i} 
= R_i \cdot \nabla_i f
\end{align}
Therefore, pairing $R_i$ with the full gradient computes the derivative of $f$ under a simultaneous rotation of all rotors since
\begin{align}
\sum_i \partial_{\theta_i}f = \sum_i R_i \cdot \nabla_i f
\end{align}
In particular, since $\bar{s}$ is invariant under a simultaneous rotation of all rotors,
\begin{align}
\label{eq:app_R_tangent}
    \sum_i R_i\cdot \nabla_i \bar{s} = 0, \quad \text{i.e.} \quad R\cdot \nu = 0 \,\,\text{on}\,\, \partial\mathcal{B}
\end{align}
so the rotation field $R$ is tangent to the wall. Furthermore, integrating by parts with respect to the restricted Gibbs measure over $\mathcal{B}$ gives us the following identity,
\begin{align}
\left\langle \sum_i \partial_{\theta_i}f \right\rangle_{\mu_{s_*}} 
= \sum_i \int_{\mathcal{B}} R_i \cdot \nabla_i f\, d\mu_{s_*}  = \frac{1}{Z_{s_*}}\oint_{\partial\mathcal{B}} f e^{-\beta H}(R\cdot n)d\sigma + \beta \int_{\mathcal{B}} f\left(\sum_i \partial_{\theta_i}H \right) d\mu_{s_*} = 0
\end{align}
Here the surface term vanishes by \eqref{eq:app_R_tangent}, we have used that $\nabla_i \cdot R_i =0$, and the last term vanishes because the Hamiltonian is invariant under global $\mathrm{U}(1)$ rotations. Note that here the surface term vanishes because of a property of the wall (it is a level set of a rotation-invariant function), not because of any property of $f$, so no boundary condition on $f$ is required. Then, combining the above identities,
\begin{align}
\left\langle \sum_i R_i \cdot \nabla_i (\bar{\theta}-f) \right\rangle_{\mu_{s_*}} = \left\langle \sum_i R_i \cdot \nabla_i \bar{\theta} \right\rangle_{\mu_{s_*}} - \left\langle \sum_i R_i \cdot \nabla_i f \right\rangle_{\mu_{s_*}} = 1
\end{align}
Physically, on the basin, the $\mathrm{U}(1)$ orbit of any configuration is a closed loop lying entirely in $\mathcal{B}$, around which $\bar{\theta}$ winds by $2\pi$ while every single-valued $f$ returns to its starting value. Restricting to the basin does not weaken this argument, because the loops that carry the winding are exactly the ones the basin keeps. Then by Cauchy-Schwarz,
\begin{align}
1 = \left\langle \sum_i R_i \cdot \nabla_i (\bar{\theta}-f) \right\rangle_{\mu_{s_*}} \leq \sqrt{\left\langle \sum_i |R_i|^2 \right\rangle_{\mu_{s_*}}} \sqrt{\left\langle \sum_i |\nabla_i (\bar{\theta}-f)|^2 \right\rangle_{\mu_{s_*}}}
\end{align}
Then, since $|R_i|^2 = r_i^2$, squaring both sides yields the following inequality,
\begin{align}
\left\langle \sum_i |\nabla_i (\bar{\theta}-f)|^2 \right\rangle_{\mu_{s_*}} \geq \frac{1}{N\left\langle Q \right\rangle_{\mu_{s_*}}}, \quad Q := \frac{1}{N}\sum_i r_i^2
\end{align}
Since this inequality holds for every admissible single-valued function
$f$, we may take the infimum over $f$. Using the variational
representation \eqref{eq:variational_diffusion_constant_app}, we obtain
\begin{align}
    D_{\theta}^{(s_*)}=\gamma T\inf_f\left\langle\sum_i|\nabla_i(\bar{\theta}-f)|^2\right\rangle_{\mu_{s_*}}\geq\frac{\gamma T}{N\left\langle Q\right\rangle_{\mu_{s_*}}}.
\end{align}
The upper bound follows directly from the same variational formula by
choosing $f=0$,
\begin{align}
    D_{\theta}^{(s_*)}\leq\gamma T\left\langle
    \sum_i|\nabla_i\bar{\theta}|^2
    \right\rangle_{\mu_{s_*}}=\frac{\gamma T}{N}\left\langle\frac{1}{\bar{s}^2}\right\rangle_{\mu_{s_*}}.
\end{align}
Therefore, we obtain an upper and lower bound for the diffusion constant,
\begin{align}
\frac{\gamma T}{N\left\langle Q \right\rangle_{\mu_{s_*}}}\leq D_{\theta}^{(s_*)} \leq \frac{\gamma T}{N}\left\langle \frac{1}{\bar{s}^2} \right\rangle_{\mu_{s_*}}
\end{align}
These bounds are exact at finite $N$ and finite $s_*$. To connect them with the mean-field quantities of the main text, consider any fixed $s_*>0$ separated by an $\mathcal O(1)$ distance from $\bar{s}_{\mathrm{eq}}$. The restricted Gibbs measure is exponentially concentrated near the same ordered saddle as the unrestricted measure, and a standard saddle-point expansion gives
\begin{align}
\left\langle \frac{1}{\bar{s}^2} \right\rangle_{\mu_{s_*}} = \frac{1}{\bar{s}_{\text{eq}}^2} + \mathcal{O}\left(\frac{1}{TN}\right), \qquad \frac{1}{\left\langle Q \right\rangle_{\mu_{s_*}}} = \frac{1}{Q_{\text{eq}}} + \mathcal{O}\left(\frac{1}{TN}\right)
\end{align}
Changing between two such fixed positive cutoffs modifies these static averages only through an exponentially suppressed region. This comparison is between positive cutoffs since the unrestricted expectation $\langle\bar{s}^{-2}\rangle_\mu$ is divergent. Consequently, for every fixed admissible cutoff,
\begin{align}
    \frac{\gamma T}{NQ_{\mathrm{eq}}} \leq D_\theta^{(s_*)} \leq \frac{\gamma T}{N\bar{s}_{\mathrm{eq}}^2}
\end{align}
up to $\mathcal O(N^{-2})$ corrections. Since the same leading-order bounds hold for every fixed admissible cutoff, we suppress the $s_*$ superscript in what follows, with the understanding that $D_\theta$ refers to the reflected-process diffusion constant at an arbitrary fixed cutoff. The self-consistent equilibrium values are 
\begin{align}
\bar{s}_{\text{eq}}^2 &= \frac{b^2-2T(2a^2-5a+3)-b\sqrt{b^2+4aT(a-1)}}{2(a-1)^2} \\
Q_{\text{eq}} &= \frac{2(a-1)(9-4a)T+5b^2-5b\sqrt{b^2+4aT(a-1)}}{6(a-1)^2}
\end{align}
Taking the infinite temperature limit yields, 
\begin{align}
0\leq \limsup_{T\to\infty}  \mathcal{C} \leq \frac{2a}{9-4a}
\end{align}
Thus the finite-$N$ bounds establish $D_\theta=\Theta(N^{-1})$ for every fixed admissible cutoff, and the corresponding renormalization fraction remains uniformly bounded in the high-temperature limit.

\section{Classification of admissible momentum-conserving models}
\label{sec:Appendix_allowed_models}
In Section \ref{subsec:Introducing_SF_model}, we showed that by explicitly adding momenta to the theory, a centrifugal term is generated in the effective potential which causes the theory to diverge. Therefore, we must modify the potential to include stronger radial confinement in order to make the $\Omega$-ensemble well-defined. However, we must ensure that the modified potential has a physically viable entropically ordered phase. Since momenta do not affect the entropic ordering mechanism, this discussion will focus on candidate potentials of the form
\begin{align}
    V = \frac{1}{2N}\sum_{i\neq j} \left(ar_i^nr_j^n - r_i^{n-1}r_j^{n-1}(\vec{s}_i\cdot \vec{s}_j)\right) + b\sum_i r_i^m
\end{align}
in $d\geq 2$ dimensions. In the ordered phase, we choose the ordering direction to lie along the $x$-axis.  The corresponding mean-field single particle potential is 
\begin{align}
    V_{\text{mf}} = aR_n r^n + br^m -Q_n r^{n-1}x
\end{align}
where the order parameters
\begin{align}
    R_n = \frac{1}{N}\sum_i r_i^n, \qquad Q_n = \frac{1}{N}\sum_i r_i^{n-1}x_i, \qquad R_m = \frac{1}{N}\sum_i r_i^m
\end{align}
are determined self-consistently as averages over the single-particle Boltzmann weight,
\begin{align}
\label{eq:self_consistency}
    R_n = \langle r^n\rangle, \qquad Q_n = \langle r^{n-1}x\rangle .
\end{align}
The ordered phase can first appear when the disordered state becomes unstable to a nonzero $Q_n$. Expanding the self-consistency equation for $Q_n$ around the rotationally invariant state gives
\begin{align}
    Q_n = \langle r^{n-1}x\rangle \approx  \langle r^{n-1}x\rangle_0 + \beta Q_n  \langle r^{2n-2}x^2\rangle_0 +\cdots 
\end{align}
where $\langle \cdots \rangle_0$ denotes an average with respect to the disordered distribution. Rotational invariance gives $\langle r^{n-1}x\rangle_0 =0$. Additionally, the radial and angular sectors decouple in the disordered distribution, so
\begin{align}
    \langle r^{2n-2}x^2\rangle_0 = \frac{1}{d}\langle r^{2n}\rangle_0 
\end{align}
Thus a nonzero solution for $Q_n$ first appears when
\begin{align}
\label{eq:Critical_condition}
    \langle r^{2n}\rangle_0 = \frac{d}{\beta} 
\end{align}
Since we are interested in entropic order that persists to arbitrarily high temperature, this instability must remain accessible as $\beta\to 0$. Equivalently, the disordered distribution must be able to support fluctuations with $\langle r^{2n}\rangle_0 \sim \beta^{-1}$.
\subsection{$m>2n$}
If $m>2n$, the $\beta b r^m$ term controls the high-temperature radial scale of the disordered distribution. Requiring the leading term in the exponent to be of order $\mathcal{O}(1)$ gives
\begin{align}
   \beta b r^m\sim 1 \implies r\sim \beta^{-1/m},
\end{align}
and therefore
\begin{align}
    \langle r^{2n}\rangle_0\sim \beta^{-2n/m} .
\end{align}
Since $m>2n$, this grows more slowly than $\beta^{-1}$ as $\beta\to0$, so $\beta\langle r^{2n}\rangle_0\to 0$ and the critical condition \eqref{eq:Critical_condition} cannot be satisfied above some temperature. Moreover, the same scaling excludes ordered solutions of any kind, not just continuous bifurcations. Since $Q_n\lesssim \langle r^n\rangle$, the alignment term in the exponent scales as
\begin{align}
    \beta Q_n r^n \sim \beta^{1-\frac{2n}{m}} \to 0 \quad \text{as}\quad\beta\to 0 ,
\end{align}
so the weight becomes rotationally invariant for any value of $Q_n$, and no ordered solution, continuous or discontinuous, can exist at high temperature. Models with $m>2n$ are therefore excluded, and we restrict to $m\leq 2n$ in what follows.

\subsection{An exact identity for ordered fixed points}
For the remainder of this appendix we transform to the coordinates
\begin{align}
    \alpha := \frac{d}{n},\qquad \gamma:=\frac{m}{n}, \qquad u:= r^n, \qquad t:= \frac{x}{r}=\cos\theta ,
\end{align}
in which the alignment term reads $Q_n u t$ and the single-particle partition function becomes
\begin{align}
    Z = \frac{\Omega_{d-2}}{n}\int_0^{\infty}\!\!\int_{-1}^1 du\, dt\; u^{\alpha -1}(1-t^2)^{\frac{d-3}{2}}\,e^{-\beta(aR_n u + bu^{\gamma}-Q_n u t)} .
\end{align}
The self-consistency conditions \eqref{eq:self_consistency} become $R_n=\langle u\rangle$ and $Q_n=\langle ut\rangle$.

We obtain two exact relations by integrating total derivatives, one in each variable:
\begin{align}
    &\int_{-1}^{1} dt\, (1-t^2)^{\frac{d-3}{2}}\,t\int_0^{\infty} du\, \frac{\partial}{\partial u}\left(u^{\alpha} e^{-\beta V_{\text{mf}}}\right)=0 ,\label{eq:IBP_u}\\
    &\int_0^{\infty} du \, u^{\alpha -1} \int_{-1}^{1} dt\, \frac{\partial}{\partial t} \left((1-t^2)^{\frac{d-1}{2}}e^{-\beta V_{\text{mf}}}\right) =0 .\label{eq:IBP_t}
\end{align}
The boundary terms vanish for any normalizable weight: at $u=0$ and $u\to\infty$ because $u^{\alpha}e^{-\beta V_{\rm mf}}\to0$, and at $t=\pm1$ because the exponent $\frac{d-1}{2}$ is positive for $d\geq2$. Carrying out the inner derivatives, dividing by $Z$, and using $Q_n=\langle ut\rangle$, the two relations become
\begin{align}
    &\alpha \left\langle t \right\rangle = \beta a R_n Q_n + \beta \gamma b\left\langle u^{\gamma}t \right\rangle -\beta Q_n \left\langle t^2u \right\rangle\\[1em]
    &(d-1)\left\langle t \right\rangle = \beta Q_n R_n - \beta Q_n \left\langle ut^2 \right\rangle
\end{align}
Eliminating $\langle t\rangle$ and dividing by $\beta Q_n R_n\neq 0$ gives 
\begin{align}
\label{eq:fixed_point_identity}
    a = \frac{\alpha}{d-1} + \left(1-\frac{\alpha}{d-1}\right)\frac{\langle u t^2\rangle}{R_n} - \frac{\gamma\, b\,\langle u^{\gamma} t\rangle}{Q_n R_n}.
\end{align}
This identity holds at every normalizable solution of the self-consistency conditions with $Q_n\neq 0$, for any $\beta$, $b\geq 0$ and $m$. Note that the explicit $\beta$ dependence has canceled. Therefore, whatever constraints \eqref{eq:fixed_point_identity} places on $a$ apply at all temperatures. As a byproduct, replacing $u^{\alpha-1}\to u^{\alpha}$ in \eqref{eq:IBP_t} and dividing by $Q_n$ yields a second identity valid at every ordered fixed point,
\begin{align}
\label{eq:sin2_identity}
    \langle r^{2n}\sin^2\theta\rangle = \frac{d-1}{\beta}
\end{align}
which generalizes the critical condition into the ordered phase. As a sanity check, in the limit $Q_n \to 0$ the radial and angular sectors decouple,
\begin{align}
     \langle r^{2n}\sin^2\theta\rangle \to \langle r^{2n}\rangle_0\, \langle \sin^2\theta\rangle_{\text{ang.}} = \frac{d-1}{d}\,\langle r^{2n}\rangle_0 ,
\end{align}
and \eqref{eq:sin2_identity} reproduces \eqref{eq:Critical_condition}.

The rest of this subsection converts \eqref{eq:fixed_point_identity} into bounds on $a$ that hold for all ordered fixed points at every temperature. The identity contains two nontrivial averages, and we bound each in turn. The key structural fact is that, conditionally on $u$, the angular variable $t$ is distributed with weight
\begin{align}
    \omega(t)\,e^{\nu t}, \qquad \omega(t) := (1-t^2)^{\frac{d-3}{2}}, \qquad \nu:= \beta Q_n u ,
\end{align}
so every joint average reduces, by the tower property, to an average of conditional angular moments over the radial marginal. We orient the $x$-axis so that $Q_n>0$, hence $\nu>0$. The opposite orientation flips the signs of $Q_n$, $\langle t\rangle$, and $\langle u^{\gamma}t\rangle$ simultaneously and leaves all conclusions unchanged.

\paragraph{The conditional mean of $t$ is positive and increasing in $u$.}
Define
\begin{align}
   A_d(\nu):= \mathbb{E}[t\,|\,u] = \frac{1}{Z_{t}}\int_{-1}^1 dt\, t\,\omega(t)\, e^{\nu t}, \qquad Z_t = \int_{-1}^1 dt\, \omega(t)\, e^{\nu t} .
\end{align}
Splitting the exponential into even and odd parts,
\begin{align}
    e^{\nu t}=\cosh(\nu t)+\sinh(\nu t)
\end{align}
and using that $\omega$ is even, only the odd part survives in the numerator and only the even part in the denominator,
\begin{align}
    A_d(\nu) = \frac{2}{Z_t}\int_0^1 dt\, \omega(t)\,t\sinh(\nu t), \qquad Z_t = 2\int_0^1 dt\, \omega(t)\cosh(\nu t) 
\end{align}
For $\nu>0$ the numerator is an integral of a strictly positive continuous integrand, so $A_d(\nu)>0$. Moreover, $A_d$ is the mean of an exponential family, so its derivative is the conditional variance,
\begin{align}
    A_d'(\nu) = \mathbb{E}[t^2|u]-A_d(\nu)^2 = \mathrm{Var}(t|u)>0 
\end{align}
and $A_d(\nu(u))$ is a positive, strictly increasing function of $u$. Two consequences follow immediately from the tower property. First, since $\langle u^{\gamma}t\rangle = \langle u^{\gamma}A_d(\nu(u))\rangle$,
\begin{align}
\label{eq:positivity_last_term}
    \frac{\langle u^{\gamma }t\rangle }{Q_n R_n} >0 ,
\end{align}
so the last term of \eqref{eq:fixed_point_identity} is strictly negative whenever $b>0$. Second, $\langle t\rangle>0$, which we use next.

\paragraph{Bounds on the middle term.}
We now show that for every ordered fixed point,
\begin{align}
\label{eq:bounds_middle_term}
    \frac{1}{d}< \frac{\langle u t^2\rangle}{R_n}< 1 .
\end{align}
The upper bound follows directly from \eqref{eq:IBP_t}. Dividing by $\beta Q_n\neq0$,
\begin{align}
    R_n -\langle ut^2\rangle = \frac{(d-1)\langle t\rangle }{\beta Q_n} >0 
\end{align}
since $\langle t\rangle$ and $Q_n$ share the same sign. In other words, an ordered state necessarily has nonzero angular polarization $\langle t\rangle$, and \eqref{eq:IBP_t} converts that polarization into a strict deficit of $\langle ut^2\rangle$ below $R_n$.

The lower bound states that the aligning field can only increase the angular second moment above its isotropic value $1/d$. Consider the conditional average
\begin{align}
\mathbb{E}\left[\left.t^2-\frac{1}{d}\right|u\right] = \frac{2}{Z_t} \int_{0}^1dt\, \left(t^2-\frac{1}{d}\right)\omega(t)\cosh(\nu t) ,
\end{align}
where the odd part of the exponential has again dropped out by symmetry. The factor $t^2-\frac1d$ changes sign once, at $t = d^{-1/2}$, while $\cosh(\nu t)$ is increasing in $t$ on $[0,1]$. Comparing $\cosh(\nu t)$ against its value at the sign change, we have pointwise on $[0,1]$
\begin{align}
    \left(t^2-\frac{1}{d}\right)\cosh(\nu t) \geq  \left(t^2-\frac{1}{d}\right)\cosh\left(\frac{\nu}{\sqrt{d}}\right)
\end{align}
Integrating, and using that $t^2$ averages to exactly $1/d$ in the isotropic measure,
\begin{align}
\int_0^1 dt \left(t^2-\frac{1}{d}\right)\omega(t)\cosh(\nu t)\geq \cosh\left(\frac{\nu}{\sqrt{d}}\right)\int_0^1 dt\, \left(t^2-\frac{1}{d}\right)\omega(t) =0 
\end{align}
with strict inequality for $\nu\neq0$, since the pointwise inequality is then strict away from $t=d^{-1/2}$. By the tower property,
\begin{align}
    \langle ut^2\rangle - \frac{R_n}{d} = \mathbb{E}\left[u\;\mathbb{E}\left[\left.t^2-\frac{1}{d}\right|u\right]\right] > 0 
\end{align}
and dividing by $R_n>0$ gives the lower bound in \eqref{eq:bounds_middle_term}.

\paragraph{Bounds on $a$.}
Since the last term of \eqref{eq:fixed_point_identity} is non-positive,
\begin{align}
    a \leq \frac{\alpha}{d-1} + \frac{\langle ut^2\rangle}{R_n}\left(1-\frac{\alpha}{d-1}\right) 
\end{align}
Which bound in \eqref{eq:bounds_middle_term} is operative depends on the sign of the coefficient, i.e. on whether $\alpha=d/n$ is smaller or larger than $d-1$. If
\begin{align}
   n > \frac{d}{d-1} \implies  1-\frac{\alpha}{d-1} >0 
\end{align}
the upper bound of \eqref{eq:bounds_middle_term} applies and
\begin{align}
    a < \frac{\alpha}{d-1} + \left(1-\frac{\alpha}{d-1}\right) =1 
\end{align}
If instead
\begin{align}
    n< \frac{d}{d-1} \implies  1-\frac{\alpha}{d-1} <0 
\end{align}
the lower bound of \eqref{eq:bounds_middle_term} applies and
\begin{align}
    a< \frac{\alpha}{d-1} + \frac{1}{d}\left(1-\frac{\alpha}{d-1}\right)  = \frac{d+n}{nd} 
\end{align}
For $d\geq2$, the first case is $n\geq2$ and the second is $n=1$, with a single borderline case, $(d,n)=(2,2)$, where the coefficient vanishes identically and \eqref{eq:fixed_point_identity} collapses to
\begin{align}
\label{eq:d2n2_case}
    a = 1 - \frac{\gamma b\,\langle u^{\gamma}t\rangle}{Q_nR_n} ,
\end{align}
so that $a<1$ for $b>0$, while $b=0$ would force $a=1$ exactly. We show below that for $(d,n)=(2,2)$ the point $(a,b)=(1,0)$ has a divergent partition function and is not an admissible model. In summary, every ordered fixed point of an admissible model satisfies, at every temperature,
\begin{align}
\label{eq:bounds_on_a}
    &n =1 \implies a< \frac{d+1}{d} ,\\
    &n\geq 2 \implies a<1 .
\end{align}

\subsection{$m<2n$}
For $m<2n$ the Hamiltonian is bounded below only if
\begin{align}
    a\geq 1 \quad \text{and} \quad b\geq 0 
\end{align}
as can be seen from aligned configurations, on which the potential reduces to
\begin{align}
\label{eq:aligned_V}
    V_{\rm aligned} = \frac{a-1}{2N}\left[\Big(\sum_i u_i\Big)^{2}-\sum_i u_i^2\right] + b\sum_i r_i^m 
\end{align}
If $a<1$ this is unbounded below along equal, growing radii, and if $b<0$ it is unbounded below for a single escaping particle. Both pieces of the boundary need a comment. On the edge $b=0$ with $a>1$, the zero-energy configurations are those in which a single rotor is away from the origin. A rotor at radius $R$ confines each of the others to a region of volume $\propto R^{-d}$, so such configurations carry weight $\propto R^{d-1-d(N-1)}$ and the partition function is finite for $N\geq3$. This edge is therefore admissible. At the corner $(a,b)=(1,0)$, by contrast, expression \eqref{eq:aligned_V} vanishes identically in the radii, so the potential is zero on the whole manifold of aligned configurations. Rescaling all radii by a common factor $\lambda$, the measure grows as $\lambda^{Nd}$ while the fluctuations of the $(N-1)(d-1)$ relative angles contribute $\lambda^{-n(N-1)(d-1)}$, so the partition function diverges whenever $Nd\geq n(N-1)(d-1)$. This includes $n=1$, and the case $(d,n)=(2,2)$ for which the corner was invoked above.

The bound \eqref{eq:bounds_on_a} now closes the case $n\geq2$ completely. Every ordered fixed point has $a<1$, while admissibility requires $a\geq1$. Hence for $m<2n$, $n\geq 2$, $b\geq0$, and $d\geq 2$, the admissible region contains no ordered phase, no ordered metastable state, and no transition of any order, at any temperature. For $n=1$ the only case with $m<2n$ is $(n,m)=(1,1)$. This is the overdamped soft rotor model which we have shown exhibits an entropically ordered phase at arbitrarily high temperatures. Furthermore, in Appendix \ref{sec:Appendix_free_energy_overdamped}, we showed that this transition is continuous. 

\subsection{$m=2n$}
The remaining possibility is the marginal case $m=2n$, for which the aligned configurations \eqref{eq:aligned_V} have both terms of the same degree. There are now two ways in which the potential can fail to be bounded below. A single rotor escaping while the others stay at the origin costs $b\,u^2$, which requires $b\geq0$. A uniform expansion of all the rotors costs $\frac{N}{2}(a+2b-1)u^2$ to leading order in $N$, which requires $a+2b\geq1$. On the line $a+2b=1$ the extensive stiffness of the uniform expansion vanishes and the size of the cloud is no longer set at leading order (for $n=1$ this is the collective threshold of Section \ref{subsubsec:critical_angular_velocities}, at which the radius diverges). We therefore take as the admissible region
\begin{align}
\label{eq:m2n_admissible}
    b\geq 0 \quad\text{and}\quad a+2b> 1 ,
\end{align}
which excludes the corner $(a,b)=(1,0)$. The natural quantity to bound is therefore $a+2b$. Adding $2b$ to both sides of \eqref{eq:fixed_point_identity}, using $\gamma=2$, we obtain
\begin{align}
\label{eq:a2b_identity}
    a+2b = \frac{\alpha}{d-1} + \frac{\langle ut^2\rangle }{R_n}\left(1-\frac{\alpha}{d-1}\right) - 2b\left(\frac{\langle u^2 t\rangle }{Q_n R_n}-1\right)
\end{align}
The first two terms are exactly those bounded in \eqref{eq:bounds_middle_term}, so it remains to show that the bracket in the last term is positive. Conditioning on $u$ as before,
\begin{align}
    \langle u^2t\rangle - Q_n R_n = \langle u\cdot uA_d(\nu)\rangle - \langle uA_d(\nu)\rangle\langle u\rangle = \text{Cov}\big(u,\,uA_d(\nu)\big) 
\end{align}
For $Q_n>0$, both $u$ and $uA_d(\nu(u))$ are strictly increasing functions of $u$, and by Chebyshev's association inequality, the covariance of two increasing functions of the same random variable is positive. Explicitly, with $u'$ an independent copy of $u$,
\begin{align}
    2\,\text{Cov}\big(f(u),g(u)\big) = \mathbb{E}\Big[\big(f(u)-f(u')\big)\big(g(u)-g(u')\big)\Big] \geq 0 ,
\end{align}
since the two factors always share a sign, with strict inequality unless one function is constant. (For $Q_n<0$ the covariance is negative, but so is $Q_nR_n$, and the ratio is unchanged.) Hence, at every ordered fixed point and every temperature,
\begin{align}
    \frac{\langle u^2t\rangle}{Q_n R_n} -1>0 
\end{align}
so the last term of \eqref{eq:a2b_identity} is strictly negative for $b>0$, and the analysis of \eqref{eq:bounds_on_a} applies verbatim to $a+2b$:
\begin{align}
\label{eq:bounds_on_a2b}
    &n=1 \implies a+2b < \frac{d+1}{d} \\[1em]
    &n\geq 2 \implies a+2b < 1 
\end{align}
The borderline case $(d,n)=(2,2)$ is again settled by \eqref{eq:d2n2_case} and the corner exclusion. Comparing with \eqref{eq:m2n_admissible}, no self-consistent ordered solution exists in the admissible region for $n\geq 2$, $m=2n$, $d\geq2$, at any temperature. The only remaining case is $(n,m)=(1,2)$, and since $\beta$ can be absorbed by rescaling $r$ when $m=2n$, its phase diagram in the $(a,b)$ plane is temperature independent. For this model we take $b>0$. At $b=0$ it coincides with the linear model \eqref{eq:linear_model} at $b=0$, which, as shown in the previous subsection, is well defined for $a>1$ but admits no rotation, since for any $\Omega\neq0$ the centrifugal term is then unopposed for a single rotor. In the rotating ensemble of Section \ref{subsubsec:critical_angular_velocities} this edge is the end point $\kappa\to0$. We now determine the physical window where the model orders.

In the disordered phase, the radial distribution is
\begin{align}
\label{eq:disordered_weight_quadratic}
    P(r)\propto r^{d-1} e^{-Ar -Br^{2}}, \qquad A := \beta a R_1, \qquad B:= \beta b .
\end{align}
Integrating the total derivative
\begin{align}
    \int_0^{\infty} dr\,\frac{d}{dr}\left(r^{d}e^{-Ar-Br^2}\right)= 0
\end{align}
and dividing by the radial partition function yields
\begin{align}
\label{eq:critical_IBP}
    d = \beta a R_1^2 + 2\beta b R_{2} .
\end{align}
At the critical point, \eqref{eq:Critical_condition} fixes the second moment,
\begin{align}
    R_{2} = \langle r^{2}\rangle_0 = \frac{d}{\beta} ,
\end{align}
and combining the two relations gives
\begin{align}
\label{eq:critical_relation}
   (1-2b)\, R_{2} = a\, R_1^2 .
\end{align}
If $a=0$, \eqref{eq:critical_relation} requires $b=\tfrac12$, which lies on the excluded line $a+2b=1$, so $a\neq0$ at a critical point. Equation \eqref{eq:critical_relation} states that the critical value of $a$ is controlled by the shape of the radial distribution through the dimensionless ratio
\begin{align}
    \rho := \frac{R_2}{R_1^2} = \frac{a}{1-2b} 
\end{align}
and the classification reduces to bounding $\rho$. Non-negativity of the radial variance gives the lower bound,
\begin{align}
    \text{Var}(r) = R_{2}-R_1^2 \geq 0 \implies \rho\geq 1 
\end{align}
For the upper bound, define the radial moments
\begin{align}
    I_k = \int_0^{\infty} dr \, r^{d+ k-1}e^{- Ar-B r^2}, \qquad R_1 = \frac{I_1}{I_0}, \qquad R_{2} = \frac{I_2}{I_0} .
\end{align}
Integrating the total derivative 
\begin{align}
    \int dr \frac{d}{dr}\left(r^{d+k}e^{-Ar-Br^2}\right) =0
\end{align}
gives the following recurrence relation
\begin{align}
    (d +k)I_k =A I_{k+1} + 2B I_{k+2} .
\end{align}
Multiplying the $k=0$ relation by $I_2$, the $k=1$ relation by $I_1$, and subtracting,
\begin{align}
    (d +1)I_1^2 -d\, I_0I_2 = 2B \left(I_1I_3-I_2^2\right) > 0 
\end{align}
where the positivity uses $B>0$ and the Cauchy-Schwarz inequality $I_2^2<I_1I_3$. Dividing by $I_1^2$,
\begin{align}
    \rho = \frac{I_0I_2}{I_1^2} < 1 + \frac{1}{d} .
\end{align}
Combining the two bounds with \eqref{eq:critical_relation},
\begin{align}
\label{eq:rho_window}
    1\leq \rho = \frac{a}{1-2b}< 1+\frac{1}{d} 
\end{align}
Although \eqref{eq:rho_window} as an algebraic condition would also allow $a<0$ together with $1-2b<0$, this regime is incompatible with the boundedness requirement and the moment inequality. The admissible region \eqref{eq:m2n_admissible} requires $a+2b>1$, and for $a<0$ this forces $\rho<1$, which contradicts the lower bound in \eqref{eq:rho_window}. Thus no critical point occurs for $a<0$. For $a>0$, condition \eqref{eq:rho_window} requires $b<\frac12$ and delimits the window in which the critical point, and hence, by \eqref{eq:bounds_on_a2b} and the continuity of the transition, the ordered phase, can occur:
\begin{align}
\label{eq:n1m2_window}
    1-2b< a < \left(1+ \frac{1}{d}\right)(1-2b), \qquad 0< b<\frac{1}{2} ,
\end{align}
where the lower bound is the condition $a+2b>1$ of \eqref{eq:m2n_admissible}.

\subsubsection{The exact ordered region of the marginal model}
The window \eqref{eq:n1m2_window} is the set of couplings in which the critical point can occur. It was derived from the necessary conditions at criticality, and is therefore an envelope of the critical manifold rather than the ordered region itself. We now determine the ordered region exactly for the single admissible case $(1,2)$. Since every term in the potential $V_{(1,2)}$ is quadratic, under $\mathbf{s}_i = \sqrt{T}\tilde{\mathbf{s}}_i$, one has 
\begin{align}
    -\beta V_{(1,2)}(\mathbf{s}) = -V_{(1,2)} (\tilde{\mathbf{s}})
\end{align}
Consequently, the phase diagram for the quadratic model is exactly independent of temperature for every $T>0$. In what follows, we therefore set $\beta =1$ without loss of generality.

We first introduce the relevant family of disordered distributions in dimensionless form. For a parameter $\sigma \in [0,\infty)$, define
\begin{align}
\label{eq:app_disordered_fam_dist}
P_{\sigma}(\xi) \propto \xi^{d-1}e^{-\xi^2 -\sigma \xi}, \quad \xi \in (0,\infty), \quad \mu_k(\sigma) = \langle \xi^k\rangle_{\sigma}
\end{align}
Integration by parts gives the moment recurrence
$(d+k)\mu_k = \sigma \mu_{k+1}+2\mu_{k+2}$, whose $k=0$ case,
\begin{align}
\label{eq:family_identity}
    d = \sigma\mu_1(\sigma) + 2\mu_2(\sigma),
\end{align}
holds identically in $\sigma$ and is analogous to \eqref{eq:critical_IBP}.  Differentiation in $\sigma$ produces covariances, so
\begin{align}
    \mu_1'=-\mathrm{Var}(\xi)<0, \qquad \mu_2'=-\mathrm{Cov}(\xi,\xi^2)<0
\end{align}
the latter by Chebyshev's inequality. Therefore, both moments are strictly decreasing in $\sigma$.

We now relate the family of distributions to the physical radial distribution of the quadratic model. Rescaling $\xi = \sqrt{b}\,r$ maps the disordered weight \eqref{eq:disordered_weight_quadratic} onto $P_{\sigma}$ with 
\begin{align}
    \sigma = \frac{aR_1}{\sqrt{b}}, \qquad \langle r^k\rangle_0 = b^{-\frac{k}{2}}\mu_k(\sigma)
\end{align}
The self-consistency condition $R_1=\langle r\rangle_0$ then eliminates $R_1$ entirely. Then, the member of \eqref{eq:app_disordered_fam_dist} occupied by the system is the root $\sigma_*$ of the single scalar equation
\begin{align}
\label{eq:sigma_selfconsistency}
    b\sigma  = a\,\mu_1(\sigma)
\end{align}
The left side is strictly increasing in $\sigma$ and the right side strictly decreasing, so $\sigma_*$ exists and is unique. Moreover, $\sigma_*$ depends on the parameters only through $a/b$ and is strictly increasing in it. So $\sigma$ itself is a dummy label of the family \eqref{eq:app_disordered_fam_dist}, and all dependence on $(a,b)$ enters only through which member $\sigma_*(a/b)$ the self consistency selects. 

By \eqref{eq:Critical_condition}, the disordered state is unstable to ordering precisely when $\langle r^2\rangle_0 > d$, i.e., when
\begin{align}
\label{eq:exact_criterion}
    \mu_2(\sigma_*) > d\,b
    \quad\Longleftrightarrow\quad
    \frac{a}{1-2b} < \rho(\sigma_*),
    \qquad \rho(\sigma):=\frac{\mu_2(\sigma)}{\mu_1(\sigma)^2},
\end{align}
where the equivalence follows from \eqref{eq:sigma_selfconsistency} and \eqref{eq:family_identity}. This is the window inequality \eqref{eq:rho_window} with $\rho$ evaluated at the self-consistent member rather than bounded over all members. For $b\geq\tfrac12$ no instability is possible, since $\mu_2(\sigma)\leq\mu_2(0)=d/2\leq db$, consistent with \eqref{eq:n1m2_window}.

The left side of \eqref{eq:exact_criterion} still involves $a$ through $\sigma_*$, but the monotonicity established above transforms it into a bound on $a$ alone. Since $\mu_2$ is strictly decreasing, for each $b\in(0,\tfrac12)$ there is a unique threshold member $\sigma_c(b)$, defined by the family alone with no reference to $a$,
\begin{align}
    \mu_2\left(\sigma_c(b)\right) = db 
\end{align}
and the chain
\begin{align}
    \mu_2(\sigma_*)>db \iff \sigma_*<\sigma_c(b) \iff a< \frac{b\sigma_c(b)}{\mu_1\left(\sigma_c(b)\right)}    
\end{align}
holds, the last step because \eqref{eq:sigma_selfconsistency} makes $a$ a strictly increasing function of $\sigma_*$ at fixed $b$. The window \eqref{eq:n1m2_window} is therefore sharpened to the exact statement: the ordered phase occupies
\begin{align}
\label{eq:n1m2_exact_window}
    1-2b < a < a_c(b) := \frac{b\sigma_c(b)}{\mu_1(\sigma_c(b))}, \qquad 0< b<\frac{1}{2},
\end{align}
The graph of $a_c$ admits a clean parametric description. By running the threshold member over the family, $\sigma \in(0,\infty)$, and using $db=\mu_2(\sigma)$ together with \eqref{eq:sigma_selfconsistency}, the phase boundary is the curve
\begin{align}
\label{eq:boundary_parametrization}
    a_c(\sigma) = \frac{\sigma \mu_2(\sigma)}{d\,\mu_1(\sigma )}, \qquad b_c(\sigma) = \frac{\mu_2(\sigma)}{d}
\end{align}
which is single-valued over $b\in(0,\tfrac12)$ since $\mu_2'<0$. Evaluating the moments yields the following closed form expression in terms of parabolic cylinder functions $D_{\nu}$, 
\begin{align}
    a_c(\sigma) = \frac{\sigma(d+1)}{d\sqrt{2}}\frac{D_{-(d+2)}\left(\frac{\sigma}{\sqrt{2}}\right)}{D_{-(d+1)}\left(\frac{\sigma}{\sqrt{2}}\right)}, \qquad b_c(\sigma) = \frac{d+1}{2}\frac{D_{-(d+2)}\left(\frac{\sigma}{\sqrt{2}}\right)}{D_{-d}\left(\frac{\sigma}{\sqrt{2}}\right)}
\end{align}
Here $\sigma$ is the curve parameter, so the graph $a_c(b)$ of \eqref{eq:n1m2_exact_window} is recovered by inverting $b_c = \mu_2(\sigma)/d$. Although this is always well-defined, it cannot be done in terms of standard functions, so a closed form expression of the upper boundary, $a_c(b)$, is not available. Therefore, we first extract the exact features of the curve that can be stated in closed form.

The closed-form features follow directly from the identity \eqref{eq:family_identity}, which on the boundary gives $1-2b_c=\sigma\mu_1/d$ and hence
\begin{align}
    \frac{a_c}{1-2b_c} = \rho(\sigma), \qquad a_c + 2b_c - 1 = \frac{\sigma \mathrm{Var}(\xi)}{d\,\mu_1} > 0
    \quad (\sigma>0)
\end{align}
The second relation shows that the boundary lies strictly inside the admissible wedge, so the region \eqref{eq:n1m2_exact_window} is nonempty for every $b\in(0,\tfrac12)$.  We now consider the endpoints of the curve which are exact. As $\sigma \to0$ the family becomes Gaussian and $(a_c,b_c)\to(0,\tfrac12)$, while as $\sigma \to\infty$ the Gaussian factor is irrelevant, $P_\sigma$ approaches a Gamma distribution with $\rho\to1+\tfrac1d$, and $(a_c,b_c)\to\big(1 + \tfrac1d,0\big)$. Expanding the moments about these two limits gives the behavior of the graph at its ends,
\begin{align}
\label{eq:ac_asymptotics}
    a_c(b) =
    \begin{cases}
        \dfrac{d+1}{d} - \dfrac{2(d+2)}{d}\,b + O(b^2),
        & b\to 0,\\[2ex]
        \rho(0)\,(1-2b) + O\big((1-2b)^2\big),
        & b\to \tfrac12,
    \end{cases}
\end{align}
with 
\begin{align}
    \rho(0) = \frac{\Gamma\left(\frac{d}{2}\right)\Gamma\left(\frac{d}{2}+1\right)}{\Gamma\left(\frac{d+1}{2}\right)^2}
\end{align}
and at $d=2$, $\rho(0) = 4/\pi$. Finally, since the bound $\rho<1+\tfrac1d$ of \eqref{eq:rho_window} is strict whenever $B>0$, the ratio 
\begin{align}
    \frac{a_c}{1-2b_c} = \rho(\sigma) < 1 + \frac{1}{d}
\end{align}
stays strictly below the upper bound in \eqref{eq:rho_window} for all $b>0$. Therefore, the exact boundary lies strictly below the upper edge of the window \eqref{eq:n1m2_window}, so a finite portion of the window near that edge contains no ordered phase.

For $d=2$, the case used in the main text, these results give $a_c(b)\approx\tfrac32-4b$ at small $b$, and $\tfrac{4}{\pi}<\rho(\sigma)<\tfrac32$ along the curve, so the phase boundary lies between the lines $a=\tfrac{4}{\pi}(1-2b)$ and $a=\tfrac32(1-2b)$. It crosses $a=1$ at $b_*\approx0.142$, which separates the two sectors of \eqref{eq:sector_split}.

Finally, we justify why \eqref{eq:n1m2_exact_window} is the full phase boundary, and not merely an instability locus. Below the curve the disordered state is linearly unstable, so the system necessarily orders. Above the curve one must additionally rule out ordered states coexisting with the stable disordered one, as would occur at a first-order transition. This is done for $d=2$ in Appendix \ref{subsec:Appendix_generalized_kernel_W}, where it is shown that the self consistency equations admit an ordered solution if and only if the instability criterion \eqref{eq:exact_criterion} holds, i.e. $a<a_c(b)$, and that this solution is unique with the order parameter vanishing continuously as $a\to a_c(b)$ from below. For $a>a_c(b)$, the disordered state is therefore the only solution so \eqref{eq:boundary_parametrization} is the complete phase boundary and the transition across it is continuous. Additionally, the critical line determined numerically in Fig. \ref{fig:app_detB_region} coincides with the present curve under $b\to \kappa$.

\section{Fixed-angular-momentum thermodynamics}
\label{sec:Appendix_free_energy_density}
In this appendix we compute the free energy density of the momentum-conserving model of Section~\ref{sec:Superfluid_model} at fixed angular momentum density $\ell = L/N$, by a saddle-point evaluation of the constrained partition function at large $N$. The resulting free energy \eqref{eq:app_fixed_ell_free_energy} is the starting point of Section~\ref{subsec:Condensate}, and the conjugate angular velocity $\Omega(\ell)=\ell/Q$ fixes the drift term of the effective theory in Appendix~\ref{sec:Appendix_single_cluster_SF}. The saddle point found here is the homogeneous one, in which the second moment $Q$ is shared by all the rotors. It describes every $\ell$ when $a\leq1$, and $|\ell|\leq\ell_{\mathrm{c}}$ when $a>1$. Beyond $\ell_{\mathrm{c}}$ the excess angular momentum is carried by the condensate of Section~\ref{subsec:Condensate}.

 The fixed-$L$ ensemble is defined by the constrained partition function
\begin{align}
    Z_L = \int d\Lambda\, \delta\big(L-L(\Lambda)\big)\, e^{-\beta H(\Lambda)},
    \qquad
    F(L) = -\frac{1}{\beta}\log Z_L,
\end{align}
where $d\Lambda$ is the full phase-space measure over all positions and momenta. Since the constraint fixes the total angular momentum, it is natural to work in polar coordinates, in which the microscopic Hamiltonian reads
\begin{align}
    H = \frac{1}{2}\sum_i\left( p_{r_i}^2 + \frac{p_{\theta_i}^2}{r_i^2}\right) + V(\{r_i,\theta_i\}),
    \qquad
    V = \frac{1}{2N}\sum_{i\neq j}r_ir_j\big(a-\cos(\theta_i-\theta_j)\big) + b\sum_i r_i^2,
\end{align}
and the total angular momentum is $L = \sum_i p_{\theta_i}$. Since the constraint involves only the momenta, the momentum integrals factor out at fixed positions,
\begin{align}
    Z_L &= \int \prod_i dr_i\, d\theta_i\; e^{-\beta V(\{r_i,\theta_i\})}\, K_L(\{r_i\}), \\
    K_L(\{r_i\}) &= \int \prod_i dp_{r_i}\, dp_{\theta_i}\;
    \delta\Big(L-\sum_i p_{\theta_i}\Big)
    \exp\bigg[-\frac{\beta}{2}\sum_i\left( p_{r_i}^2 + \frac{p_{\theta_i}^2}{r_i^2}\right)\bigg].
\end{align}
We exponentiate the constraint using the Fourier representation of the delta function,
\begin{align}
    \delta \left(L - \sum_i p_{\theta_i}\right) = \int \frac{d\lambda'}{2\pi}\,e^{i\lambda' \left(L - \sum_i p_{\theta_i}\right) }
\end{align}
At fixed positions the momentum integrals are Gaussian, and carrying them out together with the $\lambda'$ integral gives
\begin{align}
    K_L(\{r_i\}) = \left(\frac{2\pi}{\beta}\right)^{N-\frac{1}{2}}
    \frac{\prod_i r_i}{\sqrt{\sum_i r_i^2}}\,
    \exp\bigg[-\frac{\beta L^2}{2\sum_i r_i^2}\bigg].
\end{align}
Next we introduce the collective variables $\bar{r}$, $\bar{x}$, and $Q$ via
\begin{align}
    1 = N^3\int d\bar{r}\,d\bar{x}\,dQ\;
    \delta\Big(N\bar{r}-\sum_i r_i\Big)
    \delta\Big(N\bar{x}-\sum_i r_i\cos\theta_i\Big)
    \delta\Big(NQ-\sum_i r_i^2\Big).
\end{align}
In terms of these collective variables, and with $\bar{y}=N^{-1}\sum_i r_i\sin\theta_i$ the transverse component of the order parameter, the potential is exactly
\begin{align}
    V = \frac{N}{2}\left(a\bar{r}^2 -\bar{x}^2 -\bar{y}^2 + 2b Q\right) + \frac{1-a}{2}\,Q.
\end{align}
As in Appendix~\ref{sec:Appendix_free_energy_overdamped}, the global U(1) symmetry allows us to fix the ordered frame at the saddle point, where $\bar{y}$ and its conjugate multiplier vanish. Since the integral over orientations only changes the subleading prefactor, we omit the transverse constraint and set $\bar{y}=0$ in what follows. The last term is $\mathcal{O}(1)$, so it contributes $\mathcal{O}(1/N)$ to the free energy density and is therefore dropped as well. We exponentiate the three constraints with conjugate multipliers,
\begin{align}
    \delta\Big(N\bar{r}-\sum_i r_i\Big) &= \int \frac{d\rho}{2\pi}\, e^{i\rho\left(N\bar{r}-\sum_i r_i\right)}, \\
     \delta\Big(N\bar{x}-\sum_i r_i\cos\theta_i\Big)&=\int \frac{d\mu}{2\pi}\,e^{i\mu\left(N\bar{x}-\sum_i r_i\cos\theta_i\right)}, \\
     \delta\Big(NQ-\sum_i r_i^2\Big)&= \int \frac{d\lambda}{2\pi}\, e^{i\lambda\left(NQ-\sum_i r_i^2\right)}.
\end{align}
Writing $L = N\ell$ and collecting the multiplier and collective-variable measures and the constants,
\begin{align}
    d\Lambda' = d\bar{r}\,d\bar{x}\,dQ\, d\rho\, d\mu\, d\lambda,
    \qquad
    C = \left(\frac{2\pi}{\beta}\right)^{N-\frac{1}{2}}\left(\frac{N}{2\pi}\right)^3,
\end{align}
the partition function becomes
\begin{align}
    Z_L &= C \int d\Lambda'\, (NQ)^{-1/2}
    \exp\bigg[-\frac{N\beta}{2}\left(a\bar{r}^2 -\bar{x}^2 + 2bQ + \frac{\ell^2}{Q}\right) + iN \left(\rho\bar{r} + \mu\bar{x} + \lambda Q \right)\bigg] \notag \\
    &\quad\times\prod_i \int dr_i\, d\theta_i\; r_i\,
    e^{-i\rho r_i - i\mu r_i\cos\theta_i -i\lambda r_i^2}.
\end{align}
The site integrals factorize into $N$ copies of the effective single-site integral
\begin{align}
    z_1 = \int dr\,d\theta \; r\, e^{-i\rho r -i\mu r\cos\theta -i\lambda r^2}.
\end{align}
The prefactor $(NQ)^{-1/2}$ contributes $\mathcal{O}(\log N/N)$ to the free energy density and is likewise dropped. To leading order, the partition function then takes the saddle-point form,
\begin{align}
    Z_L = C \int d\Lambda'\, e^{N\Phi},
    \qquad
    \Phi = -\frac{\beta }{2}\left(a\bar{r}^2 -\bar{x}^2 + 2bQ + \frac{\ell^2}{Q}\right) + i\rho \bar{r}+i\mu\bar{x}+i\lambda Q +  \log z_1.
\end{align}
In the large-$N$ limit the integral is dominated by its saddle point, so, up to an $\ell$-independent additive constant, the fixed-$\ell$ free energy density $f = F/N$ is
\begin{align}
    f(\ell) = -\frac{1}{\beta}\Phi_*(\ell),
\end{align}
where $\Phi_*$ denotes the value of $\Phi$ at the stationary saddle point. Stationarity with respect to the multipliers requires the collective variables to be reproduced by the effective single-site problem,
\begin{align}
    \bar{r}_*=\langle r\rangle_1, \qquad \bar{x}_*=\langle r\cos\theta\rangle_1, \qquad Q_*=\langle r^2\rangle_1,
\end{align}
while stationarity with respect to the collective variables moves the multipliers onto the imaginary axis, as usual in steepest descent,
\begin{align}
    i\rho_* = \beta a\bar{r}, \qquad i\mu_* = -\beta \bar{x}, \qquad i\lambda_* = \beta \left(b-\frac{\ell^2}{2Q^2}\right).
\end{align}
The averages $\langle \cdots\rangle_1$ are taken with respect to the single-site weight evaluated at the saddle,
\begin{align}
    z_1 = \int dr\,d\theta\; r\,\exp\bigg[-\beta \left(a\bar{r}r-\bar{x}r\cos\theta + \left(b- \frac{\ell^2}{2Q^2}\right)r^2\right)\bigg].
\end{align}
Substituting the saddle values of the multipliers back into $\Phi$ gives
\begin{align}
\label{eq:app_f_ell_saddle}
    f(\ell) = \frac{\ell^2}{Q} - \frac{1}{2}\left(a\bar{r}^2-\bar{x}^2\right)-\frac{1}{\beta}\log z_1,
\end{align}
with all quantities evaluated at the self-consistent saddle. Rescaling $r \to (1+\epsilon) r$ in $z_1$ and expanding to first order in $\epsilon$ yields the virial identity
\begin{align}
    a\bar{r}^2-\bar{x}^2 + 2\left(b-\frac{\ell^2}{2Q^2}\right) Q = \frac{2}{\beta},
\end{align}
where we used the saddle conditions to replace the single-site averages by the collective variables. Eliminating $a\bar{r}^2-\bar{x}^2$ from \eqref{eq:app_f_ell_saddle} brings the free energy density to the form quoted in the main text,
\begin{align}
\label{eq:app_fixed_ell_free_energy}
    f(\ell) = bQ + \frac{\ell^2}{2Q} -\frac{1}{\beta}\log z_1 -\frac{1}{\beta}.
\end{align}
Finally, we compute the conjugate angular velocity from the stationary expression \eqref{eq:app_f_ell_saddle}. By the envelope theorem, the collective variables may be held fixed at their saddle values when differentiating with respect to $\ell$, so
\begin{align}
\label{eq:app_Omega_ell}
    \Omega(\ell) = \frac{\partial f}{\partial \ell} = \frac{2\ell}{Q}- \frac{1}{\beta}\frac{\partial \log z_1}{\partial \ell} = \frac{2\ell}{Q} - \frac{\ell}{Q} = \frac{\ell}{Q},
\end{align}
where we used
\begin{align}
    \frac{\partial \log z_1}{\partial \ell} = \beta \frac{\ell}{Q^2}\langle r^2\rangle_1 = \beta \frac{\ell}{Q}.
\end{align}

\section{Single-cluster effective theory}
\label{sec:Appendix_single_cluster_SF}
In this section, we derive the coefficients of the effective MSR Lagrangian in the hydrodynamic regime, 
\begin{align}
\label{eq:app_L_slow_SF}
   \mathcal{L}_{\text{slow}} = \bar{\pi}_{\theta}\dot{\bar{\theta}} + \bar{\sigma}_{\theta}\dot{L} - \bar{\pi}_{\theta} \left\langle \frac{\partial H}{\partial L}\right\rangle + iD_{\theta} \bar{\pi}_{\theta}^2 
\end{align}
We first examine the structure of the projected mobility, which shows why the phase mobility is renormalized by the fast modes. Changing variables from the microscopic phase-space coordinates $a_i\in\{x_i,y_i,p_{x_i},p_{y_i}\}$ to the slow pair and a set of fast transverse coordinates, $q^I = (\bar{\theta},L,\xi_A)$, the dissipative part of the microscopic MSR Lagrangian is governed by the projected mobility
\begin{align}
    M^{IJ} = \sum_i \frac{\partial q^I}{\partial a_i}\Gamma_{a_i b_i}\frac{\partial q^J}{\partial b_i}
\end{align}
Since $\Gamma_{a_ib_i}\partial_{b_i}L=0$ exactly, every entry involving $L$ vanishes, $M^{LI}=0$, so the response field $\bar{\sigma}_{\theta}$ drops out of the dissipative sector. The mixed entries $M^{\theta A}$ do not vanish, however. To see this, use the $\mathrm{U}(1)$ covariance of the problem to evaluate at the aligned configuration $\vec{\bar{s}} = (\bar{s},0)$, where $\partial\bar{\theta}/\partial a_i = (0,1,0,0)/(N\bar{s})$. With $\Gamma = Q^TGQ$ from \eqref{eq:mobility_matrix}, the noise vector projected onto the phase is, rotor by rotor,
\begin{align}
    \left(\Gamma\nabla\bar{\theta}\right)_i = \frac{1}{N\bar{s}}\,Q_i^T G \begin{pmatrix} p_{y_i} \\ y_i \\ x_i\end{pmatrix}
\end{align}
whereas the tangent to the Goldstone orbit, i.e. a simultaneous rotation of all positions and momenta, is 
\begin{align}
    (-y_i,x_i,-p_{y_i},p_{x_i}) = Q_i^T e_3
\end{align}
The two are parallel only if $G(p_{y_i},y_i,x_i)^T\propto e_3$, i.e. only if $\gamma_1 p_{y_i} = 0$ and $\gamma_2 y_i+\gamma_4 x_i = 0$ for every rotor. These conditions fail on typical configurations, since $y_i$ and $p_{y_i}$ are fluctuating fast coordinates (and $x_i\neq0$ when $\gamma_4\neq0$). Hence $\Gamma\nabla\bar{\theta}$ generically has a component off the slow manifold, and in any complete set of rotation-invariant transverse coordinates some of the mixed entries $M^{\theta A}$ are nonzero. This is the momentum-conserving analogue of the nonlinear coupling between the phase and the transverse fluctuations in the overdamped model, and it is the origin of the factor $1-\mathcal{C}_H$ in \eqref{eq:slow_MSR_single_cluster_momenta}. Integrating out $\xi_A$ at fixed $(\bar\theta,L)$ therefore yields the slow Lagrangian \eqref{eq:app_L_slow_SF} with a phase mobility of the form
\begin{align}
\label{eq:app_renormalized_mobility}
    D_{\theta} = \left(1-\mathcal{C}_H\right)D_{\theta}^{(0)} + o\left(\frac{1}{N}\right), \qquad D_{\theta}^{(0)}:= \langle M^{\theta\theta}\rangle,
\end{align}
where $D_{\theta}^{(0)}$ is the rigid-manifold (or direct-noise) contribution, obtained by averaging the projected noise over the fast modes at fixed $(\bar{\theta},L)$, and $\mathcal{C}_H$ collects the correction generated by the mixed coupling $M^{\theta A}$ together with the fast-mode-induced drift of $\bar{\theta}$. The latter is $\mathcal{O}(N^{-1/2})$ with an expected $\mathcal{O}(1)$ correlation time, exactly as in Section \ref{subsec:Diffusion_Goldstone_mode}, so its time-integrated effect is $\mathcal{O}(N^{-1})$ and $\mathcal{C}_H$ is an order-one number. In the overdamped model, reversibility of the Langevin dynamics allowed us to compute this correction exactly by the corrector method of Appendix \ref{sec:Appendix_finite_N_phase_diffusion} and to bound it between $0$ and $1-\bar{s}^2/Q$. The kinetic generator of the present model contains the antisymmetric Hamiltonian flow in addition to the symmetric dissipative part, so that argument does not apply as stated. We therefore leave $\mathcal{C}_H$ as a non-universal coefficient. The lower bound derived in Section \ref{subsec:Appendix_rotation_bound} below gives $\mathcal{C}_H<1$, and the exact scale invariance of the quadratic model at $\Omega=0$ shows that $\mathcal{C}_H$ depends on $(a,b,G)$ only. 
 
We now compute the rigid-manifold coefficient, expanding about the non-rotating background $(\Omega =0)$,
\begin{align}
   M^{\theta\theta} = \sum_i \frac{\partial \bar{\theta}}{\partial a_i}\Gamma_{a_ib_i}\frac{\partial \bar{\theta}}{\partial b_i}
\end{align}
The mobility matrix $\Gamma$ is given by \eqref{eq:mobility_matrix} in Section \ref{subsubsec:constructing_effective_kinetic_eq}. Using
\begin{align}
    \frac{\partial \bar{\theta}}{\partial a_i} = \left(-\frac{\sin\bar{\theta}}{N\bar{s}},\frac{\cos\bar{\theta}}{N\bar{s}} ,0,0\right)
\end{align}
a direct calculation yields
\begin{align}
   M^{\theta\theta} &= \frac{1}{N^2\bar{s}^2}\sum_i\left[\cos^2\bar{\theta}\left(p_{y_i}^2\gamma_1 +y_i^2\gamma_2 +x_i^2\gamma_3 +2x_iy_i\gamma_4\right) + \sin^2\bar{\theta}\left(p_{x_i}^2\gamma_1 +x_i^2\gamma_2 + y_i^2\gamma_3 -2x_iy_i\gamma_4\right)\right. \notag\\
    &\quad -\left.\sin(2\bar{\theta})\left(p_{x_i}p_{y_i}\gamma_1 +x_iy_i(\gamma_2-\gamma_3)+(x_i^2-y_i^2)\gamma_4\right)\right]
\end{align}
Choosing the ordered frame to align with the $x$-axis implies $\sin\bar{\theta}=0$. Then, 
\begin{align}
    M^{\theta\theta} &= \frac{1}{N^2\bar{s}^2}\sum_i\left(p_{y_i}^2\gamma_1 +y_i^2\gamma_2 +x_i^2\gamma_3 +2x_iy_i\gamma_4\right) 
\end{align}
Averaging over the fast modes at fixed $(\bar{\theta},L)$ gives, to leading order in $1/N$, the rigid-manifold coefficient
\begin{align}
    D_{\theta}^{(0)} = \langle M^{\theta\theta}\rangle = \frac{1}{N\bar{s}^2}\left\langle p_y^2\gamma_1 +y^2\gamma_2 +x^2\gamma_3 +2xy\gamma_4 \right\rangle 
\end{align}
In the ordered frame $\vec{\bar{s}}=(\bar{s},0)$, by symmetry, all averages odd in $y$ vanish and the equilibrium self-consistency equations give $\langle y^2\rangle = 1/\beta$. These simplifications, together with the identity,
\begin{align}
    \langle p_y^2\rangle = \frac{1}{\beta}+\Omega^2\langle x^2\rangle
\end{align}
reduce the coefficient to 
\begin{align}
    D_{\theta}^{(0)} = \frac{1}{N\bar{s}^2}\left(\frac{1}{\beta}(\gamma_1+\gamma_2)+ \langle x^2\rangle (\gamma_3 + \Omega^2\gamma_1)\right)
\end{align}
Expanding about the non-rotating saddle ($\Omega =0$),
\begin{align}
    D_{\theta}^{(0)} = \frac{\gamma_3}{N\bar{s}^2}\langle x^2\rangle + \frac{\gamma_1+\gamma_2}{\beta N\bar{s}^2} + \mathcal{O}(\Omega^2)
\end{align}
At the ordered saddle, 
\begin{align}
    Q=\langle r^2\rangle = \langle y^2\rangle +\langle x^2\rangle = \langle x^2\rangle + \frac{1}{\beta}
\end{align}
Therefore, the rigid-manifold coefficient can equivalently be expressed as 
\begin{align}
     D_{\theta}^{(0)} = \frac{\gamma_3 Q}{N\bar{s}^2}+ \frac{\gamma_1+\gamma_2-\gamma_3}{\beta N\bar{s}^2} + \mathcal{O}(\Omega^2)
\end{align}
Under the simplification 
\begin{align}
     \gamma_1 + \gamma_2 = \gamma_3, \quad \gamma_3 = \gamma >0
\end{align}
we obtain the final expression 
\begin{align}
     D_{\theta}^{(0)} = \frac{\gamma Q}{N\bar{s}^2}+  \mathcal{O}(\Omega^2)
\end{align}
which is manifestly temperature independent, since $Q\propto T$ and $\bar{s}^2\propto T$. For the remaining coefficient, the drift of the slow pair is governed by the constrained free energy \eqref{eq:app_f_ell_saddle}, with which we identify $H_{\text{eff}}$:
\begin{align}
    \left\langle \frac{\partial H}{\partial L}\right\rangle= \frac{\partial H_{\rm eff}}{\partial L} = f'(\ell)
\end{align}
On the homogeneous branch, \eqref{eq:app_Omega_ell} gives $f'(\ell)=\Omega(\ell)=\ell/Q$. Hence, using $\ell = L/N$, 
\begin{align}
     \frac{\partial H_{\text{eff}}}{\partial L} = \frac{L}{NQ}
\end{align}

\subsection{Lower bound on $D_\theta$}
\label{subsec:Appendix_rotation_bound}
Although the antisymmetric Hamiltonian flow prevents us from using the method of Appendix \ref{sec:Appendix_finite_N_phase_diffusion} to obtain bounds, strict positivity of $D_\theta$ can be obtained directly. For the constant positive-definite damping matrix \eqref{eq:G_time_reversal}, define
\begin{align}
    \gamma_{\mathrm{rot}} := \gamma_3-\frac{\gamma_4^2}{\gamma_2} = \frac{1}{(G^{-1})_{33}}>0 .
\end{align}
Completing the square in the lower block of $G$ gives, for every $v\in\mathbb{R}^3$,
\begin{align}
\label{eq:app_G_square}
    v^TGv = \gamma_1v_1^2+\gamma_2\left(v_2+\frac{\gamma_4}{\gamma_2}v_3\right)^2+\gamma_{\mathrm{rot}}v_3^2 \geq \gamma_{\mathrm{rot}}v_3^2 .
\end{align}
The third row of $Q$ in \eqref{eq:mobility_matrix} is the vector field that rotates the position and momentum of a single rotor. Let $\mathcal{R}_i$ denote this field acting on rotor $i$, so that $\mathcal{R}:=\sum_i\mathcal{R}_i$ generates a simultaneous rotation of the whole cluster. With $\Gamma_i=Q_i^TGQ_i$ the mobility of rotor $i$, \eqref{eq:app_G_square} and Cauchy-Schwarz give, pointwise on phase space and for every smooth $f$,
\begin{align}
    \sum_i(\nabla_if)^T\Gamma_i\nabla_if \geq \gamma_{\mathrm{rot}}\sum_i(\mathcal{R}_if)^2 \geq \frac{\gamma_{\mathrm{rot}}}{N}(\mathcal{R}f)^2 .
\end{align}
The microscopic diffusion tensor is therefore the sum of a global-rotation piece with constant coefficient $\gamma_{\mathrm{rot}}/N$ and a positive-semidefinite remainder. Since $\mathcal{R}$ is divergence free and $\mathcal{R}H=0$, the first piece contributes
\begin{align}
    \mathcal{L}_{\mathrm{rot}} = \frac{\gamma_{\mathrm{rot}}}{N}\mathcal{R}^2
\end{align}
to the backward generator of the $N$-body process. The remainder of the generator, including the Hamiltonian flow, is rotation covariant and so commutes with $\mathcal{L}_{\mathrm{rot}}$. Consequently, if $Y_t$ is the process generated by the remainder and $B_t$ is a standard Brownian motion independent of $Y$, then
\begin{align}
    X_t = \mathrm{Rot}_{\varphi_t}(Y_t), \qquad \varphi_t:=\sqrt{2\gamma_{\mathrm{rot}}/N}\,B_t ,
\end{align}
where $\mathrm{Rot}_\varphi$ is the global rotation by the angle $\varphi$, has the generator of the original process. The construction preserves the total angular momentum. A global rotation adds its angle to the lifted phase, so
\begin{align}
    \mathrm{Var}\left[\mathrm{\Delta}\bar{\theta}(X_t)\right] = \frac{2\gamma_{\mathrm{rot}}}{N}t+\mathrm{Var}\left[\mathrm{\Delta}\bar{\theta}(Y_t)\right] \geq \frac{2\gamma_{\mathrm{rot}}}{N}t ,
\end{align}
and therefore
\begin{align}
\label{eq:app_rotation_bound}
    D_\theta \geq \frac{\gamma_{\mathrm{rot}}}{N} .
\end{align}
This decomposition uses only the rotation symmetry, so the bound holds for $D_\theta$ as it is understood in Section \ref{subsec:MSR_formalism_soft_rotor}, namely as the coefficient that governs smooth fluctuations about the ordered state. Under the simplification $\gamma_1+\gamma_2=\gamma_3=\gamma$, \eqref{eq:app_rotation_bound} gives, at leading order in $1/N$,
\begin{align}
    \mathcal{C}_H \leq 1-\frac{\gamma_{\mathrm{rot}}\,\bar{s}^2}{\gamma\,Q} < 1 .
\end{align}
This shows that the fast modes cannot cancel the leading $1/N$ coefficient. It does not give $\mathcal{C}_H\geq0$, and it does not give the $\mathcal{O}(N^{-1})$ upper scaling, which rests on the expected $\mathcal{O}(1)$ correlation time of the fast modes.

\section{Spectral properties of the momentum-conserving model}
\label{sec:Appendix_spectral_properties_SF_model}
\subsection{Generalized Kernel of W}
\label{subsec:Appendix_generalized_kernel_W}
We want to show that the generalized kernel of $W$ is 
\begin{align}
    \text{gen}\ker W = \text{span}\{1,y,\phi_L\}
\end{align}
In other words, the full root space stabilizes at order two, 
\begin{align}
    \bigcup_{n\geq 1}\ker W^n = \ker W^2 = \text{span}\{1,y,\phi_L\}
\end{align}
\subsubsection{Exact Kernel}
\label{subsubsec:Appendix_exact_kernel_SF}
In this section, we show
\begin{align}
    \ker W = \text{span}\{1,y\}
\end{align}
As shown in Section \ref{subsubsec:zero_modes_with_momenta},
\begin{align}
    \text{span}\{1,y\}\subseteq \ker W, \qquad
    \phi \in \ker W \iff \phi +\delta \bar{H}(\phi) = \text{const.}
\end{align}
These are precisely the two facts from which the overdamped kernel
computation of Appendix \ref{subsec:Appendix_kernel_W} proceeds, and the
argument from this point on carries over verbatim. Therefore, there are additional zero modes if and only if 
\begin{align}
    \det B = 0
\end{align}
with $B=\mathbb{I}+DM$ as in \eqref{eq:matrices_ker_overdamped}. The only difference from the overdamped case enters through the coefficients. The entries of the Gram matrix $M$ are now equilibrium covariances with respect to the mean-field Hamiltonian of the momentum-conserving model. Since the basis functions are purely positional, the Gaussian momentum integrals cancel from every entry, and the rotation enters the coefficients only through the centrifugal shift $b\to\kappa$ of
\eqref{eq:rotating_quad_potential}. All averages below are therefore taken with respect to the positional weight $e^{-\beta V^{\Omega}}$,
\begin{align}
\label{eq:app_potential}
    V^{\Omega} = \kappa r^2 + a\bar{r}r -\bar{x}x ,\qquad \kappa := b-\frac{\Omega^2}{2}
\end{align}
We do not repeat the construction, and pass directly to the evaluation of
\begin{align}
\label{eq:app_SF_detB}
     \det B = (1 + aM_{rr})(1 - M_{xx}) + aM_{xr}^2,
\end{align}
where the entries of the Gram matrix $M$ are 
\begin{align}
      M_{rr} := \beta\,\text{Var}(r), \quad
     M_{xx} := \beta\,\text{Var}(x), \quad
     M_{xr} := \beta\,\text{Cov}(x,r) 
\end{align}
Unlike the overdamped case, these covariances cannot be evaluated in closed form, so establishing $\det B\neq 0$ requires a different strategy. We note at the outset that the same determinant appears as the denominator of the corrector coefficients that define the generalized zero mode $\phi_L$ in Section \ref{subsubsec:zero_modes_with_momenta}, so $\det B\neq0$ simultaneously rules out additional exact zero modes and guarantees that $\phi_L$ is well defined. 

Two structural facts considerably simplify the calculation. First, every term in \eqref{eq:app_potential} is quadratic in $(r,\bar{r},\bar{x})$ and scales as $\beta^{-1}$. Therefore, the $\beta$ dependence cancels in all normalized averages so $\det B$ depends on $(\beta, a, b, \Omega)$ only through the pair $(a,\kappa)$ and is exactly independent of $\beta$. We therefore set $\beta =1$ without loss of generality. Second, the allowed region reduces to the window \eqref{eq:n1m2_window} of Appendix \ref{sec:Appendix_allowed_models} (with $d=2$ and $b\to\kappa$),
\begin{align}
    0<a<\frac{3}{2}, \quad \kappa_{\min}(a)<\kappa <\frac{3-2a}{6}, \quad \kappa_{\min}(a) := \max\{0,\frac{1-a}{2}\}
\end{align}
So the entire four-parameter family collapses, with no redundancy and no loss, onto a two-dimensional wedge, and all the analysis happens there.

The self consistency problem is a fixed point in two variables $(\bar{r},\bar{x})$, and cannot be solved in closed form. To reduce it to one dimension, we introduce an auxiliary field $h$, and define for each fixed $h\geq 0$ the family of the measures
\begin{align}
    \mu_h \propto e^{-\kappa r^2 -a\bar{r}_*(h)r+hx}
\end{align}
where $\bar{x} = h$ while the radial field
alone is self-consistent, i.e. $\bar{r}_*(h)$ is the solution to 
\begin{align}
    \langle r\rangle_{\mu_h} = \bar{r}
\end{align}
Since averages respond to fields through covariances,
\begin{align}
\label{eq:app_partials}
    \frac{\partial \langle r\rangle}{\partial \bar{r}} = -aM_{rr}, \quad \frac{\partial \langle r\rangle}{\partial h} = M_{xr}, \quad \frac{\partial \langle x\rangle}{\partial \bar{r}} = -aM_{xr}, \quad \frac{\partial \langle x\rangle }{\partial h} = M_{xx}
\end{align}
To see that this reduction is well posed, let us define
\begin{align}
    G(\bar{r}):= \langle r\rangle -\bar{r}
\end{align}
This function is strictly decreasing since by non-negativity of variance $M_{rr}>0$, together with $a>0$, 
\begin{align}
    \frac{dG}{d\bar{r}} = -aM_{rr}-1\leq -1
\end{align}
Additionally, $G(0)=\langle r\rangle_{\mu_h}|_{\bar{r}=0}>0$ and $G\to -\infty$ as $\bar{r}\to \infty$, so there is exactly one root $\bar{r}_*(h)$ and it is smooth in $h$ by the implicit function theorem. Now define the response
\begin{align}
    m(h):= \langle x\rangle_{\mu_h}
\end{align}
The physical ordered states are exactly the solutions of 
\begin{align}
    m(h) = h, \quad h>0
\end{align}
Solutions are the points at which $h=\bar{x}$ and $\mu_h$ coincides with the equilibrium measure. Away from those points $h$ is a continuation parameter that converts the two-dimensional fixed point into a one-dimensional crossing problem.

Differentiating the response,  
\begin{align}
    m'(h) = \frac{\partial \langle x\rangle }{\partial \bar{r}}\frac{d \bar{r}_*}{dh} +\frac{\partial \langle x\rangle }{\partial h} = M_{xx}-\frac{aM_{xr}^2}{1+aM_{rr}}
\end{align}
Multiplying by $(1+aM_{rr})$ and comparing with \eqref{eq:app_SF_detB}, 
\begin{align}
    \det B = (1+aM_{rr})(1-m'(h))
\end{align}
Since $1+aM_{rr}>1$, the sign of $\det B$ is entirely determined by the sign of $1-m'(h)$. Moreover, $\det B \neq 0$ if and only if $m(h)$ crosses the diagonal transversally.

The crossing condition acquires a physical meaning through a Ward
identity. Consider $N(\epsilon) = \langle y\rangle$ computed with the field direction rotated by a small angle $\epsilon$. Then, 
\begin{align}
    N(\epsilon) = \frac{1}{Z_{\epsilon}}\int dxdy \, y e^{-\beta(\kappa r^2+a\bar{r}r)}e^{\beta h(x\cos\epsilon + y\sin\epsilon)}
\end{align}
Differentiating at $\epsilon =0$ yields
\begin{align}
   \left. \frac{d}{d\epsilon}N(\epsilon)\right|_{\epsilon =0} = \beta h \langle y^2\rangle_{\mu_h}
\end{align}
Additionally, define new integration variables $(x',y')$ as the coordinates of the same point in a frame rotated by $+\epsilon$. Then, 
\begin{align}
    N(\epsilon) = \langle x\rangle_{\mu_h}\sin\epsilon \implies  \left. \frac{d}{d\epsilon}N(\epsilon)\right|_{\epsilon =0} =\langle x\rangle_{\mu_h} = m(h)
\end{align}
Equating both derivatives yields the identity,
\begin{align}
\label{eq:app_ward}
    \beta \langle y^2\rangle_{\mu_h} = \frac{m(h)}{h}
\end{align}
At a self consistent ordered state $m(h)=h$, and using $\langle y|y\rangle = \beta \langle y^2\rangle$ we obtain the equilibrium Goldstone mode normalization
\begin{align}
    \langle y|y\rangle = 1 
\end{align}
used in Section \ref{subsec:linear_MF_PDE}. We define
\begin{align}
    g(h) := \langle y|y\rangle(h)
\end{align}
This curve is the transverse fluctuation strength, i.e. the norm of the would-be Goldstone mode at field $h$. The crossing condition $g=1$ is the Goldstone normalization and is equivalent to self consistency itself. Now the sign of the determinant acquires a physical reading since,
\begin{align}
\label{eq:app_slope}
    \left.hg'(h)\right|_{\text{crossing}} = m'(h)-1 = -\frac{\det B}{1+aM_{rr}}
\end{align}
The fixed point condition $m(h)=h$ is the level condition $g(h)=1$, so ordered states are the crossings of the curve $g$ with the horizontal level $g=1$. Then, $\det B>0$ at an ordered state if and only if the curve $g(h)$ crosses the level $g=1$ from above.

The endpoints of the curve can be computed exactly, and they reproduce the ordering threshold of Appendix \ref{sec:Appendix_allowed_models} and the collective threshold of Section \ref{subsubsec:critical_angular_velocities}. At $h\to0^+$ the measure is isotropic, so $\langle
y^2\rangle=\langle x^2\rangle$ and
\begin{align}
    g(0^+) = \beta\langle x^2\rangle_0 =: \chi_0(a,\kappa)
\end{align}
which is the critical condition \eqref{eq:Critical_condition} of Appendix \ref{sec:Appendix_allowed_models} at $d=2$ since isotropy gives $\langle r^2\rangle_0 = 2\langle x^2\rangle_0$, so $g(0^+)=1$ is precisely $\langle r^2\rangle_0 = 2/\beta$. At $h\to\infty$, the measure concentrates in a $\mathcal{O}(1)$ neighborhood of the potential minimum, $\theta=0$, $r_0 = (h-a\bar{r}_*)/(2\kappa)$. The radial width $(2\beta\kappa)^{-1/2}$ and the tangential width $\sqrt{r_0/(\beta h)}$ both remain finite while $r_0$ increases. Imposing the radial self-consistency gives $\bar{r}_* = h/(2\kappa+a) + \mathcal{O}(1)$ and, since the angular spread vanishes,
\begin{align}
    \lim_{h\to\infty} g(h) = \frac{1}{2\kappa+a} 
\end{align}
The asymptote lies below the Goldstone level exactly when
\begin{align}
    2\kappa +a > 1 \iff \kappa>\kappa_{\min}(a) \iff \Omega<\Omega_1
  \,\,  (a\leq 1)
\end{align}
recovering the mean-field critical angular velocity of
\eqref{eq:critical_velocities}, while for $a>1$ the condition is automatic for any $\kappa>0$. Two conclusions follow immediately. First, whenever $\chi_0>1$ the continuous curve starts above the level and ends below it, so an ordered solution exists. The ordered branch occupies exactly the subregion $\chi_0(a,\kappa)>1$ of the wedge, which is characterized precisely in Appendix \ref{sec:Appendix_allowed_models}. Second, the loci where $\det B$ can degenerate are read off from the endpoints, and they are the ordering transition and the collective threshold $\Omega_1$ of Section \ref{subsubsec:critical_angular_velocities}:
\begin{itemize}
    \item As $\chi_0\to 1^+$, the crossing slides to $h\to 0$, as in the orange curve in Fig. \ref{fig:app_gcurve}. The ordered state shrinks continuously into the isotropic state, with $\bar{x}\propto \sqrt{\chi_0 -1}$. The slope of the curve at the crossing goes to zero along with it so $\det B \propto (\chi_0-1)$. 
    \item As $\kappa\to\kappa_{\min}^+$ with $a\leq 1$
    ($\Omega\to\Omega_1$), the asymptote $(2\kappa +a)^{-1}$ rises toward 1, and the crossing is pushed to increasingly larger $h$, as in the green curve of Fig. \ref{fig:app_gcurve}. Here, the ordered state does not vanish by losing its amplitude. Instead $\bar{r}$ and $\bar{x}$ grow without bound as described in Section \ref{subsec:homogeneous_saddle}. Since the curve approaches the level nearly flat at the crossing, $\det B \propto (\kappa-\kappa_{\min})$. 
    \item For $a>1$ and $\kappa\to0^+$ ($\Omega\to\Omega_2$), by contrast,
    \begin{align}
        \frac{1}{2\kappa + a}\leq \frac{1}{a}<1 ,
    \end{align}
    so the asymptote is uniformly bounded away from the level and $\det B$ remains $\mathcal{O}(1)$ at the threshold.
\end{itemize}
The distinct critical behavior for $a\leq 1$ and $a>1$ mirrors the sector split \eqref{eq:sector_split} exactly. Physically, $\det B$ is the stiffness of the collective amplitude sector $\text{span}\{\tilde r,\tilde x\}$. The threshold $\Omega_1$ is a collective instability, i.e. the entire cloud softens, so the amplitude sector goes marginal and $\det B\to0$. The threshold $\Omega_2$ is the escape of individual rotors to large radius (the condensate of Section \ref{subsec:Condensate}), during which the collective mode remains stiff so $\det B$ does not vanish there. In other words, $\det B\to0$ occurs exactly where the amplitude sector degenerates against the Goldstone normalization, and both such edges lie on the boundary of the ordered phase itself, excluded by the strict inequalities defining it.

The preceding chain of identities is exact, and it reduces the claim $\det B\neq0$ to a single monotonicity property of one curve in one variable, i.e. $g(h)$ should cross the level $g=1$ strictly from above at every ordered state. In fact the numerics of the next section verifies the stronger statement that $g'(h)<0$ everywhere on the wedge, which simultaneously yields uniqueness of the ordered state, its existence precisely for $\chi_0>1$, and $\det B>0$ at it.

\subsubsection{Numerical verification}
\label{subsubsec:Appendix_detB_numerics}
It remains to verify that $g(h)$ is strictly decreasing on the wedge. By completing the angular integral, all
quantities reduce to one-dimensional integrals. Using,
\begin{align}
    \int_0^{2\pi}d\theta\, e^{hr\cos\theta}\cos(n\theta) = 2\pi I_n(hr)
\end{align}
the partition function becomes, up to an overall factor of $2\pi$ that cancels in every average,
\begin{align}
    Z = \int dr W(r) I_0(hr), \quad W(r):=r\,e^{-\kappa r^2 - a\bar{r} r}
\end{align}
and every covariance in \eqref{eq:app_SF_detB} is a radial integral,
\begin{align}
    &\langle r\rangle = \frac{1}{Z}\int dr W(r)r I_0(hr), \quad \langle x\rangle = \frac{1}{Z}\int dr W(r)r I_1(hr), \quad \langle r^2\rangle = \frac{1}{Z}\int dr W(r) r^2 I_0(hr) \\[1em]
    &\langle rx\rangle = \frac{1}{Z}\int dr W(r)r^2 I_1(hr), \quad \langle x^2\rangle = \frac{1}{2Z}\int dr W(r)r^2 \left(I_0(hr)+I_2(hr)\right)
\end{align}
and $\langle y^2\rangle = \langle r^2\rangle - \langle x^2\rangle$. We
evaluate these by Gauss--Legendre quadrature ($N=800$--$1400$ nodes) on an
adaptively chosen domain, with exponentially scaled Bessel functions and
stabilized exponents so that all integrands are $\mathcal{O}(1)$; the
self-consistency for $\bar{r}_*(h)$, and for the full two-field problem, is
solved by Newton and hybrid methods with residuals below $10^{-14}$.
Node-doubling changes no reported quantity at relative level $10^{-14}$,
and the Bessel-reduced integrals agree with brute-force two-dimensional
quadrature at the same level. As independent end-to-end checks, all five
covariances were reproduced within statistical error by Metropolis Monte
Carlo both in the reduced positional measure ($4\times10^6$ samples) and
in the full four-dimensional phase space including the
$\Omega(xp_y-yp_x)$ coupling ($6\times10^6$ samples), directly validating
the momentum reduction; applied to the overdamped model, the same pipeline
reproduces the closed-form determinant \eqref{eq:determinant_of_B} and the
saddle-point relations of Appendix \ref{subsec:Appendix_kernel_W} to
relative accuracy $10^{-12}$; and solving the self-consistency
independently at $21$ temperatures spanning $T\in[10^{-2},10^{2}]$
confirms the exact temperature independence of $\det B$ to a spread of
$\sim10^{-13}$.

\begin{figure}[t]
    \centering
    \includegraphics[width=0.62\textwidth]{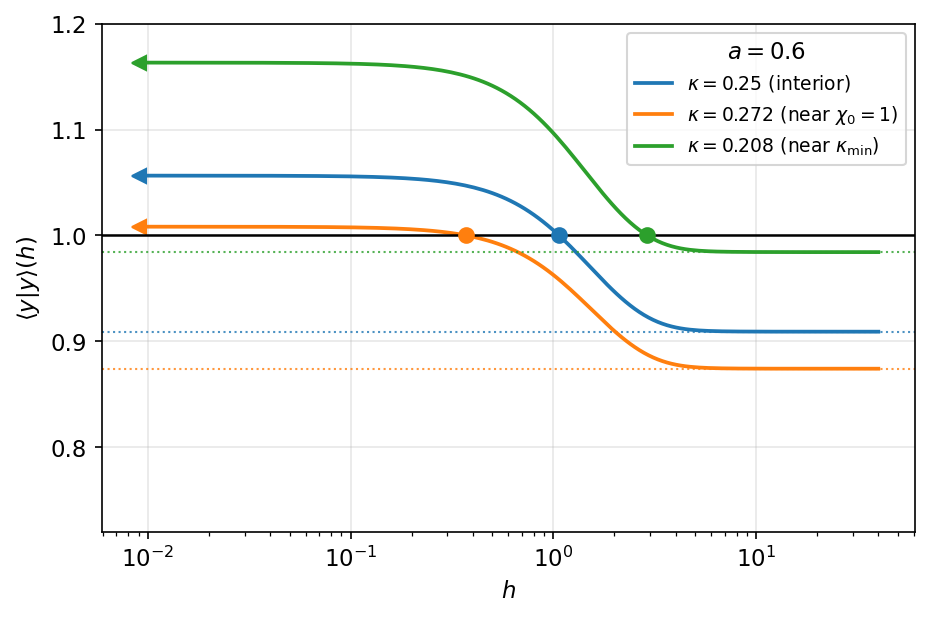}
    \caption{The transverse fluctuation strength
    $g(h)=\langle y|y\rangle(h)$ at $a=0.6$ for three values of $\kappa$.
    Arrows mark the $h\to0$ limits $\chi_0$, dotted lines the asymptotes
    $1/(2\kappa+a)$, and dots the crossings of the Goldstone level $g=1$,
    i.e., the ordered states. The curve is strictly decreasing, so each
    crossing is transversal and unique: $\det B>0$ by
    \eqref{eq:app_slope}. Near the critical line the crossing slides to
    $h\to0$; near $\kappa_{\min}$ the asymptote pinches against the level
    and the crossing escapes to $h\to\infty$; in both limits the crossing
    flattens, which is the linear vanishing of $\det B$ at the marginal
    edges.}
    \label{fig:app_gcurve}
\end{figure}

\begin{figure}[t]
    \centering
    \includegraphics[width=0.66\textwidth]{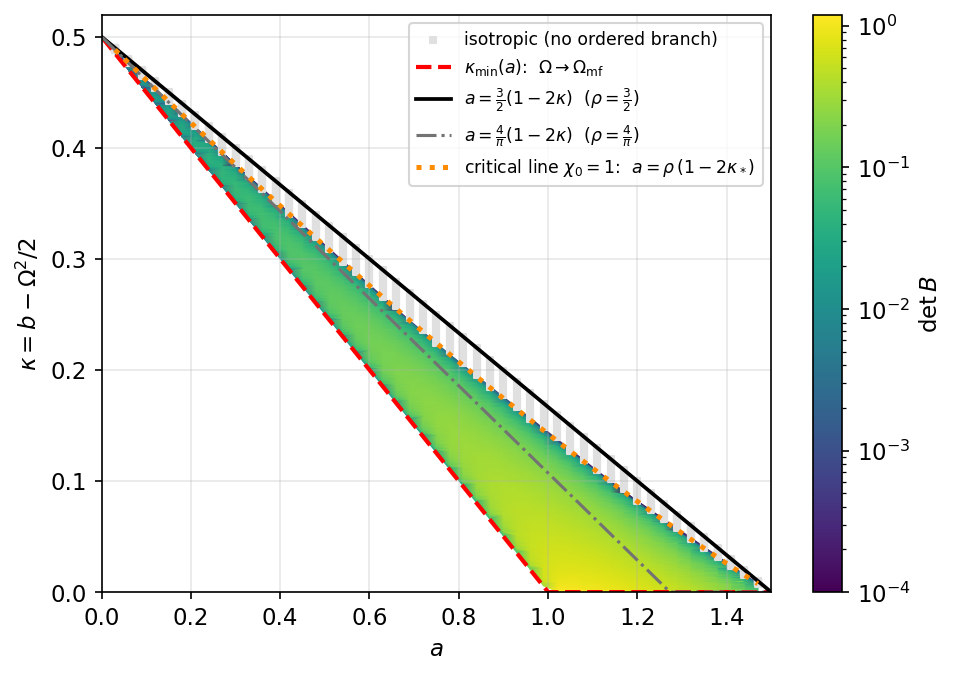}
    \caption{$\det B$ on the ordered branch over the wedge
    $(a,\kappa)$, exactly independent of temperature. Grey squares: no
    ordered branch ($\chi_0<1$). The ordered region is bounded below by
    $\kappa_{\min}(a)$ (red dashed; $\Omega\to\Omega_{1}$ for $a\leq1$)
    and above by the exact critical line $\chi_0=1$ (orange dotted), which
    satisfies $a=\rho\,(1-2\kappa_*)$ with
    $\rho\in(4/\pi,\tfrac{3}{2})$ and is confined between the constant-$\rho$
    lines shown (see Appendix \ref{sec:Appendix_allowed_models}).
    $\det B>0$ at every ordered point, approaching zero only at the two
    marginal edges.}
    \label{fig:app_detB_region}
\end{figure}

\begin{figure}[t]
    \centering
    \includegraphics[width=\textwidth]{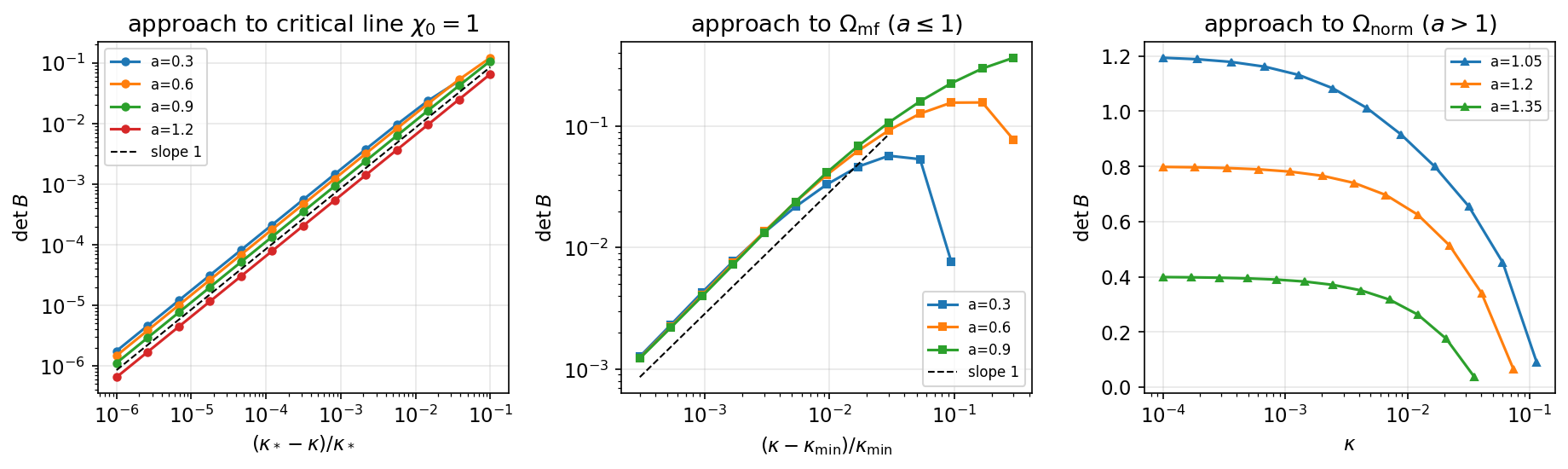}
    \caption{Approach to the three boundaries of the ordered region.
    Left: critical line, $\det B\propto(\kappa_*-\kappa)$ down to relative
    distance $10^{-6}$. Middle: collective threshold $\Omega\to\Omega_1$
    ($a\leq1$), $\det B\propto(\kappa-\kappa_{\min})$; the downturn at the
    right of each curve is the approach to the critical line. Right:
    single-rotor threshold $\Omega\to\Omega_2$ ($a>1$): $\det B$ remains
    $\mathcal{O}(1)$, since the collective amplitude sector stays stiff at
    the condensate transition.}
    \label{fig:app_detB_edges}
\end{figure}

The derivative of $g$ is evaluated through the exact identity obtained by
combining \eqref{eq:app_ward} with the chain rule,
\begin{align}
\label{eq:app_gprime_identity}
    h\,g'(h) = \Big(M_{xx} - \langle y|y\rangle\Big) - \frac{aM_{xr}^2}{1+aM_{rr}} ,
\end{align}
with all covariances taken in $\mu_h$; the identity was checked against
finite differences of $g$ at the $10^{-10}$ level. Scanning $15\times10$
values of $(a,\kappa)$ covering the wedge and $48$ logarithmically spaced
fields $h\in[0.02,30]$ ($7200$ evaluations) we find $g'(h)<0$ at
every point (Fig.~\ref{fig:app_gcurve}). Independently, solving the full
two-field self-consistency on a grid of $\sim2000$ points covering the
wedge yields $\det B>0$ at every ordered point
(Fig.~\ref{fig:app_detB_region}), with $\det B\geq 0.03$ outside a $2\%$
neighborhood of the boundaries and clean linear vanishing on approach to
the two marginal edges, with fitted exponents $0.997$--$1.000$ at the
critical line (where also $\bar{x}\propto(\chi_0-1)^{1/2}$, the mean-field
pitchfork) and $0.91$--$1.01$ at $\Omega\to\Omega_1$, while at
$\Omega\to\Omega_2$ ($a>1$) the determinant remains $\mathcal{O}(1)$
(Fig.~\ref{fig:app_detB_edges}), in agreement with the endpoint analysis
above. We conclude that
\begin{align}
    \det B \neq 0
\end{align}
at every point of the ordered branch in the allowed region, at all
temperatures. Hence $B$ is invertible, there are no additional exact zero
modes, and
\begin{align}
    \ker W = \text{span}\{1,y\} .
\end{align}

\subsubsection{The Kernel of $W^2$ and Stabilization of the Kernel Chain}
\label{subsubsec:genker_W}
The remaining statements, $\ker W^2 = \text{span}\{1,y,\phi_L\}$ and $\ker W^n = \ker W^2$ for $n\geq 2$, follow from a general result, proven in a companion paper \cite{companion}. We state the result in the notation of this paper and then verify its hypotheses for the momentum conserving model. 

Consider a kinetic equation on phase space of the form
\begin{align}
\label{eq:app_metriplectic}
    \partial_tP = \{\bar{H}[P],P\} + \partial_a \left(\mathcal{M}_{ab}[P]\partial_b \frac{\delta F}{\delta P}\right), \quad F[P] = \frac{1}{\beta}\int dz \,P \log P + E[P], \quad \bar{H}[P]:=\frac{\delta E}{\delta P}
\end{align}
with $\mathcal{M}[P]$ symmetric and positive semidefinite, linearized about a self consistent Gibbs state $P_* = e^{-\beta \bar{H}[P_*]}/Z_*$. The weighted $L^2(\mu)$ inner product with respect to the Gibbs measure $d\mu := P_*dz$ is defined as in Section \ref{subsec:linear_MF_PDE}, as well as $\delta \bar{H}(\phi)$, $u[\phi]$, and the linearized operator $W$. Finally, the mobility matrix $\Gamma$ is defined through $\mathcal{M}[P_*]=\beta \Gamma P_*$. The hypotheses are:
\begin{enumerate}[label=(A\arabic*)]
    \item \label{A1} \textbf{Variational structure.} $E$ is twice continuously differentiable, and $\delta\bar{H}$ has finite rank.
    \item \label{A2}\textbf{Conserved charge.} There is a smooth function $\mathcal{Q}(z)$ whose Hamiltonian flow $\Phi_s$ is complete, and $E[P\circ\Phi_{-s}] = E[P]$ for all $P$ and $s$.
    \item \label{A3}\textbf{Admissible dissipation.} $\mathcal{M}_{ab}[P]\,\partial_b\mathcal{Q}=0$ for every $P$, and $\langle\partial_a f|\Gamma_{ab}\partial_b f\rangle=0$ if and only if $f=F(\mathcal{Q})$. For $\mathcal{M}[P]=\beta\Gamma(z)P$, the latter holds if $\ker\Gamma=\text{span}\{\nabla\mathcal{Q}\}$ for almost every $z$ and the level sets of $\mathcal{Q}$ are connected.
    \item \label{A4}\textbf{Ordered state.} The Goldstone mode $\phi_G:=\{\bar{H}[P_*],\mathcal{Q}\}$ is nonzero. Moreover $\mathcal{Q},\phi_G\in L^2$, $\langle\partial_a\mathcal{Q}|\Gamma_{ab}\partial_b\mathcal{Q}\rangle<\infty$, and $h(q):=\mathbb{E}[\phi_G^2\,|\,\mathcal{Q}=q]>0$ for almost every $q$.
    \item \label{A5}\textbf{Linear thermodynamic stability.} There is $c>0$ such that $\langle\phi|u[\phi]\rangle\geq c\,\langle\phi|\phi\rangle$ for all $\phi$ with $\langle\phi|1\rangle=\langle\phi|\phi_G\rangle=0$.
\end{enumerate}
\begin{thm}[\cite{companion}]
\label{thm:app_generalized_kernel}
Assume \ref{A1}--\ref{A5}. Then, modulo the constant mode:
\begin{enumerate}[label=(\roman*)]
    \item $\ker W=\text{span}\{\phi_G\}$
    \item there is a unique $\phi_{\mathcal{Q}}$ with $u[\phi_{\mathcal{Q}}]=\mathcal{Q}+\text{const.}$ and $\langle\phi_{\mathcal{Q}}|\phi_G\rangle=0$, and it satisfies $W\phi_{\mathcal{Q}}=-\phi_G$
    \item $\ker W^2=\text{span}\{\phi_G,\phi_{\mathcal{Q}}\}$
    \item $\ker W^n=\ker W^2$ for all $n\geq 2$
\end{enumerate}
\end{thm}
We now verify the hypotheses for \eqref{eq:rotating_FPE_momenta}, with $P_*=P_{\mathrm{eq}}^{\Omega}$ and $\mathcal{Q}=L$. 

\textbf{(A1).} With $\mathcal{M}[P] = \beta \Gamma P$ and 
\begin{align}
    E^{\Omega}[P] = \int dz\, P\left(\frac{p_x^2}{2}+\frac{p_y^2}{2}+br^2-\Omega L\right) + \frac{1}{2}\left(a\bar{r}[P]^2-\bar{x}[P]^2-\bar{y}[P]^2\right)
\end{align}
one has $\delta E^{\Omega}/\delta P=\bar{H}^{\Omega}[P]$, and \eqref{eq:app_metriplectic} is exactly \eqref{eq:rotating_FPE_momenta}. Since $E^{\Omega}$ is a quadratic polynomial in $P$, it is smooth, and $\delta\bar{H}$ has rank three.

\textbf{(A2).} The flow of $L$ is the simultaneous rotation of $(x,y)$ and $(p_x,p_y)$, which is complete and leaves $E^{\Omega}$ invariant.

\textbf{(A3).} $\Gamma$ is independent of $P$, with $\Gamma_{ab}\partial_bL=0$ and \eqref{eq:kernel_of_Gamma} by construction. Write $p_r$ and $p_t$ for the radial and tangential components of the momentum, so that $L=rp_t$. (Here $p_t$ is the tangential Cartesian component, not the canonical momentum $p_\theta=L$ of Appendix \ref{sec:Appendix_free_energy_density}.) Every point of a level set $\{L=\ell\}$ with $\ell\neq0$ has $r>0$, and the level set is the graph of $p_t=\ell/r$ over $(r,\theta,p_r)\in(0,\infty)\times S^1\times\mathbb{R}$; it is therefore connected. The single level $\ell=0$ has probability zero.

\textbf{(A4).} From Section \ref{subsubsec:zero_modes_with_momenta}, $\phi_G=\{\bar{H}_{\mathrm{eq}}^{\Omega},L\}=\bar{x}_{\mathrm{eq}}\,y\neq 0$, since in the ordered phase, $\bar{x}_{\text{eq}}\neq 0$. Next, since $\kappa>0$, $P_{\mathrm{eq}}^{\Omega}$ has Gaussian tails, so $L,y\in L^2(\mu)$, while $\langle\partial_aL|\Gamma_{ab}\partial_bL\rangle=0$. Finally, on the level set $L=\ell$ the factor $e^{\beta\Omega\ell}$ is constant and cancels, and integrating out the momenta at fixed $p_t=\ell/r$ gives
\begin{align}
h(\ell)=\bar{x}_{\mathrm{eq}}^2\,\mathbb{E}[y^2\,|\,L=\ell]=\frac{\bar{x}_{\mathrm{eq}}}{\beta}\,\frac{\int_0^{\infty}dr\,r\,w_\ell(r)\,I_1(\beta\bar{x}_{\mathrm{eq}}r)}{\int_0^{\infty}dr\,w_\ell(r)\,I_0(\beta\bar{x}_{\mathrm{eq}}r)}>0, \quad w_\ell(r):=e^{-\beta\left(a\bar{r}_{\mathrm{eq}}r+br^2+\ell^2/2r^2\right)}
\end{align}

\textbf{(A5).} For $\langle\phi|1\rangle=\langle\phi|y\rangle=0$, $u[\phi]$ differs from $\phi$ only through the component of $\phi$ in $\text{span}\{\tilde{r},\tilde{x}\}$, where it acts as the matrix $B$ of Appendix \ref{subsubsec:Appendix_exact_kernel_SF}. The similarity argument of Appendix \ref{subsubsec:Appendix_spectrum_B} applies verbatim, so the eigenvalues $\lambda_\pm$ of $B$ are real. Therefore, \ref{A5} holds with $c=\min\{1,\lambda_-\}$ provided $\lambda_->0$. Since $a>0$, $\langle\phi|u[\phi]\rangle\geq\langle\phi|\phi\rangle$ whenever $\langle\phi|\tilde{x}\rangle=0$, so at most one of $\lambda_\pm$ can be nonpositive. From Appendix \ref{subsubsec:Appendix_detB_numerics}, $\det B=\lambda_+\lambda_->0$ which then implies $\lambda_\pm>0$.

Theorem \ref{thm:app_generalized_kernel} therefore applies with $\phi_G=\bar{x}_{\mathrm{eq}}\,y$ and $\phi_{\mathcal{Q}}=\phi_L$, since $u[\phi_L]=L + \text{const.}$ and $\langle\phi_L|y\rangle=0$. Finally, restoring the constant mode, we obtain
\begin{align}
 \text{gen}\ker W=\ker W^2=\text{span}\{1,y,\phi_L\}
\end{align}

\subsection{Hypocoercive spectral gap}
\label{subsec:Appendix_Spectral_Gap_momenta}
We now show that the generalized kernel contains the only gapless modes. In other words, there is $\lambda_{\mathrm{gap}}>0$ such that
\begin{align}
\label{eq:app_gap_statement}
    \sigma(W)\cap\{z\in\mathbb{C}:\Re z<\lambda_{\mathrm{gap}}\}=\{0\}
\end{align}
where $\sigma(W)$ denotes the spectrum of $W$. Two features of $W$ obstruct a direct coercivity estimate. First, the transport term is antisymmetric and drops out of $\Re\langle f|W|f\rangle$, so $W$ is non-normal. Second, the dissipative term is degenerate since, by \eqref{eq:kernel_of_Gamma}, it vanishes on the infinite-dimensional space of functions of $L$, which the finite-rank response $\delta\bar{H}$ cannot control. These modes are nevertheless not stationary, since the transport term maps them into directions that are dissipated,
\begin{align}
\label{eq:app_transport_of_FL}
    \{\bar{H}_{\mathrm{eq}}^{\Omega},F(L)\}=\bar{x}_{\mathrm{eq}}F'(L)\,y
\end{align}
This is the mechanism of hypocoercivity \cite{Villani2009,Dolbeault2015}. The companion paper \cite{companion} adapts it to the setting of this model and proves the spectral gap under four inputs \ref{H1}--\ref{H4} stated below. It also contains a complete verification of these inputs for the present model. Here we state the result and outline the verification, retaining the formulas that make each step recognizable.

From now on, $L_0^2(\mu)$ is complexified, 
\begin{align}
    \langle f|g\rangle = \beta \int d\mu \, \bar{f}g
\end{align}
Define the symmetric and antisymmetric parts of the linearized operator at frozen mean field (i.e. $\nabla \delta \bar{H}(\phi)=0$),
\begin{align}
\langle\Psi|\mathcal{S}|\phi\rangle:=\langle\partial_a\Psi|\Gamma_{ab}\partial_b\phi\rangle, \qquad \langle\Psi|\mathcal{J}|\phi\rangle:=-\langle\Psi|\{\bar{H}[P_*],\phi\}\rangle, \qquad W^{(0)}:=\mathcal{S}+\mathcal{J}
\end{align}
so that $\langle\Psi|W|\phi\rangle=\langle\Psi|W^{(0)}|u[\phi]\rangle$. Let $\Pi f:=\mathbb{E}[f\,|\,\mathcal{Q}]$ be the orthogonal projection onto functions of $\mathcal{Q}$, let $\mu_q$ be the conditional measure on the level set $\{\mathcal{Q}=q\}$, and let $\rho_{\mathcal{Q}}$ be the pushforward measure of $\mu$ under $\mathcal{Q}$. By \ref{A3}, $\mathcal{S}\Pi=0$, while $\mathcal{J}F(\mathcal{Q})=-F'(\mathcal{Q})\phi_G$, so that
\begin{align}
\label{eq:app_SL}
    \|\mathcal{J}\Pi F\|^2=\beta\int d\rho_{\mathcal{Q}}\,h\,|F'|^2, \qquad (\mathcal{J}\Pi)^{\dagger}(\mathcal{J}\Pi)F=\mathcal{L}_hF:=-\frac{1}{\rho_{\mathcal{Q}}}\left(h\rho_{\mathcal{Q}}F'\right)'
\end{align}
with $h$ as in \ref{A4}. Finally, let 
\begin{align}
\label{eq:app_definition_A}
    \mathcal{A}:=\left(\mathbb{I}+(\mathcal{J}\Pi)^{\dagger}(\mathcal{J}\Pi)\right)^{-1}(\mathcal{J}\Pi)^{\dagger}
\end{align}
We now state the hypotheses of Theorem \ref{thm:app_gap}:
\begin{enumerate}[label=(H\arabic*)]
    \item \label{H1}\textbf{Regularity.} $-W^{(0)}$ generates a contraction semigroup on $L_0^2$, and $C_c^\infty$ is a core on which $W^{(0)}=\mathcal{S}+\mathcal{J}$. The range of $\delta\bar{H}$, and functions $F(\mathcal{Q})$ with $F$ smooth and compactly supported, are limits of $C_c^\infty$ functions in the norm $\|f\|^2+\langle f|\mathcal{S}|f\rangle+\|\mathcal{J}f\|^2$; in particular they lie in the form domain of $\mathcal{S}$ and the domain of $\mathcal{J}$. $\Pi$ maps $C_c^\infty$ into the domain of $\mathcal{J}$.
    \item \label{H2}\textbf{Microscopic coercivity.} There is $\lambda_m>0$ such that $\langle f|\mathcal{S}|f\rangle\geq\lambda_m\|(\mathbb{I}-\Pi)f\|^2$. Equivalently, the conditional measures satisfy a uniform Poincar\'e inequality, $\text{Var}_{\mu_q}(f)\leq\lambda_m^{-1}\int d\mu_q\,\partial_a\bar{f}\,\Gamma_{ab}\,\partial_bf$ for almost every $q$.
    \item \label{H3}\textbf{Macroscopic coercivity.} There is $\lambda_M>0$ such that $\|\mathcal{J}\Pi f\|^2\geq\lambda_M\|\Pi f\|^2$ on $L_0^2$. By \eqref{eq:app_SL}, this holds with $\lambda_M=h_0/C_P$ if $h\geq h_0>0$ and $\rho_{\mathcal{Q}}$ satisfies a Poincar\'e inequality with constant $C_P$.
    \item \label{H4}\textbf{Coupling bounds.} $\mathcal{A}\mathcal{J}(\mathbb{I}-\Pi)$ and $\mathcal{A}\mathcal{S}^{1/2}$ are bounded.
\end{enumerate}

\begin{thm}[\cite{companion}]
\label{thm:app_gap}
Assume \ref{A1}--\ref{A5}, and let the linearization have the regularity, coercivity and coupling properties \ref{H1}--\ref{H4}. Then there is $\lambda_{\mathrm{gap}}>0$ such that, on $L_0^2(\mu)$ (the mean-zero subspace of $L^2(\mu)$), \eqref{eq:app_gap_statement} holds and $0$ is an isolated eigenvalue with generalized eigenspace $\text{span}\{\phi_G,\phi_{\mathcal{Q}}\}$. If $\Pi_0$ denotes the associated Riesz projection, then $\|e^{-tW}(\mathbb{I}-\Pi_0)\|\leq C e^{-\lambda' t}$ for some $\lambda'>0$ and all $t\geq0$.
\end{thm}

In the twisted energy $\Re\langle f|(\mathbb{I}+\epsilon(\mathcal{A}+\mathcal{A}^{\dagger}))W^{(0)}f\rangle$, \ref{H2} damps the microscopic part of $f$ directly, \ref{H3} damps the macroscopic part at order $\epsilon$ through the twist, and \ref{H4} controls the two cross terms. Finally, \ref{H1} is what allows the estimate, done on $C_c^\infty$, to be extended to the whole domain, and what allows the mean-field correction $W-W^{(0)}=W^{(0)}\delta\bar{H}$ to be treated as a finite-rank perturbation.

\subsubsection{Outline of the verification}

Following the previous subsection \ref{subsubsec:genker_W}, $P_* = P_{\text{eq}}^{\Omega}$, $\mathcal{Q}=L$, and $\phi_G =\bar{x}_{\text{eq}}y$. We write $g_{\min}$ and $g_{\max}$ for the smallest and largest eigenvalues of $G$, and recall from \eqref{eq:mobility_matrix} that $\Gamma=Q^TGQ$, where the rows of $Q$ are the vector fields
\begin{align}
w_1=p_a\partial_{x_a}+x_a\partial_{p_a}, \qquad w_2=x_a\partial_{x_a}-p_a\partial_{p_a}, \qquad w_3=x\partial_y-y\partial_x+p_x\partial_{p_y}-p_y\partial_{p_x}
\end{align}
These are mutually orthogonal, orthogonal to $\nabla L$, and all four have norm $|z|$, so
\begin{align}
\label{eq:app_Gamma_bounds}
g_{\min}\sum_{i=1}^3|w_if|^2\leq\partial_a\bar{f}\,\Gamma_{ab}\,\partial_bf\leq g_{\max}\sum_{i=1}^3|w_if|^2=g_{\max}|z|^2|\nabla_{\Sigma}f|^2
\end{align}
where $\nabla_\Sigma$ is the gradient along the level set of $L$ and each $w_i$ is divergence free and tangent to the level sets $\Sigma_\ell=\{L=\ell\}$. As discussed in the section above, for $\ell\neq0$ on $\Sigma_\ell$ we have that $r>0$, and $(r,\theta,p_r)$, with $p_t=\ell/r$, are global coordinates in which
\begin{align}
\label{eq:app_fiber_chart}
    d\mu_\ell\propto \text{exp}\left( -\beta\left(\frac{p_r^2}{2}+\frac{\ell^2}{2r^2}+br^2+a\bar{r}_{\mathrm{eq}}r-\bar{x}_{\mathrm{eq}}r\cos\theta\right)\right)dr\,d\theta\,dp_r
\end{align}
In these coordinates, 
\begin{align}
\label{eq:app_fields_chart}
w_1=p_r\partial_r+\frac{\ell}{r^2}\partial_\theta+\left(r+\frac{\ell^2}{r^3}\right)\partial_{p_r}, \qquad w_2=r\partial_r-p_r\partial_{p_r}, \qquad w_3=\partial_\theta
\end{align}

Before we begin outlining the verification of each hypothesis, we make the following remark. The interaction term $a\bar{r}_{\text{eq}}r$ has a conical singularity on the axis $r=0$, which appears in both $\bar{H}_{\mathrm{eq}}^{\Omega}$ and $P_{\mathrm{eq}}^{\Omega}$. It plays no role in the physics, but it is an obstruction to the regularity input \ref{H1}. For the analysis we therefore round it off, replacing $r$ by $r_\delta:=\sqrt{r^2+\delta^2}$ in the interaction with $\delta$ small. Nothing in Sections \ref{subsubsec:zero_modes_with_momenta}--\ref{subsubsec:genker_W} depends on the cusp, the inputs \ref{H2}--\ref{H4} do not see the axis, and the quantities from \ref{H1} that enter the constants of the gap involve first derivatives of $r_\delta$ only and converge as $\delta\to0$. 

\textbf{(H1).} Three facts are needed to verify \ref{H1}: the frozen linear equation $\partial_t\phi = -W^{(0)}\phi$ is well-posed on $L_0^2(\mu)$ with non-increasing norm, that $C_c^\infty$ is a core for $W^{(0)}$, and that the functions $r_{\delta},x,y$ and $F(L)$ are limits of $C_c^\infty$ functions in the norm $\|f\|^2+\langle f|\mathcal{S}|f\rangle+\|\mathcal{J}f\|^2$. 

Since $W^{(0)}$ is a differential expression, to make it an operator on $L^2(\mu)$, we start by defining $\dom W^{(0)}= C_c^{\infty}(\mathbb{R}^4)$. However, this is too restrictive as solutions are not usually compactly supported, so we upgrade it to the closure. In other words, $f$ is in the closed domain if there are test functions $f_n\to f$ in $L^2(\mu)$ with $W^{(0)}f_n$ converging in $L^2(\mu)$. Then $W^{(0)}f$ is defined as that limit. This is well defined whenever the operator is closable. Since $\Re\langle \phi| W^{(0)}|\phi\rangle = \langle \phi|\mathcal{S}|\phi\rangle\geq 0 $, we have that $W^{(0)}$ is accretive. A densely defined accretive operator is closable, and its closure is again accretive. Therefore, it is well defined, hence from now on $W^{(0)}$ means this closure and $\dom W^{(0)}$ its domain. Since we construct $W^{(0)}$ as the closure from $C_c^{\infty}(\mathbb{R}^4)$, the test functions are a core by definition. The following content is then that this closure is the right operator, i.e. that it is maximally accretive. 

$W^{(0)}$ is maximally accretive, if additionally, the equation $W^{(0)}f+\lambda f=g$ has a solution for every $g$ and $\lambda >0$. By the Lumer--Phillips theorem \cite{LumerPhillips1961}, maximally accretive is exactly the condition for $-W^{(0)}$ to generate well-posed contracting dynamics. Since $W^{(0)}$ is a closed accretive operator, for $\lambda >0$, the range of $W^{(0)}+\lambda$ is a closed subspace. Therefore, maximal accretivity reduces to showing that a weak solution $g\in L^2(\mu)$ of the adjoint equation $(\mathcal{S}-\mathcal{J}+\lambda)g=0$ vanishes. The operator $\mathcal{S}-\mathcal{J}$ is of H\"ormander type, $-\sum_kY_k^2+Y_0$ with $Y_k$ a basis of $\text{span}\{w_i\}$, and the bracket condition holds everywhere. Indeed, off the origin the $w_i$ span the complement of $\nabla L$, and the identities
\begin{align}
    \{L,\bar{H}_{\mathrm{eq}}^{\Omega}\}=-\bar{x}_{\mathrm{eq}}\,y, \qquad w_3y=x, \qquad w_1y=p_y, \qquad w_1w_3y=p_x
\end{align}
show that some commutator of the $w_i$ with the transport field has a nonzero component along $\nabla L$ at every point. At the origin a separate three-bracket computation applies. By H\"ormander's theorem \cite{Hormander1967} the weak solution is smooth, and a cut-off argument, using only $g\in L^2(\mu)$ and the growth bounds \eqref{eq:app_Gamma_bounds}, forces $g=0$. 

The remaining parts of \ref{H1} are approximation statements. Since $P_{\mathrm{eq}}^{\Omega}$ has Gaussian tails, the functions $r_{\delta},x,y$, and smooth compactly supported functions $F(L)$ can be approximated by spatial cutoffs in the norm appearing in \ref{H1}. Finally, for $f\in C_c^\infty$, $\Pi f=\mathbb{E}[f|L]$ is smooth for $\ell\neq0$, and its derivative has only a logarithmic singularity as $\ell\to0$. This singularity is square integrable against $h(\ell)\rho_L(\ell)d\ell$, so $\mathcal{J}\Pi f=-(\Pi f)'(L)\phi_G\in L^2$. These facts give the regularity input for the rounded problem and the technical details are deferred to \cite{companion}.

\textbf{(H2).} Microscopic coercivity says that dissipation damps every perturbation which is not a function of $L$, with a rate that is uniform over the value of $L$. Since the dissipation is tangent to the level sets of $L$, this is naturally a Poincar\'e inequality on each fiber $\Sigma_\ell$. Disintegrating the equilibrium measure $\mu$ over the level sets, 
\begin{align}
    d\mu = d\mu_{\ell}(z)d\rho_L(\ell), \qquad \mu_{\ell}:= \mu(\cdot|L=\ell)
\end{align}
where $\mu_\ell$ is the conditional measure on the fiber $\Sigma_\ell$ and $\rho_L$ is the pushforward measure of $\mu$ under $L$. Then, since $\Pi f=\mathbb{E}[f|L]$ is the conditional mean, the two sides of \ref{H2} are
\begin{align}
    \|(\mathbb{I}-\Pi)f\|^2=\beta\int d\ell\,\rho_L(\ell)\,\text{Var}_{\mu_\ell}(f), \qquad
    \langle f|\mathcal{S}|f\rangle=\beta\int d\ell\,\rho_L(\ell)\int d\mu_\ell\,\partial_a\bar{f}\,\Gamma_{ab}\,\partial_bf
\end{align}
By \eqref{eq:app_Gamma_bounds}, the integrand of the second is at least $g_{\min}\sum_i|w_if|^2$. Let $\lambda_\ell$ be the spectral gap of the fiber, i.e. the best constant in the Poincar\'e inequality
\begin{align}
    \int d\mu_\ell\sum_{i=1}^3|w_if|^2\ \geq\ \lambda_\ell\,\text{Var}_{\mu_\ell}(f)
\end{align}
Integrating this against $\rho_L$ gives microscopic coercivity with $\lambda_m=g_{\min}\inf_{\ell\neq0}\lambda_\ell$. For every fixed $\ell\neq0$, this gap is positive since the three fields in \eqref{eq:app_fields_chart} span the tangent space of the connected fiber, and the effective potential in \eqref{eq:app_fiber_chart} confines in the $r$, $p_r$, and angular directions. The only nontrivial point is uniformity as $\ell\to0$ and $|\ell|\to\infty$. 

As $\ell\to0$, the fiber develops a narrow neck near the origin. The key observation is that the dissipative geometry is scale invariant there. Let $\mathcal{L}^{\mathrm{fib}}_\ell$ denote the generator of the diffusion on the fiber, that is, the operator whose Dirichlet form is
\begin{align}
  -\int d\mu_{\ell}\,f \mathcal{L}^{\mathrm{fib}}_\ell f= \int d\mu_\ell\sum_{ij}G_{ij}(w_if)(w_jf)
\end{align}
so that $\mathcal{L}^{\mathrm{fib}}_\ell V$ is the expected rate of change of $V$ along the fiber dynamics. With $\upsilon=|z|^2/2$, a Lyapunov function of the form $V=\upsilon^{-\alpha}$ satisfies, for sufficiently small $\alpha>0$,
\begin{align}
\mathcal{L}^{\mathrm{fib}}_\ell V\leq-cV 
\end{align}
inside a fixed neighborhood of the origin, with $c>0$ independent of $\ell$. Thus the narrowing neck cannot support an increasingly slow mode. Away from the neck, the confining potential provides the corresponding Lyapunov estimate \cite{Khasminskii2012, BakryCattiauxGuillin2008} at large $|z|$, and a local Poincar\'e inequality controls the remaining compact region. For large $|\ell|$, rescaling $r$ and $p_r$ by $|\ell|^{1/2}$ makes the structure transparent. The radial measure concentrates near the minimum of
\begin{align}
\frac{1}{2r^2}+br^2,
\end{align}
while the angular distribution is a von Mises weight. The transverse directions become increasingly stiff, whereas the angular Poincar\'e constant remains of order one. Consequently the fiber gap does not close at large $|\ell|$. Making the Lyapunov, compact-region, and large-$|\ell|$ estimates uniform is the technical content of the microscopic-coercivity proof and is deferred to \cite{companion}. The important physical point is that neither the pinching of the small-$|\ell|$ fibers nor the growth of their typical radius at large $|\ell|$ produces an additional soft mode.

\textbf{(H3).} Macroscopic coercivity controls the complementary sector, i.e. functions of $L$. These functions are invisible to $\mathcal{S}$, so transport must move them into dissipative directions at a rate bounded away from zero. By \eqref{eq:app_SL}, it is enough to show that $h(\ell)=\mathbb{E}[\phi_G^2|L=\ell]$ has a positive lower bound and that the distribution $\rho_L$ satisfies a one-dimensional Poincar\'e inequality.

Both quantities can be computed explicitly from \eqref{eq:app_fiber_chart}. Integrating over $p_r$ and $\theta$ gives
\begin{align}
\label{eq:app_rho_h}
\rho_L(\ell)\propto e^{\beta\Omega\ell}\int_0^\infty dr\,w_\ell(r)I_0(\beta\bar{x}_{\mathrm{eq}}r),\qquad
h(\ell)=\frac{\bar{x}_{\mathrm{eq}}}{\beta}\frac{\int_0^\infty dr\,r\,w_\ell(r)I_1(\beta\bar{x}_{\mathrm{eq}}r)}{\int_0^\infty dr\,w_\ell(r)I_0(\beta\bar{x}_{\mathrm{eq}}r)},
\end{align}
where
\begin{align}
w_\ell(r):=\exp\left[-\beta\left(a\bar{r}_{\mathrm{eq}}r+br^2+\frac{\ell^2}{2r^2}\right)\right].
\end{align}
The second expression is the average of an increasing function of $r$. Increasing $|\ell|$ strengthens the centrifugal barrier and shifts the radial measure toward larger $r$, so $h$ is even and non-decreasing in $|\ell|$. Hence
\begin{align}
h(\ell)\geq h(0)=:h_0>0.
\end{align}

It remains to show that $\rho_L$ satisfies a one-dimensional Poincar\'e inequality. Differentiating the first expression in \eqref{eq:app_rho_h} gives
\begin{align}
-\partial_\ell\log\rho_L=\beta\left(\ell\left\langle \frac{1}{r^2}\right\rangle_\ell-\Omega\right).
\end{align}
At large $|\ell|$, the radial measure is concentrated where the centrifugal term balances the harmonic trap, so
\begin{align}
\ell\left\langle \frac{1}{r^2}\right\rangle_\ell\longrightarrow\pm\sqrt{2b}, \qquad \ell\to\pm\infty.
\end{align}
Therefore $\rho_L$ has exponential tails precisely when
\begin{align}
\sqrt{2b}>|\Omega| \quad\Longleftrightarrow\quad \kappa=b-\frac{\Omega^2}{2}>0,
\end{align}
which is exactly the confinement condition for the rotating equilibrium. Since $\rho_L$ is continuous and strictly positive on compact intervals, it remains only to control the tails. The exponential asymptotics above imply that the one-dimensional Muckenhoupt criterion \cite{Muckenhoupt1972} is satisfied on both tails. Hence $\rho_L$ obeys a Poincar\'e inequality, 
\begin{align}
\text{Var}_{\rho_L}(F)\leq C_P\int d\ell\,\rho_L(\ell)|F'(\ell)|^2,
\end{align}
and \eqref{eq:app_SL} gives \ref{H3} with
\begin{align}
\lambda_M=\frac{h_0}{C_P}>0
\end{align}
This is also the physical content of macroscopic coercivity. In the ordered phase the Goldstone mode has nonzero weight on every angular-momentum sector, while confinement prevents the probability distribution of $L$ from developing arbitrarily broad tails.

\textbf{(H4).} The coupling bounds control the cross terms generated by the hypocoercive twist. Since boundedness is equivalent to boundedness of the adjoint, both operators in \ref{H4} reduce to a one-dimensional resolvent estimate. Let $F$ be a mean-zero function of $L$ and define
\begin{align}
\varphi=(\mathbb{I}+\mathcal{L}_h)^{-1}F,
\end{align}
so that $\varphi+\mathcal{L}_h\varphi=F$. Since $\varphi$ is a function of $L$,
\begin{align}
\mathcal{J}\varphi=-\bar{x}_{\mathrm{eq}}\varphi'(L)y,\qquad
\mathcal{J}^2\varphi=\bar{x}_{\mathrm{eq}}\left(\bar{x}_{\mathrm{eq}}\varphi''(L)y^2-\varphi'(L)(p_y-\Omega x)\right)
\end{align}
Since $\Gamma\nabla L=0$, derivatives of $\varphi'(L)$ drop out of the dissipative form. After conditioning on $L$, the two adjoint bounds are therefore controlled by
\begin{align}
\label{eq:app_B4_reduction}
\|\mathcal{S}^{1/2}\mathcal{J}\varphi\|^2+\|(\mathbb{I}-\Pi)\mathcal{J}^2\varphi\|^2
\leq C\beta\int d\ell\,\rho_L(\ell)h(\ell)^2\left(|\varphi'|^2+|\varphi''|^2\right).
\end{align}
The same large-$|\ell|$ concentration used above gives
\begin{align}
\mathbb{E}[|z|^2|L=\ell]\lesssim1+|\ell|, \qquad h(\ell)^2\gtrsim1+|\ell|,
\end{align}
therefore these conditional moments are controlled by the same weight $h^2$. The problem is then reduced to the one-dimensional estimate
\begin{align}
\int d\ell\,\rho_Lh^2\left(|\varphi'|^2+|\varphi''|^2\right)\leq C\|F\|^2.
\end{align}
To see why this estimate holds, write
\begin{align}
\mathcal{L}_h\varphi=-h\varphi''+\mathfrak{b}\varphi', \qquad \mathfrak{b}:=-\frac{(h\rho_L)'}{\rho_L}.
\end{align}
Expanding $\|F\|^2=\|(\mathbb{I}+\mathcal{L}_h)\varphi\|^2$ and integrating by parts gives
\begin{align}
&\|F\|^2=\|\varphi\|^2+2\beta\int d\ell\,\rho_Lh|\varphi'|^2+\|\mathcal{L}_h\varphi\|^2,\\
&\|\mathcal{L}_h\varphi\|^2=\beta\int d\ell\,\rho_L\left(h^2|\varphi''|^2+h\mathfrak{b}'|\varphi'|^2\right)
\end{align}
The only apparent difficulty is the cusp of $\rho_L$ at $\ell=0$. However, the combination that enters here is $h\rho_L$. Its radial integral contains two additional powers of $r$, so its first nonanalytic term is proportional to $|\ell|^3$ rather than $|\ell|$. Thus $h\rho_L$ is twice continuously differentiable at the origin and $\mathfrak{b}'$ remains bounded. At large $|\ell|$, the same radial concentration implies that $\mathfrak{b}'$ is bounded and tends to zero. The second identity therefore controls the weighted second derivative, while a Hardy inequality using the exponential tails of $h^2\rho_L$ controls the weighted first derivative. This gives the two bounded operators required in \ref{H4}. The detailed uniform moment and tail estimates are deferred to \cite{companion}.

The complete mathematical proofs, including the uniform estimates suppressed here, are deferred to the companion paper \cite{companion}. With these inputs, Theorem \ref{thm:app_gap} applies. Restoring the constant mode, $0$ is an isolated eigenvalue with generalized eigenspace
\begin{align}
\text{gen}\ker W=\ker W^2=\text{span}\{1,y,\phi_L\},
\end{align}
and, if $\Pi_0$ denotes the corresponding Riesz projection,
\begin{align}
\|e^{-tW}(\mathbb{I}-\Pi_0)\|\leq Ce^{-\lambda't}
\end{align}
for some $\lambda'>0$. Thus the Goldstone mode, the mode associated with the conserved-charge $L$, and normalization are the only gapless modes and every other perturbation relaxes exponentially. This statement is for fixed parameters in the interior of the homogeneous ordered region. The gap need not remain bounded away from zero as the ordering transition or the collective threshold $\Omega_1$ is approached, where $\det B\to0$.

\end{appendix}

\phantomsection
\section*{References}
\addcontentsline{toc}{section}{References}
\makeatletter
\let\bibsection\relax
\makeatother
\bibliography{references}
\end{document}